\documentclass[11pt,a4paper]{article}
\usepackage{jheppub}
\usepackage{amsmath,amsfonts,epsfig,color}
\usepackage{amssymb}
\usepackage[english, USenglish]{babel}
\usepackage{epsfig,psfrag}

\usepackage{hyperref}

\csname @addtoreset\endcsname{equation}{section}

\makeatletter
\def\timenow{\@tempcnta\time
  \@tempcntb\@tempcnta
  \divide\@tempcntb60
  \ifnum10>\@tempcntb0\fi\number\@tempcntb
  \multiply\@tempcntb60
  \advance\@tempcnta-\@tempcntb
  :\ifnum10>\@tempcnta0\fi\number\@tempcnta}
\makeatother

\def\oonoo#1#2#3{\vbox{\ialign{##\crcr
	\hfil\hfil\hfil{$#3{#1}$}\hfil\crcr\noalign{\kern1pt\nointerlineskip}
	$#3{#2}$\crcr}}}
\def\oon#1#2{\mathchoice{\oonoo{#1}{#2}{\displaystyle}}
	{\oonoo{#1}{#2}{\textstyle}}{\oonoo{#1}{#2}{\scriptstyle}}
	{\oonoo{#1}{#2}{\scriptscriptstyle}}}

\def\dt#1{\oon{\hbox{\bf .}}{#1}}  
\def\ddt#1{\oon{\hbox{\bf .\kern-1pt.}}#1}
\def\slap#1#2{\setbox0=\hbox{$#1{#2}$}
	#2\kern-\wd0{\hfuzz=1pt\hbox to\wd0{\hfil$#1{/}$\hfil}}}

\title{$\cN=4$ super-Yang-Mills in LHC superspace\\  Part II: Non-chiral correlation functions\\ of the stress-tensor multiplet}

\author[a]{Dmitry Chicherin}
\author[a,b]{and Emery Sokatchev}

\affiliation[a]{LAPTH $^{*}$\note{Laboratoire d'Annecy-le-Vieux de Physique Th\'{e}orique, UMR 5108}, Universit\'{e} de Savoie, CNRS, B.P. 110,  F-74941 Annecy-le-Vieux, France}
\affiliation[b]{Theoretical Physics Department, CERN, CH -1211, Geneva 23, Switzerland}

\abstract{
We study the multipoint super-correlation functions of the full non-chiral 
stress-tensor multiplet in $\cN=4$ super-Yang-Mills theory in the Born approximation.
We derive effective supergraph Feynman rules for them. Surprisingly, 
the Feynman rules for the non-chiral correlators are obtained from those for the chiral correlators by a simple Grassmann shift of the space-time variables.
We rely on the formulation of the theory in Lorentz harmonic chiral (LHC) 
superspace elaborated in the twin paper {arXiv:1601.06803}. In this approach only the chiral half of the supersymmetry is manifest. The other half is realized by nonlinear and nonlocal transformations of the LHC superfields. However, at Born level only the simple linear part of the transformations is relevant. It corresponds to effectively working in the self-dual sector of the theory.
Our method is also  applicable to a wider class of supermultiplets like all the half-BPS operators and the Konishi multiplet.}

\emailAdd{chicherin@uni-mainz.de}
\emailAdd{emeri.sokatchev@cern.ch}

\keywords{Supersymmetric Gauge Theory, Superspaces, Extended Supersymmetry, Chern-Simons Theories}

\arxivnumber{1601.06804}

\begin{document}
\renewcommand{\thefootnote}{\fnsymbol{footnote}}


\newcommand{\norm}[1]{{\protect\normalsize{#1}}}
\newcommand{\p}[1]{(\ref{#1})}
\newcommand{\half}{{\ts \frac{1}{2}}}
\newcommand \vev [1] {\langle{#1}\rangle}
\newcommand \ket [1] {|{#1}\rangle}
\newcommand \bra [1] {\langle {#1}|}

\newcommand{\cM}{{\cal M}} 
\newcommand{\cR}{{\cal R}} 
\newcommand{\cS}{{\cal S}} 
\newcommand{\cK}{{\cal K}}
\newcommand{\cL}{{\cal L}} 
\newcommand{\cF}{{\cal F}}
\newcommand{\cN}{{\cal N}}
\newcommand{\cA}{{\cal A}}
\newcommand{\cB}{{\cal B}}
\newcommand{\cG}{{\cal G}}
\newcommand{\cO}{{\cal O}}
\newcommand{\cY}{{\cal Y}}
\newcommand{\cX}{{\cal X}}
\newcommand{\cT}{{\cal T}}
\newcommand{\cW}{{\cal W}}
\newcommand{\cP}{{\cal P}}
\newcommand{\cQ}{{\cal Q}}
\newcommand{\nt}{\notag\\} 
\newcommand{\pa}{\partial}
\newcommand{\ep}{\epsilon}
\newcommand{\bep}{\bar\epsilon}
\renewcommand{\a}{\alpha}
\renewcommand{\b}{\beta}
\newcommand{\g}{\gamma}
\newcommand{\s}{\sigma}
\newcommand{\la}{\lambda}
\newcommand{\da}{{\dot\alpha}}
\newcommand{\db}{{\dot\beta}}
\newcommand{\dg}{{\dot\gamma}}
\newcommand{\dd}{{\dot\delta}}
\newcommand{\q}{\theta}
\newcommand{\bq}{\bar\theta}
\newcommand{\bQ}{\bar Q}
\newcommand{\tx}{\tilde{x}}
\newcommand{\tr}{\mbox{tr}}
\newcommand{\+}{{\dt+}}
\renewcommand{\-}{{\dt-}}

\maketitle
\flushbottom

\setcounter{page}{1}\setcounter{footnote}{0}
\renewcommand{\thefootnote}{\arabic{footnote}}


\section{Introduction}

In this paper we study the multipoint super-correlation functions of the full stress-tensor supermultiplet $T(x,\q,\bq)$ in $\cN= 4$ super-Yang-Mills (SYM) theory. This multiplet is of special interest since it comprises all the conserved currents of the theory. Its chiral truncation $\cT = T(\bq = 0)$ contains in particular the protected scalar half-BPS operator $\cO_{\bf 20'}$ and the $\cN= 4$ SYM Lagrangian.

The chiral correlators of $\cT$ have been intensively studied in the past.
Their (super)-light-cone limits reproduce all the tree-level scattering superamplitudes in $\cN=4$ SYM, as well as the  integrands of their loop corrections to all orders in perturbation theory \cite{Alday:2010zy,Eden:2010zz,Eden:2011yp,Eden:2011ku,Adamo:2011dq}. The native superconformal symmetry of the correlators, inherited from the action, is translated into the hidden dual super-conformal symmetry of the amplitudes \cite{Drummond:2008vq,Berkovits:2008ic,Beisert:2008iq}. 

The multi-point chiral supercorrelators  of $\cT$ in the Born-level approximation have been constructed in \cite{Chicherin:2014uca}. There a twistor diagram technique has been developed relying on the twistor approach to $\cN = 4$ SYM \cite{Mason:2005zm,Boels:2006ir,Adamo:2011cb}. The correlators are built from a special class of nilpotent off-shell superconformal R-invariants. In this approach no space-time integrations are involved, so the Born-level correlators are manifestly rational functions.  On shell the nilpotent building blocks are identified with the familiar R-invariants of the superamplitude. Thus,  the amplitude--correlator duality becomes manifest. 

An independent approach to the chiral correlators of $\cT$ exploits only their superconformal symmetry and basic analytic properties. This has led to an explicit result for the NMHV-like six-point correlator  \cite{Chicherin:2015bza}.

Much less is known about the full non-chiral correlators of $T$. The stress-energy tensor $T_{\mu \nu}(x)$, which gives the name to the multiplet, appears in the non-chiral sector. The two- and three-point functions of $T$ are protected \cite{Penati:1999ba,Penati:2000zv,D'Hoker:1998tz,Lee:1998bxa,Howe:1998zi} and so can be easily constructed as free-field theory objects. The four-point function is the first example of a correlator with non-trivial dynamical content. Superconformal symmetry imposes severe restrictions on it \cite{Eden:2000bk,Dolan:2001tt}, making it possible to reconstruct the entire super-correlator starting from its bottom component $\vev{\cO_{\bf 20'}\cO_{\bf 20'}\cO_{\bf 20'}\cO_{\bf 20'}}$ \cite{Drummond:2006by,Belitsky:2014zha,Korchemsky:2015ssa}. From this non-chiral supercorrelator one can extract many interesting components with conserved currents and Lagrangians.

The stress-tensor supermultiplet is half-BPS. This property is most conveniently formulated in harmonic/analytic superspace as Grassmann analyticity \cite{Galperin:1984av,Galperin:1984bu,Hartwell:1994rp,Heslop:2003xu}. This superspace involves auxiliary complex coordinates $y^{a}_{a'}$ (with $a,a'=1,2$), which parametrize the coset $SU(4)/(SU(2)\times SU(2)'\times U(1)) \sim GL(4)/(\text{Borel subgroup})$ . The indices $a,a'$ together form the R-symmetry index $A=1,\ldots,4$ of the odd variables $\q^A$ and
$\bq_A$. The stress-tensor supermultiplet is then defined as a Grassmann analytic superfield $T(x,\q_+,\bq_+,y)$, depending only on  half of the odd variables,  $\q^{a}_{+} = \q^{a} + y^a_{b'} \q^{b'}$ and $\bq_{+a'} = \bq_{a'} - y^{b}_{a'} \bq_b\,$. 

In this paper we propose a generalization of the chiral diagrammatic technique from \cite{Chicherin:2014uca} to the full (non-chiral) $n-$point supercorrelators of $T$ in the Born approximation, i.e. at the lowest perturbative order. They are rational functions of the space-time variables $x_i$ with $i=1,\ldots,n$. The R-symmetry $SU(4)$ imposes restrictions on the expansion of the supercorrelators in the Grassmann variables $\q_i, \bq_i$.
The difference between the numbers of $\q$ and $\bq$ for each term in the expansion must be  a multiple of four, $4p$. We call this number the Grassmann degree of the correlator. In particular, at the Grassmann level $(\q)^{4p}$ we find $p$ copies of the Lagrangian. According to the Lagrangian insertion procedure,  this is the integrand of the $p-$loop correction to the correlator with $n-p$ points. So, if we can construct all the Born-level correlators, then in principle we know all the loop corrections. For a component of the correlator with Grassmann degree $4p$, the Born-level 
correlator is of order $O(g^{2p})$ in the gauge coupling constant $g$. This perturbative order corresponds to the number of Lagrangian insertions.

\begin{figure}[!h]
\centerline {
\includegraphics[height = 2.5 cm]{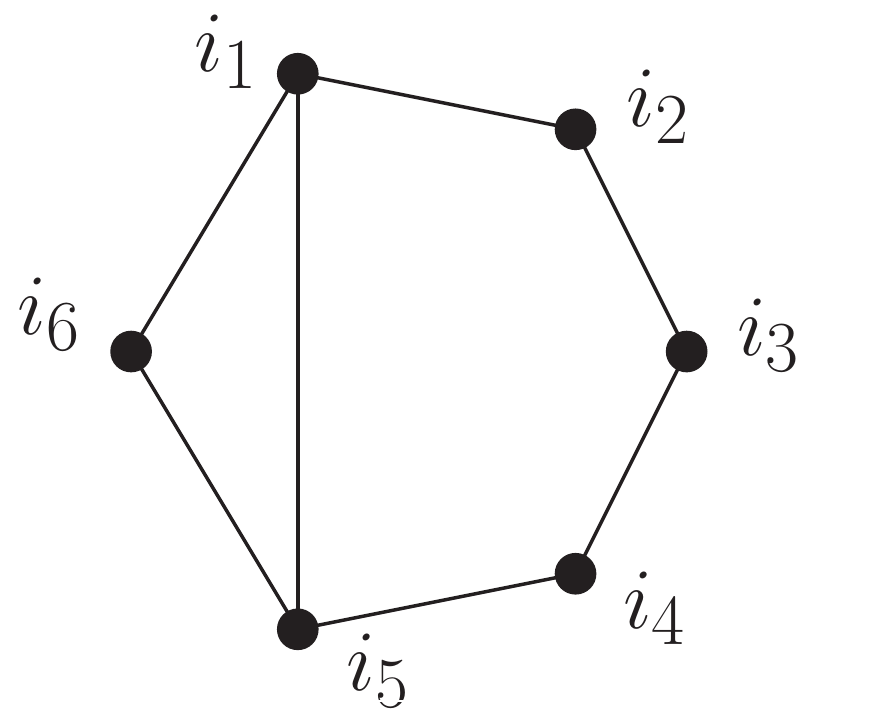} \quad
    \includegraphics[height = 2.5 cm]{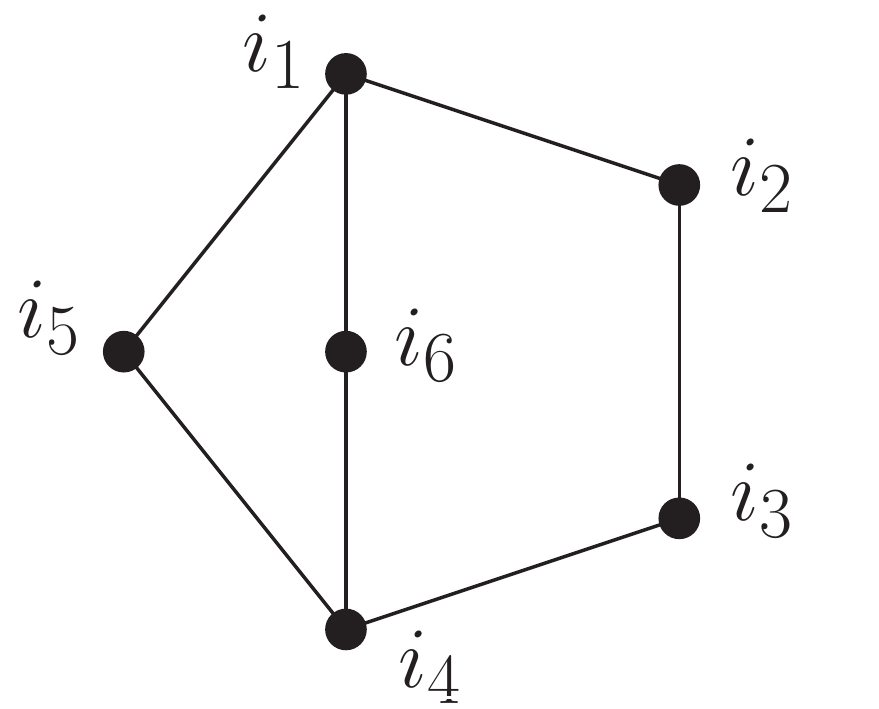} \quad \includegraphics[height = 2.5 cm]{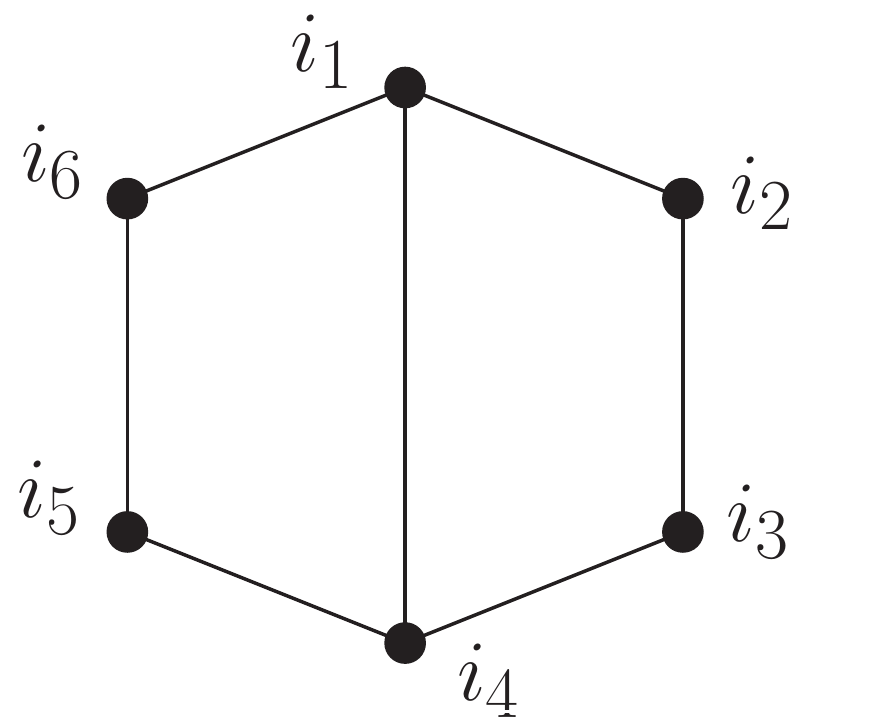} 
		\quad \includegraphics[height = 2.5 cm]{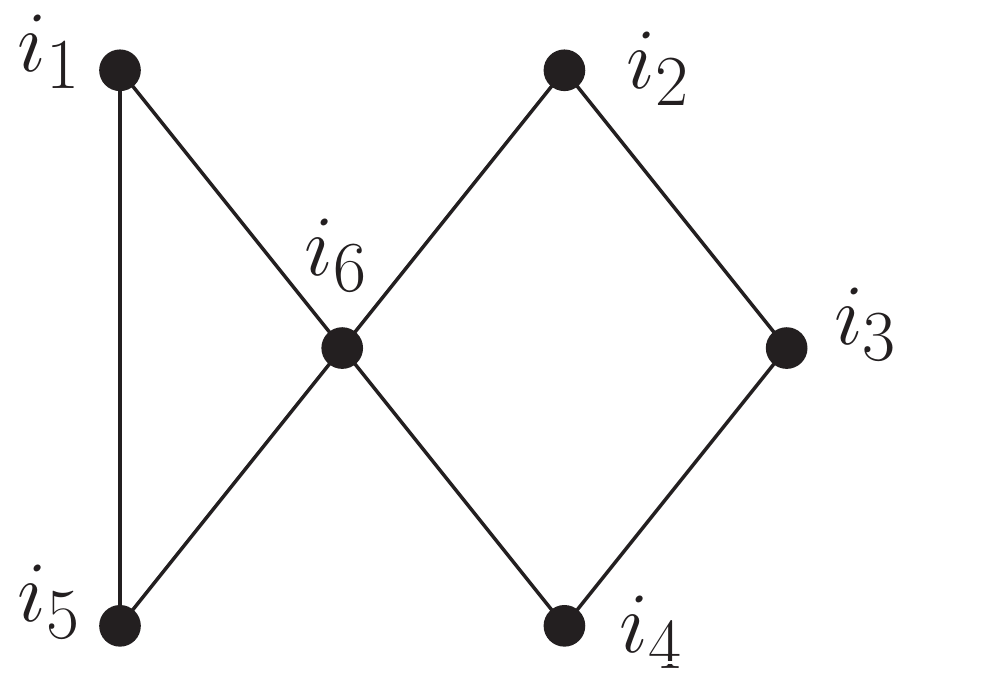} }
\caption{Supergraphs contributing to the six-point Born-level supercorrelator at order $O(g^2)$.
All nonequivalent permutations of the external points $i_1,\ldots, i_6 \in S_6$ are to be taken into account (360, 180, 180 and 360 permutations, respectively).}		
\label{NMHV6fig}
\end{figure}

Our main result is surprisingly simple. We find very little difference between the full non-chiral and the chiral cases. 
The off-shell chiral nilpotent invariants need only a slight modification.
Let us explain how this works on the example of the six-point Born-level supercorrelator $\langle T(1) \cdots T(6)\rangle_{g^2}$  of Grassmann degree 4 at order $O(g^2)$.  It can be evaluated by means of the following \emph{effective} Feynman rules. They do not follow directly from the $\cN=4$ SYM action, but are the result of partial simplifications of the original ones. 
\begin{itemize}
\item We draw (connected) vacuum graphs where the vertices correspond to the $T$'s in the correlator (six in our example). The vertices are of arbitrary valence $\geq 2$. They are connected by edges, see Fig.~\ref{NMHV6fig}.
To a vertex of valence $p$ we assign the coupling $g^{p-2}$. So, in our example we are allowed to have either one 4-valent vertex or
a pair of 3-valent vertices.
\item The scalar supersymmetrized propagator $y_{ij}^2 / \hat x_{ij}^2 $ corresponds to an edge connecting vertices $i$ and $j$, where  (we use the two-component notation $x^{\da\a} =  (\sigma_\mu)^{\da\a} x^\mu$)
\begin{align*}
\hat x_{ij} \equiv x_{ij} + \q_{ij+} y_{ij}^{-1} \bq_{ij+}\,, \,\,\,\,\, \q_{ij+} \equiv \q_{i+} - \q_{j+} \,, 
\,\,\,\,\, \bq_{ij+} \equiv \bq_{i+} - \bq_{j+} \,, \,\,\,\,\, y^2_{ij} \equiv \det(y_i - y_j)\,.
\end{align*}
\item Each 3-valent vertex gives rise to the 4-point superinvariant
\begin{align*}
\hat{R}(j_1;j_2,j_3,j_4)= - \frac{\delta^2\Big((\hat x^{\+}_{j_1 j_3} \hat x^{\+}_{j_1 j_4})(\hat x^{\+}_{j_1 j_2} \q_{j_1 j_2+})(y_{j_1 j_2})^{-1} + {\rm cycle}(j_2,j_3,j_4) \Big)}{(\hat x^{\+}_{j_1 j_2} \hat x^{\+}_{j_1 j_3})(\hat x^{\+}_{j_1 j_3} \hat x^{\+}_{j_1 j_4})(\hat x^{\+}_{j_1 j_4} \hat x^{\+}_{j_1 j_2})} \,,
\end{align*}
which depends on the auxiliary spinor (gauge-fixing parameter) $\xi^{\+}_\da$: $\hat x^{\+}_{ij} \equiv [ \xi^{\+} | \hat x_{ij}$.
\item The superinvariants assigned to higher-order vertices factorize in a product of 4-point superinvariants. 
For example,  the 4th graphs in Fig.~\ref{NMHV6fig} gives
\begin{align*}
\hat R(i;j_1,j_2,j_3,j_4) = \hat R(i;j_1,j_2,j_3) \hat R(i;j_1,j_3,j_4) \,.
\end{align*}
\item The gauge-fixing parameter $\xi^{\+}$ drops out of the sum of all diagrams.
\end{itemize}

The chiral Feynman rules of \cite{Chicherin:2014uca} are obtained from the above by setting all $\bq_+ = 0$. Thus, our main result is that the full non-chiral Born-level correlators of the stress-tensor multiplets are obtained from the chiral ones simply by putting hats on each coordinate difference, $x_{ij} \to \hat x_{ij}$. This is a very non-trivial statement. Indeed, starting with five points, $\cN=4$ superconformal symmetry allows the existence of nilpotent non-chiral invariants, which could alter the above simple result. The fact that this does not happen is likely to reveal some new deep property of the correlators, which remains to be discovered. 

We derive these effective Feynman rules by restoring the $\bq_+-$dependence of the chiral truncated correlator    by means of $\bQ$ supersymmetry transformations. The twistor formulation   of $\cN=4$ SYM \cite{Mason:2005zm,Boels:2006ir,Adamo:2011cb} or the alternative Lorentz-harmonic chiral (LHC) formulation developed in the twin paper \cite{PartI} are purely chiral, in them $\bQ$ supersymmetry is not manifest. 

The results of the present paper are based on the realization of  $\bQ$ supersymmetry elaborated in \cite{PartI}. The theory is formulated in terms of two gauge connections $A^{++}(x,\q^+,u)$ and $A^{+}_\da(x,\q^+,u)$. They depend on Lorentz harmonics (LHs) $u^\pm_\a$ parametrizing the coset $SU(2)_L/U(1) \sim S^2$ of the chiral half of the Euclidean Lorentz group $SO(4) \sim SU(2)_L \times SU(2)_R$. The connections depend on half of the chiral Grassmann variables, $\q^{+A} = u^+_\a \q^{a A}$, i.e. on  a quarter of the odd variables of $\cN=4$ superspace. The LHs allow us to handle efficiently infinite towers of auxiliary fields and pure gauges. Due to them the interaction vertices disappear from the Born-level supercorrelators, but the operator $T$ takes a complicated non-local and non-polynomial form as compared to the conventional component expressions.
In this formulation only the $Q$ half of supersymmetry is manifest. The $\bQ$ half is \emph{non-linear} and the algebra closes on shell.  The $\bQ$ variations of the chiral $\cT$ give rise to the complete $\bq_+$ expansion of the full non-chiral $T$. 
Fortunately, in the calculation of the Born-level correlators we only need the  simple \emph{linear} part of the $\bQ$ transformations, corresponding to the self-dual theory.
The complicated non-self-dual part of $\bQ$ is irrelevant in the Born approximation. The $\bq$-deviations from the chiral sector of the supercorrelator are mild and they are effectively captured by the self-dual  $\cN= 4$ SYM theory.

The plan of the paper is as follows. In Section \ref{s2} we briefly review the LHC formalism and the main results of \cite{PartI} which we  need here. In Section \ref{s5} we introduce the stress-tensor supermultiplet and 
explain how to construct it from the two  LHC gauge connections. We use chiral `semi-superfields' since in theLHC framework there 
are no $\bq$-variables. We successively apply  $\bQ$-transformations to the chiral  $\cT$ to construct the components in the expansion of $T$ in terms of $\bq_+$. In this way we reconstruct the full supermultiplet, although some of its pieces are irrelevant in the Born-level correlator calculations. 
In Section \ref{s7} we describe the classes of Born-level correlators and reproduce the results of \cite{Chicherin:2014uca} for the chiral correlators in the  LHC formalism. In Section \ref{s8} we consider the correlators of the various semi-superfield components of $T$ and prove that they are combined together in an object with full supersymmetry by the simple shift $x_{ij} \to \hat x_{ij}$ in the chiral Feynman rules for $\cT$. 
In Section \ref{s9} we show that this shift corresponds to a change of basis from the chiral to the Grassmann analytic (or half-BPS) basis. We display some technical details and calculations in several Appendices. In Appendix~\ref{proofTW} we show the equivalence of two different definitions of $\cT$. In Appendix~\ref{BPS} we explain how the half-BPS condition is realized on $\cT$. In Appendix~\ref{loop} we give an example of an LHC supergraph calculation at loop level. We compute the one-loop  four-point correlator of $\cO_{\bf 20'}$ and demonstrate explicitly the absence of the gauge-fixing parameter $\xi$ in the final answer. In Appendix~\ref{nolambda} we discuss the gauge invariance of the effective Feynman rules. In Appendix~\ref{apE} we compare some of our results with the conventional Feynman graph calculation of the components of the non-chiral correlators. In Appendix F we present an alternative calculation of the NMHV-like correlators, still in the LHC formalism but this time in a covariant Landau-type gauge. The result is manifestly Lorentz covariant and free from spurious poles, but is much more bulky than the equivalent light-cone gauge result.

\section{$\mathcal{N}=4$ SYM  in Lorentz harmonic chiral superspace}\label{s2}

In this Section we  summarize the Lorentz harmonic chiral (LHC) formulation of the $\cN=4$ SYM action 
developed in \cite{PartI}. This approach involves infinite towers of auxiliary and pure gauge higher-spin 
component fields. In order to handle them efficiently we combine them in \emph{ harmonic
superfields} having  infinite harmonic expansions on the sphere $S^2$.

We introduce 
harmonic variables on the chiral half of the Euclidean Lorentz group $SO(4) \sim SU(2)_L \times SU(2)_R$. The harmonics $u^{\pm \a}$ are defined as two $SU(2)_L$ spinors forming
an $SU(2)_L$ matrix:
\begin{equation}\label{2}
u^{\pm \a} \in SU(2)_L \ : \qquad u^{+\a} u^-_{\a} \equiv  u^{+\a} \ep_{\a\b} u^{-\b}= 1 \,, \quad (u^{+\a})^* = u^-_{\a}\, .  
\end{equation}
The index $\pm$ refers to the charge of these variables with respect to
 $U(1)_L \subset SU(2)_L$.  Thus, the harmonic
variables defined in this way describe the compact coset
$S^2 \sim SU(2)_L/U(1)_L$. 
The harmonic functions on the coset, homogeneous under the action of $U(1)_L$ with charge $q$, are defined by their harmonic expansion (here $q \geq 0$)
\begin{equation}\label{3}
f^{(q)}(u) = \sum^\infty_{n=0} f^{\a_1 \ldots \a_{2n+q}}
u^+_{(\a_1} \ldots u^+_{\a_{n+q}} u^-_{\a_{n+q+1}}
\ldots u^-_{\a_{2n+q})} \,.
\end{equation}
The coefficients $f^{\a_1 \ldots \a_{2n+q}}$ are higher-spin symmetric tensors of rank $2n+q$, i.e. they carry the representation $(n+q/2,0)$ of the Lorentz group.
We extensively apply the 
rules of harmonic calculus introduced in \cite{Galperin:1984av}. 
We need the covariant harmonic derivatives
\begin{align}\label{4}
& \pa^{++}= u^{+\a}{\pa\over\pa u^{-\a}} \ : \ \ \
\pa^{++}u^{+\a} =0\,,\  \ \pa^{++} u^{-\a} = u^{+\a}\nt
&\pa^{--}= u^{-\a}{\pa\over\pa u^{+\a}} \ : \ \ \
\pa^{--}u^{+\a} =u^{-\a}\,,\  \ \pa^{--} u^{-\a} = 0\,.
\end{align}
Together   with the charge operator $\pa^0$ (which counts the $U(1)_L$ charge of the harmonics, $\pa^0 u^{\pm\a}= \pm u^{\pm\a}$) 
the harmonic derivatives form the algebra of $SU(2)$ realized on the indices $\pm$ of the harmonics, i.e. $[\pa^{++}, \pa^{--}] = \pa^0$.
The derivatives $\pa^{++}$ and $\pa^{--}$ have the meaning of the raising and lowering operators of $SU(2)_L$.
Harmonic integration amounts to projecting out the singlet part of
a chargeless integrand,  according to the  rule
\begin{equation}\label{6}
\int du\; f^{(q)}(u) =  \left\{ \begin{array}{ll}
0,  \ q \neq 0 \\ f_{\rm singlet} ,  \ q= 0 \end{array} \right. .
\end{equation}
It is compatible with integration by parts for the harmonic derivatives $\pa^{++}$ and $\pa^{--}$.

We would like to describe the theory by superfields. 
However, the   $\cN=4$ extended
superspace with coordinates
\begin{equation}\label{1}
x^{\da\a}\,,  \  \q^{\a A}\,, \ \bq^\da_A \,,
\end{equation}
where $A=1,\ldots,4$ is an $SU(4)$ (R-symmetry) index, is too big. 
We are not attempting to formulate the theory in terms of \emph{ unconstrained} superfields on this space, $\Phi(x^{\da\a},\q^{\a A},\bq^\da_A)$. 
Instead, we deal with  \emph{ chiral-analytic} `semi-superfields' which depend only on a quarter of the Grassmann variables,
\begin{align}\label{chan}
\Phi(x^{\da\a},\q^{+ A}, u)\,, \qquad {\rm where} \ \q^{+ A} \equiv u^+_{\a} \q^{\a A}\,.
\end{align}
They are also harmonic functions having an infinite expansion on $S^2$ in terms of ordinary fields on space-time. We call such objects `semi-superfields' because only $Q$-supersymmetry is realized linearly on them.
The realization of $\bar{Q}$-supersymmetry is non-linear. They are chiral because they do not depend on $\bq$,
and they are Lorentz- (or L-)analytic because they are annihilated by the LH-projected derivative $\pa^+_A \equiv u^{+ \a} \frac{\pa}{\pa \q^{\a A}}$\,, i.e.
$$
\pa^+_A \, \Phi(x^{\da\a},\q^{+ A}, u) = 0 \,.
$$

We formulate the $\cN=4$ SYM theory in terms of two chiral-analytic semi-superfields, 
\begin{equation}\label{09}
A^+_\da = A^+_\da (x,\q^+,u) \,, \qquad A^{++} = A^{++}(x,\q^+,u) \,.
\end{equation}
They are defined as the gauge connections for two covariant  derivatives, the harmonic-projected  space-time derivative $\pa^+_\da \equiv u^{+ \a} \frac{\pa}{\pa x^{\da\a}}$ and the harmonic derivative $\pa^{++}$:
\begin{align}\label{3.3}
\nabla^+_\da =  \pa^+_\da + g A^+_\da \,, \qquad \nabla^{++} = \pa^{++} + g A^{++}\,,
\end{align}
with respect to a gauge group with  a {chiral-analytic parameter}  $\Lambda(x,\q^+,u)$.

The action of $\mathcal{N}=4$ SYM can be written as a sum of two terms
\begin{equation} \label{N4}
S_{\cN=4} = S_{ \rm CS} + S_{\rm Z} = \int d^4x du d^4\q^+\; L_{\rm CS}(x,\q^+,u) + \int d^4x d^8\q\; L_{\rm Z}(x, \q) \,.
\end{equation}
Here the chiral-analytic Lagrangian $L_{\rm CS}$ is of the Chern-Simons type,
\begin{align}\label{CS}
L_{\rm CS} =  \tr\; (A^{++}\pa^{+\da}A^+_\da-
{1\over 2} A^{+\da}\pa^{++} A^+_\da + g A^{++} A^{+\da} A^+_\da) \,.
\end{align}
In the cubic interaction term  we see the gauge coupling constant $g$. As is customary in gauge theories, it accompanies every gauge group commutator (or equivalently, the structure constant $f_{abc}$ of the gauge group $SU(N_c)$). In other words, the Abelian theory is  the free (or $g=0$) theory.
The second non-polynomial term is chiral but not L-analytic,\footnote{This type of action term was introduced in 1987 by Zupnik \cite{Zupnik:1987vm} in the context of the formulation of $\cN=2$ SYM in harmonic superspace. There one uses $SU(2)$ harmonics parametrizing the R-symmetry group, instead of the Lorentz group in \cite{PartI} and in this article.} 
\begin{align}\label{lint}
L_{\rm Z} = \omega\ \tr\sum^\infty_{n=2}{(-1)^n g^{n-2}\over n} \int du_1\ldots du_n\; {A^{++}(x, \q^+_1,u_1) \ldots A^{++}(x, \q^+_n,u_n) \over
(u^+_1u^+_2) \ldots (u^+_nu^+_1)}\,,
\end{align} 
where $\q^{+A}_i = \q^{\a A} (u_i)^+_{\a}$ with $i=1,\ldots,n$.  This action term is local in $(x,\q)$ space but non-local in the harmonic space (each copy of $A^{++}$ depends on its own harmonic variable). The bilinear ($n=2$) term in the expansion is part of the kinetic Lagrangian since it does not vanish in the free theory ($g=0$). However, it turns out  more convenient to treat it as an `interaction' when deriving the Feynman rules.   {We leave the constant $\omega$ in \p{lint} and in what follows unfixed, although there is a canonical choice for it. It measures the deviation from the self-dual sector of the theory described by the Chern-Simons type action \p{CS} (see \cite{Sokatchev:1995nj}). As we will see in Sects.~\ref{s7} and \ref{s8}, $\omega$ does not appear in the Born-level MHV-like and non-MHV-like super-correlators. Roughly speaking, it counts the number of loop integrations, which equals zero at Born level. The reader who is only interested in our main result about the non-chiral Born-level super-correlators,  can set $\omega = 0$. However, we need to consider all the contributions with $\omega \neq 0$ to justify this result.}

Both terms in \p{N4} are invariant under a gauge group with chiral-analytic parameters $\Lambda(x,\q^+,u)$. The invariance of $S_{\rm CS}$ holds up to total derivatives, that of $S_{\rm Z}$ is based on the following property of the gauge variation of $L_{\rm Z}$:
\begin{align}\label{3.32}
\delta_{\Lambda} L_{\rm Z} =  \int du \Delta L\,, \qquad \text{where} \;\; \pa^+_A \pa^+_B   \Delta L =0\,.
\end{align}
The variations of \p{N4} with respect to the super-connections $A^{+}_{\da}$ and $A^{++}$ yield the equations of motion
\begin{align}
\label{3.6}
\pa^+_\da A^{++} - \pa^{++} A^+_\da + g[A^+_\da,A^{++}]= 0\,, \\
\label{3.14'}
\pa^{+\da} A^+_\da + gA^{+\da} A^+_\da = \omega\ (\pa^+)^4 A^{--}\,.
\end{align}
The first one is kinematical (it expresses the auxiliary fields contained in $A^+_\da$ in terms of the physical fields in $A^{++}$). The second equation is the dynamical one. In it we see the super-connection for the other harmonic derivative in \p{4}, $\nabla^{--} = \pa^{--} + A^{--}$.   
 The gauge-covariant harmonic derivatives respect the $SU(2)$ algebra,  $[\nabla^{++}, \nabla^{--}] = \pa^0$, or in detail
\begin{align}\label{3.11}
\pa^{++} A^{--} - \pa^{--} A^{++} + g[A^{++},A^{--}]=0
\end{align}
(the charge operator $\pa^0$ needs no connection because the gauge parameter is a harmonic function of charge zero, $\pa^0 \Lambda=0$). 
The (unique) solution to this nonlinear differential harmonic equation was found by Zupnik \cite{Zupnik:1987vm},
\begin{align}\label{448}
A^{--}(x,\q,u) =  \sum^\infty_{n=1} (-1)^n g^{n-1}\int du_1\ldots du_n\; { 
A^{++}(x, \q^+_1,u_1) \ldots A^{++}(x, \q^+_n,u_n) \over (u^+u^+_1)(u^+_1u^+_2) \ldots
(u^+_nu^+)}\,,
\end{align} 
where  $\q^{+A}_i = \q^{\a A} (u_i)^+_{\a}$ with $i=1,\ldots,n$, as in \p{lint}. It is local in $(x,\q)$ space but non-local in the harmonic space 
(each copy of $A^{++}$ depends on its own harmonic variable). 
Unlike the connection $A^{++}$, this one is not chiral-analytic but only chiral, 
i.e. it depends on the full chiral $\q^{\a A}$. { The non-local expression \p{448} is rather similar to $L_{\rm Z}$ \p{lint}.
So, it is not a surprise that it appears in the variation of $L_{\rm Z}$,}
\begin{align} \label{Lvar}
\delta L_{\rm Z} =  - \omega\ \tr \int d u \ \delta A^{++}(x,\q^+,u)\ A^{--}(x,\q,u)\,.
\end{align}
{From here one easily derives the field equation \p{3.14'} by varying $S_{\cN = 4}$ with respect to $A^{++}$.}

The action \p{N4} is invariant under $Q$ and $\bar{Q}$ supersymmetries. 
$Q$-supersymmetry is realized linearly, $Q^\b_B = \pa^\b_B$, on the connections $A^{++}$, $A^{+}_{\da}$, 
\begin{align}\label{4.27'}
Q^\b_B \, A(\q^+) =  \pa^\b_B \, A(\q^+) = ( u^{-\b} \pa^+_B - u^{+\b} \pa^-_B ) \,A(\q^+) = - u^{+\b} \pa^-_B \,A(\q^+)\,,
\end{align} 
where we used the L-analyticity $\pa^+_B A=0$.

$\bar{Q}$ supersymmetry is more involved. {We write $\bar{Q}$ as a sum of two generators,}
\begin{align}  \label{4.28'}
\bar Q^B_{\db} = \bar{\mathbb{Q}}^B_{\db} + (\bar Q_{\rm Z})^B_{\db}\,.
\end{align} 
{Their action on the super-connections is summarized in the following table,} 
\begin{align} \label{4.27}
\begin{array}{c|c|c}
 & A^{++} & A^{+}_{\da} \\[0.8 ex] \hline & & \\[-1.5ex]
\bar {\mathbb{Q}}^B_{\db} & \q^{+B} \bigl( \pa^-_\db A^{++} + A^+_\db \bigr) & \q^{+B} \pa^-_\db A^+_\da \\[1.0 ex] \hline & & \\[-1.5ex]
(\bar Q_{\rm Z})^B_{\db} & 0 & \omega\ (\pa^+)^4 (\q^{-B} A^{--})\ep_{\da\db}
\end{array}
\end{align} 
{We see that the $\bar{Q}$-variation  mixes the two connections. 
The important difference between $ \bar {\mathbb{Q}}$ and $\bQ_Z$ is that the former is \emph{ linear} and the latter is \emph{ non-linear} (non-polynomial in $A^{++}$) due to the presence of $A^{--}$ (see~\p{448}).
The $\bar{{\mathbb{Q}}}$-variation of $A^{+}_{\da}$ (and the first term in the $\bar{{\mathbb{Q}}}$-variation of $A^{++}$)
is a remnant of the conventional $\bar{Q}$-transformation, $\bar{Q} = \pa_{\bq}- \q \pa_x$ 
(our chiral `semi-superfields' do not depend on $\bq$). More precisely, it is $\q^{+B}\pa^-_\db$ for L-analytic superfields (the full shift $-\q^{\b B} \pa_{\b\db}$ involves  both $\theta^+$ and $\theta^-$).
The generator $\bar{\mathbb{Q}}$ leaves  the Chern-Simons action $S_{\rm CS}$  \p{N4} invariant. The $\bQ_{\rm Z}$ term in \p{4.28'} is needed because of the presence of $S_{\rm Z}$ in \p{N4}. Remark that $\bQ_{\rm Z}$ is proportional to the constant $\omega$, which measures the deviation from the self-dual sector.}

The variations   
$\delta(\ep) \equiv \delta_{\ep} + \delta_{\bar{\ep}} \equiv \ep^{\b B} Q_{\b B} + \bar\ep^\db_B \bQ^B_\db$  satisfy the supersymmetry algebra
\begin{align}
\left[ \delta(\kappa) \,,\, \delta(\ep) \right] A^{+}_{\da} 
&= (\ep^{\b B} \bar\kappa^{\db}_B - \kappa^{\b B} \bep^{\db}_B) \,\pa_{\b\db } A^{+}_{\da} +
\delta_{\Lambda}A^{+}_{\da}
\nt
\left[ \delta(\kappa) \,,\, \delta(\ep) \right] A^{++} 
&= (\ep^{\b B} \bar\kappa^{\db}_B - \kappa^{\b B} \bep^{\db}_B) \,\pa_{\b\db } A^{++} +
\delta_{\Lambda}A^{++} \,, 
\notag
\end{align}
 modulo the equations of motion \p{3.6}, \p{3.14'} and up to a compensating gauge transformation with a field-dependent chiral-analytic parameter $\Lambda$ whose explicit expression can be found in \cite{PartI}. 
{The  Abelian subalgebras formed by the $Q$-variations and by the $\bar{\mathbb{Q}}$-variations close off shell. 
However, the subalgebra formed by the complete $\bQ$-variations closes only on shell (the kinematical constraint \p{3.6} is needed for the closure on $A^{+}_{\da}$, but the field equations are not required for the closure on  $A^{++}$).}

The $\bar{Q}$-variation of the gauge connections will enable us to reintroduce the $\bq$ variable 
in our chiral setup. In the `semi-superfield' formalism only the $Q$-supersymmetry is manifest.
In order to reconstruct the full supermultiplet of some composite operators 
(which is not chiral, i.e. it involves both the $\q$ and $\bq$ variables), we apply successively the $\bar{Q}$-variations.
In Section \ref{s5}, where we work out these $\bar{Q}$-variations for the stress-tensor multiplet, we use the fact that this transformation commutes with 
the gauge variation, 
\begin{align} \label{gcom}
[\delta_{\bar\ep},\delta_{\Lambda} ] \, A^{++} = 0 \,, \qquad 
[\delta_{\bar\ep},\delta_{\Lambda} ] \, A^{+}_{\da} = 0\,,
\end{align}
where the gauge parameter transforms as a chiral-analytic semi-superfield, \\
$\delta_{\bep} \Lambda = \bep^\db_B \q^{+B} \pa^-_\db\Lambda$. 

The non-polynomial action \p{N4} takes the conventional component field form in the Wess-Zumino gauge. The gauge parameter $\Lambda$ with its infinite harmonic expansion can  eliminate almost all of the higher-spin fields from $A^{++}$. The remainder is
\begin{equation}\label{23}
A^{++}= (\q^+)^{2\, AB}\phi_{AB}(x) +
(\q^+)^3_A
u^{-\a}\psi^A_{\a}(x) + 3(\q^+)^4 u^{-\a}u^{-\b}
G_{\a\b}(x)\,,
\end{equation}
where we apply the short-hand notations 
$$
(\q^+)^{2\, AB} = {1\over 2!}\q^{+A}\q^{+B},  \ (\q^+)^3_A  =
{1\over 3!} \ep_{ABCD}\q^{+B}\q^{+C}\q^{+D},  \
(\q^+)^4 = {1\over 4!}
\ep_{ABCD}\q^{+A}\q^{+B}\q^{+C}\q^{+D}
\,.  $$
In this (non-supersymmetric) gauge  $A^{++}$ contains part of the $\cN=4$ vector multiplet:  the six scalars $\phi_{AB}$, the four chiral fermions $\psi^A_{\a}$, 
and the Lagrange multiplier $G_{\a\b}$ for the self-dual field strength tensor $F_{\a\b}$. 
After the elimination of the auxiliary fields by means of the  kinematical field equation \p{3.6}, the other connection takes the form 
\begin{equation}\label{25}
A^{+}_\da = u^{+\a}{\cal A}_{\a\da}(x) + \q^{+A}
\bar\psi_{\da A}(x) + \text{derivative terms} \,.
\end{equation}
It contains the rest of the multiplet: the gauge field ${\cal A}_{\a\da}$ and the four anti-chiral fermions $\bar\psi_{\da A}$.
In the next section the Wess-Zumino gauge will help us  reveal the physical content of some composite operators.

In the standard superspace approach to $\cN=4$ SYM (see, e.g., \cite{Sohnius:1978wk})  the scalars of the vector multiplet are identified with  the super-curvature $W_{AB}=-W_{BA}$, which appears in the anticommutator of two chiral spinor covariant derivatives,
\begin{align}\label{3.7}
\{\nabla^\a_A , \nabla^\b_B\} = \ep^{\a\b} W_{AB}\,.
\end{align}
The absence of a curvature symmetric in the Lorentz indices $\a,\b$ on the right-hand side is the defining constraint of $\cN=4$ SYM in this standard formulation. In our LHC approach we can construct the curvature $W_{AB}$ directly in terms of the harmonic gauge connection $A^{--}$ (or effectively, in terms of $A^{++}$, recall \p{448}),
\begin{align}\label{3.12}
W_{AB} = \pa^+_A \pa^+_B A^{--}\,.
\end{align}
This curvature is {gauge covariant} and is (covariantly) harmonic independent, 
\begin{align}\label{314}
\nabla^{++} W_{AB}=0\,.
\end{align}

Finally, we quantize the action \p{N4} in  the light-cone gauge\footnote{It is called ``axial" or ``CSW" gauge in \cite{Boels:2006ir}, \cite{Cachazo:2004kj}.}
\begin{align}\label{417}
\xi^{\-\da} A^+_\da =0\,, 
\end{align}
where the gauge-fixing parameter $\xi^{\-\da}$ belongs to a new set of LH variables, this time for the second (anti-chiral) factor of the Euclidean Lorentz group $SO(4) \sim SU(2)_L\times SU(2)_R$. These new harmonics satisfy the usual $SU(2)$ conditions (recall \p{2})
\begin{align}\label{6.4}
\xi^{\+\da} \xi^{\-}_\da= 1\,, \qquad (\xi^{\+ \da})^* = \xi^{\-}_{\da}\ .
\end{align}
{With the help of the two sets of harmonics we can project the space-time coordinate $x^{\da\a}$ onto its light-cone components} 
\begin{align}\label{419}
x^{\dt\pm\pm} = \xi^{\dt\pm}_{\da} x^{\da\a} u^{\pm}_{ \a}\,.
\end{align}
{After fixing the gauge we can derive the propagators for the super-connections. They are singular  distributions in both the harmonic and coordinate spaces. We need the harmonic delta function $\delta(u,v)$ defined by the property}
\begin{equation}
\int dv\; \delta(u,v)\; f^{(q)}(v) = f^{(q)}(u)\,,
\label{4.7.4}
\end{equation}
{with a test function \p{3}. We also need the complex delta function
$\delta^2(t) \equiv \delta(t,\bar t)\,$ satisfying the relation 
${\partial\over \partial\bar t}\;\frac1{t} = \pi \delta^2(t)$. {More specifically, we use $ \delta^2( x^{\+ +} ) =  \delta( x^{\+ +},x^{\- -} )$.}
The set of propagators is as follows}
\begin{align}
&\vev{A^{++}(x,\q^+,u_1) A^{++}(0,0,u_2)}= \frac{1}{\pi} \delta^2( x^{\+ +} )\, \delta (u_1, u_2)\, \delta^{(4)}(\q^+) \label{prop1}\\
&\langle A^{+}_{\da}(x,\q^+,u_1) A^{++}(0,0,u_2) \rangle= \frac{1}{\pi}\frac{\xi^{\-}_{\da}} { x^{\- +}  }\,
\delta^2( x^{\+ +} )\, 
\delta(u_1,u_2) \,\delta^{(4)}(\q^+) \label{prop2} \\
&\langle A^{+}_{\da}(x,\q^+,u_1) A^{+}_{\db}(0,0,u_2) \rangle=0\,, \label{prop3}
\end{align}
{where we have used translation invariance to set $x_2=\q_2=0$. We do not specify if the projections $x^{\dot\pm \pm}$ (see~\p{419}) and $\q^+$ are made with the harmonic $u_1$ or $u_2$, in view of the harmonic delta functions above. We also suppress the color structure $\delta_{ab}$ for the sake of brevity. In the light-cone gauge \p{417} only the quadratic terms in $S_{\rm CS}$ (see~\p{N4}, \p{CS}) survive, and these propagators correspond to them. The bivalent vertex in $S_{\rm Z}$ (the $n=2$ term in \p{lint}) is treated as an `interaction'. The genuine interaction vertices come only from the $S_{\rm Z}$ part of the action $S_{\cN = 4}$ (the terms with $n \geq 3$). The gauge superfield $A^+_{\da}$ is non-interacting in the light-cone gauge and hence can appear only at the external points of the LHC supergraphs.}

\section{The stress-tensor supermultiplet} \label{s5}

The $\cN=4$ SYM stress-tensor multiplet is a short supermultiplet.\footnote{This means that the top spin in the supermultiplet is 2 (the stress-energy tensor) unlike  a generic long multiplet with top spin 4 \cite{Andrianopoli:1998ut}.}  Its superconformal primary is the half-BPS scalar operator $\cO^{AB;CD}_{20'} = {\rm tr} (\phi^{\{AB} \phi^{CD\}})$ (where $\{\}$ denotes  the traceless part) made from the six physical scalars of the theory (recall~\p{23}). It belongs to the representation $\mathbf{20'}$ of the R-symmetry group $SU(4)$. The stress-tensor multiplet is most naturally described in $\cN=4$ harmonic (or R-analytic) superspace \cite{Hartwell:1994rp,Andrianopoli:1999vr,Heslop:2000af}. We introduce a new set of harmonics $w^{a}_{+ A}, \, w^{a'}_{- A}$ and their conjugates $\bar w^A_{-a}, \, \bar w^A_{+a'}$ on the R-symmetry group $SU(4)$, projecting the index $A=1,\ldots,4$ of the (anti)fundamental irrep onto the subgroup $SU(2)\times SU(2)'\times U(1)$ (indices $a,a',\pm$): 
\begin{align}
&(\ w^a_{+A} \ ,\ w^{a'}_{-A}\ ) \in SU(4): \notag \\ \label{5.1}
&w^{a}_{+ A}   \bar w^{A}_{- b} = \delta^{a}_{b}  \,, \quad w^{a}_{+ A}  \bar w^A_{+a'} = 0\,, \quad w^{a'}_{- A}  \bar w^{A}_{- b} =0\,, \quad w^{a'}_{- A}  \bar w^A_{+b'} = \delta^{a'}_{b'}\,;\\
&w^{a}_{+ A} \bar w^B_{-a} +  w^{a'}_{- A} \bar w^B_{+a'} =\delta^B_A \,; 
\quad \frac14 \epsilon^{ABCD} w^a_{+A} \ep_{ab} w^b_{+B}\ w^{c'}_{-C} \ep_{c'd'} w^{d'}_{-D} = 1\,. \label{compl}
\end{align}

{So far we have introduced  three species of harmonics: chiral LHs $u^{\pm}_{\a}$ (see~\p{2}), which are coordinates of the superspace and `harmonize' the $SU(2)_L$ half of the Euclidean Lorentz group; antichiral LHs $\xi^{\dt\pm}_{\da}$ (see~\p{6.4}), which are auxiliary parameters needed to fix the gauge \p{417} and `harmonize' the $SU(2)_R$ half of the Euclidean Lorentz group; RHs $(w_\pm, \bar w_\pm)$ on the R-symmetry group $SU(4)$, which are a convenient tool for dealing with $SU(4)$ tensors.}
In order to distinguish  the {RHs} from the LHs $u^\pm_\a$  we indicate the Lorentz charge of  $u^\pm$ (and of all projected quantities)  upstairs, while the R-symmetry charge of the RHs $w_\pm, \bar w_\pm$ is shown downstairs.  The antichiral LHs $\xi^{\dt\pm}_{\da}$ have a dot above the charge.

With {the help of the RHs} the half-BPS operator can be projected onto the highest-weight state of $U(1)$ charge $(+4)$:
\begin{align}\label{5.2}
\cO =   \frac14 (w^{a}_{+ A} \ep_{ab} w^{b}_{+ B}) (w^{c}_{+ C} \ep_{cd} w^{d}_{+ D}) \cO^{AB;CD}_{20'} = {\rm tr}( \phi_{++} \phi_{++}) \,.
\end{align}
The harmonic projection automatically renders the tensor traceless and hence irreducible. 

The short stress-tensor multiplet  is annihilated by half of the $\cN=4$ supercharges, namely, by the RH projections $Q^{\a}_{+ a'} =  \bar w^A_{+ a'} Q^{\a}_{A}$ and $\bar Q^a_{+\da} = w^a_{+A} \bar Q^A_\da $. Equivalently, it satisfies the R-analyticity conditions
\begin{align}\label{5.3}
\bar w^A_{+ a'} D^{\a}_{A} T =   w^a_{+A} \bar D^A_\da T=0\,. 
\end{align}
{The supercovariant derivatives verify the $\cN=4$ supersymmetry algebra  $\{\bar D^A_\da,  D_{\b B}\} = \delta^A_B \pa_{\b\da}$.}
Consequently, the projected spinor derivatives in \p{5.3} anticommute,
\begin{align}\notag
\{ \ w^a_{+A}\bar D^A_\da \ ,\ \bar w^B_{+ b'} D_{\b B} \ \} \ = \ 0\,,
\end{align}
so the two shortening conditions  are compatible. To solve the  constraints we decompose the Grassmann variables with the help of the RHs,
\begin{align}\label{4.39}
& \q_{+\a}^a = w^{a}_{+ A} \q^A_\a\,, \qquad   \q_{- \a}^{a'} =  w^{a'}_{- A} \q^A_\a\,,  \qquad  \bq^\da_{+a'} = \bar w^A_{+a'}\bq^\da_{A}\,, \qquad \bq^\da_{-a} = \bar w^A_{-a}\bq^\da_{A}\,.
\end{align}
Further, we define the R-analytic basis\footnote{More about the different bases in superspace see in Section \ref{s9}.} 
\begin{align}\label{AnB}
x^{\da\a}_{\rm an} = x^{\da\a} +  \frac1{2} \left(  \q_{+}^{\a a} \bq^\da_{-a} -   \q_{- }^{\a a'} \bq^\da_{+a'}\right)\,.
\end{align}
In this basis 
\begin{align}\label{DaB}
D_{+a'} \equiv \bar w^A_{+a'} D_A = \pa/\pa\q^{a'}_-\,, \quad \bar D^a_+ \equiv w^{a}_{+A} \bar D^A  = \pa/\pa \bq_{-a} \,.
\end{align} 
Consequently, the stress-tensor multiplet satisfying the  conditions  \p{5.3} is described by an R-analytic superfield\footnote{Not to be confused with the L-analytic semi-superfields  $A^{++}(x,\q^+,u), A^+_\da(x,\q^+,u)$ in  \p{09}!} 
\begin{align}\label{37}
T(x_{\rm an}, \q_+,\bq_{+},w) =   \cO(x_{\rm an},w) +\ldots + (\bq_+\sigma^\mu \q_+) (\bq_+\sigma^\nu \q_+)  T_{\mu\nu}(x) + \ldots \,, 
\end{align}
depending only on half of the projected Grassmann coordinates. We emphasize that this superfield is \emph{ not chiral}. Among its components we find the stress-energy tensor $T_{\mu\nu}(x)$. It is an R-symmetry singlet, hence it does not depend on the RHs.

\begin{figure}
\begin{center}
\includegraphics[width = 7 cm]{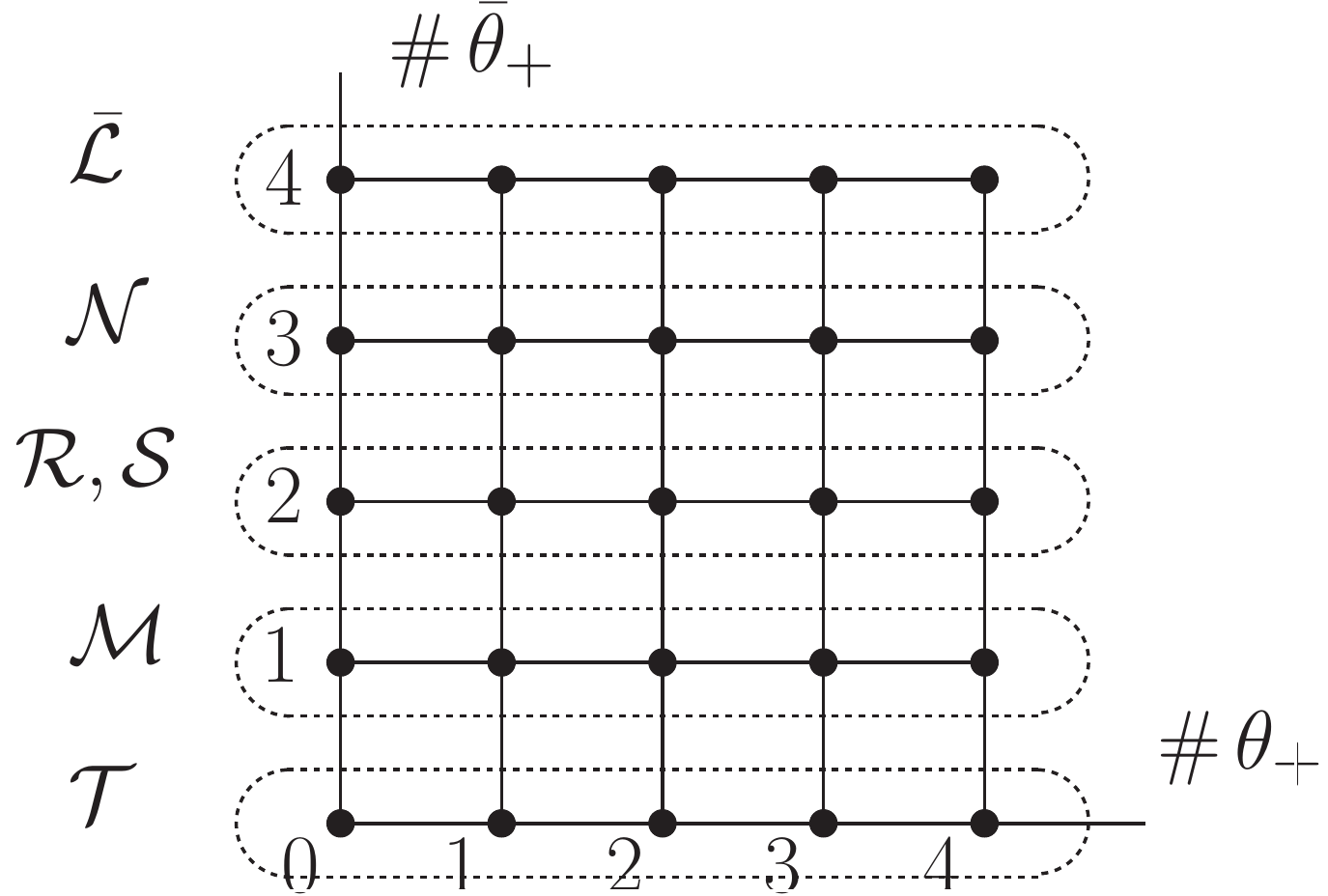}
\end{center}
\caption{Structure of the stress-tensor supermultiplet. It lives on R-analytic superspace, i.e. it depends only on  half 
 of the odd coordinates ($\q_+$ and $\bq_+$). Each horizontal layer encircled by a dashed contour corresponds to a semi-superfield, i.e. it transforms covariantly under $Q$ supersymmetry.}
\label{STMfig}
\end{figure}

Let us now see how this multiplet can be reformulated in our LHC superspace. As stated in Section \ref{s2}, our dynamical semi-superfields $A^{++}, A^+_\da$ are chiral, i.e. they do not depend on $\bq$. Therefore we need to expand $T$ in the antichiral variables $\bq_+$,
\begin{align}\label{4.33}
T = \cT+ \bq^{\da}_{+a'} \cM_\da^{a'}  + (\bq_+^2)_{(a'b')} \cR^{(a'b')}  + (\bq_+^2)_{(\da\db)} \cS^{(\da\db)}  + (\bq_+^3)_{a'}^\da \cN^{a'}_\da + (\bq_+)^4 \bar \cL\,,
\end{align}
where 
\begin{align}
&(\bq^2_+)^{(\da\db)} = \bq^{\da}_{+a'}  \ep^{a'b'}\bq^{\db}_{+b'}\;, \;\;     (\bq^2_+)_{(a'b')} = \bq^{\da}_{+a'} \ep_{\db\da} \bq^{\db}_{+b'}  \nt   & (\bq^3_+)^{\da}_{a'} = \frac13 \bq^{\da}_{+b'} \bq^{b'}_{+\db} \bq^{\db}_{+a'} \;, \;\; 
(\bq_+)^4 = \frac{1}{12} \bq^{a'}_{+\da} \bq^{\da}_{+b'} \bq^{b'}_{+\db} \bq^{\db}_{+a'} \ .  \notag
\end{align}
The `semi-superfield' components $\cT, \cM,\ldots,\bar\cL$ are gauge invariant operators depending only on $\q^{\a a}_+$. {They are summarized in Table~\ref{STMtab}, where their RH charge and scaling dimension are indicated,  and are depicted in Fig.~\ref{STMfig}.}
\begin{table}
\begin{center}
\begin{tabular}{l|c|c|c|c|c|c}
Superfield & $\cT$ & $\cM_\da^{a'}$ & $\cR^{(a'b')}$ & $\cS^{(\da\db)}$ & $\cN^{a'}_\da$ & $\bar \cL$ \\ \hline
Dimension & 2 & 5/2 & 3 & 3 & 7/2 & 4 \\ \hline
RH charge & +4 & +3 & +2 & +2 & +1 & 0
\end{tabular}
\end{center}
\caption{Semi-superfield content of the stress-tensor supermultiplet.}
\label{STMtab}
\end{table}
These  are semi-superfields in the sense that they transform covariantly under $Q$ supersymmetry, but $\bQ$ supersymmetry transforms them into each other. The bottom components of each semi-superfield are shown below (the color traces are omitted):\footnote{The field normalizations chosen here are not the conventional ones. We have also fixed $\omega = 1$.}  
\begin{align}\notag
&\cT= - \frac12 \cO +\ldots + (\q_+)^4 L(x) \nt
& \cM_\da^{a'} = 
\bar\psi^{a'}_{+\da} \phi_{++}+O(\q_+)\nt
& \cR^{(a'b')}  = -\frac14 \bar\psi_{+\dg}^{(a'}\bar\psi_{+}^{b')\dg}
- \frac{  g}{4} [\phi_-^{(a'C},\phi_{+C}^{ b')}]\phi_{++}+O(\q_+)\nt
&\cS^{(\da\db)} = 
 \frac14 \bar\psi_{+c'}^{(\da} \bar\psi^{\db )c'}_{+} 
- \frac12  \tilde F^{\dot\alpha\dot\beta}\phi_{++}  +O(\q_+)\nt 
&\cN^{a'}_\da =  \tilde F_\da^\db \bar\psi_{+\db}^{a'}
- \frac{  g}{2} [\phi^{a'}_{+B}, \phi^{BC}]\bar\psi_{C\da} +O(\q_+)\nt
&\bar \cL = \bar L(x) +O(\q_+)   \,, \label{4.48}
\end{align}
{where the operator $\cO$ is defined in \p{5.2}, $\tilde F$ is the anti-self-dual YM field strength and}
\begin{align}\notag
\phi_{++} = - \frac12 w^a_{+A} \ep_{ab} w^b_{+B} \ \phi^{AB} \;,\; \bar\psi^{\da}_{+ a'} = \bar\psi^{\da}_{A} \bar w^A_{+a'}\,. 
\end{align}
In the last line we see the anti-chiral on-shell $\cN=4$ SYM Lagrangian
\begin{align}\label{5.10}
\bar L(x) = -\frac12  \tilde F_{\dot\alpha\dot\beta}\tilde F^{\dot\alpha\dot\beta} 
 - \frac{  g}{4} \phi^{AB} \bar\psi^{\dot\alpha}_{A} \bar\psi_{\dot\alpha B} + \frac{ g^2}{32} [\phi^{AB},\phi^{CD}][\phi_{AB},\phi_{CD}]\,.
\end{align}
Its chiral counterpart $L(x)$ (see \p{LSD}) appears as the top component in the expansion of the chiral semi-superfield $\cT$ above.

Now we want to express the various semi-superfield components of the stress-tensor multiplet in terms of our chiral-analytic super-connections $A^{++}$ and $A^+_\da$. {We  use simultaneously two types of harmonics, LHs $u^{\pm}$ and RHs $w_{\pm}$, $\bar{w}_{\pm}$, with which we project the indices of  the Grassmann variables and derivatives. The semi-superfield  components $\cT, \cM,\ldots,\bar\cL$ depend on $\q^{\a a}_+$, whereas the semi-superfields $A^{+}_{\da}$ and $A^{++}$ depend on $\q^{+A}$. So the two types of superfields live in different halves of the chiral superspace. We need both types of harmonics to split the chiral superspace into four two-dimensional subspaces,}
\begin{align} \notag
\q^{A}_{\a} \quad \Rightarrow \quad 
\begin{array}{c|c} \quad \q^{+a}_+ \equiv w^a_{+A} \q^{A\a} u^{+}_{\a} \quad & \quad \q^{+a'}_- \equiv w^{a'}_{-A} \q^{A\a} u^{+}_{\a} \quad 
\\[1.3ex] \hline \\ [-1.5ex] \quad \q^{-a}_+ \equiv w^a_{+A} \q^{A\a} u^{-}_{\a} \quad & \quad \q^{-a'}_- \equiv w^{a'}_{-A} \q^{A\a} u^{-}_{\a} \quad \end{array} 
\end{align} 
{Analogously, we covariantly split the eight chiral Grassmann derivatives $\pa_{\da A}$ into four two-dimensional derivatives,}
\begin{align} \notag
\pa_{\a A} \quad \Rightarrow \quad 
\begin{array}{c|c} \quad \pa^+_{+a'} \equiv \bar{w}^A_{+a'} \pa_{\a A} u^{+\a} \quad & \quad \pa^{+}_{-a} \equiv \bar{w}^{A}_{-a} \pa_{\a A} u^{+\a} \quad 
\\[1.3ex] \hline \\ [-1.5ex] \quad \pa^{-}_{+a'} \equiv \bar{w}^{A}_{+a'} \pa_{\a A} u^{-\a} \quad & \quad \pa^{-}_{-a} \equiv \bar{w}^{A}_{-a} \pa_{\a A} u^{-\a} \quad \end{array} 
\end{align}

We define the maximal powers of the (projected) Grassmann coordinates as follows
\begin{align}\notag
(\q)^8 = \prod_{\substack{\a = 1,2 \\ A= 1 , \ldots,4}} \q^A_{\a} \;\;,\;\; 
(\q_+)^4 = \prod_{a,\a = 1,2} \q^a_{+\a} = \frac{1}{12} \q^{\a}_{+b} \q^{a}_{+\a} \q^{\b}_{+a} \q^{b}_{+\b} \;\;,\;\; 
(\q^+_+)^2 = \q^{+a}_+ \ep_{ab} \q^{+b}_+ \,,
\end{align}
{and analogously for $(\q_-)^4$, $(\q^+_-)^2$, $(\q^-_+)^2$, $(\q^-_-)^2$ and for the Grassmann derivatives. They decompose as follows}
\begin{align}
(\q)^8 = (\q_+)^4 (\q_-)^4  = (\q^+)^4 (\q^-)^4 \;,\quad (\q^\pm)^4 = \frac14 (\q^\pm_+)^2 (\q^\pm_-)^2 \; , \quad (\q_\pm)^4 = \frac14 (\q^+_\pm)^2 (\q^-_\pm)^2\,, 
\label{contr}
\end{align}
and similarly for the spinor derivatives  $(\pa)^8$, etc.

Our strategy for constructing the component semi-superfields in the expansion \p{4.33} will be as follows. Firstly, we propose a natural candidate for the chiral semi-superfield $\cT$ made out of $A^{++}$ and we show that it is gauge invariant and R-analytic. It is the starting point in the construction of the other component semi-superfield that are accompanied by $\bq_+$ in the expansion \p{4.33}. The dependence on $\bq_+$ is restored by acting repeatedly with the $\bar{Q}$ supersymmetry transformations. In our chiral formulation   this allows us to construct operators carrying dotted Lorentz indices. These operators, contracted with the $\bq_{+}^{\da}$ variables, are  identified with the components in the expansion \p{4.33}. 

\begin{figure}
\begin{center}
\includegraphics[width = 7 cm]{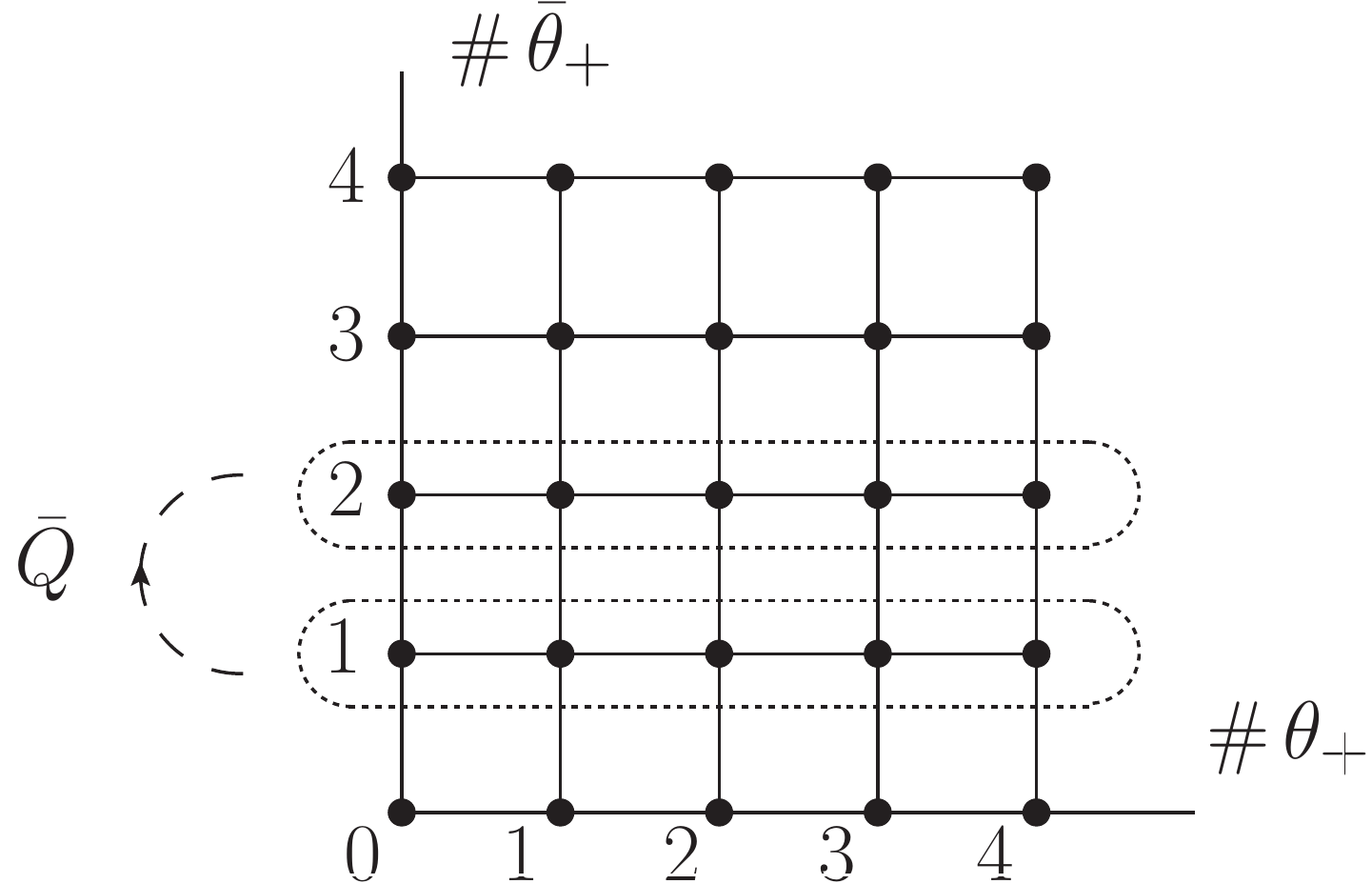}
\end{center}
\caption{The anti-chiral supersymmetry transformations combine together different R-analytic semi-superfields.
Acting with $\bar Q$  we move up to the next layer increasing the number of $\bq_+$ by one.}
\label{bQfig}
\end{figure}

 We realize the supersymmetry transformations on the semi-superfields $A^{++}$ and $A^{+}_{\da}$, so that the superspace coordinates are inert. Alternatively, supersymmetry may act on the superspace coordinates leaving the components of the superfield inert. The two realizations are related to each other by the requirement that  the superfield stay invariant. In the R-analytic basis \p{AnB} we have $\bQ^{a'}_{-\da} \equiv w^{a'}_{-A}\bQ^{A}_{\da} = \pa/\pa \bq^{\da}_{+a'}$. The invariance of the stress-tensor supermultiplet \p{4.33},  $\bQ^{a'}_{-\da} T = 0$, tells us how the semi-superfield components are transformed by the $\bQ$-variations (see \p{435}, \p{4.57}, \p{4.72}, \p{NL}). In this way we are able to reconstruct recursively the full non-chiral stress-tensor supermultiplet, starting from its chiral truncation $\cT$. The procedure is depicted schematically in Fig.~\ref{bQfig}. The generator $\bar{Q}$ mixes $A^{+}_{\da}$ and $A^{++}$ and acts nonlinearly on $A^{+}_{\da}$ (recall~\p{4.28'} and \p{4.27}). So it introduces dependence on $A^{+}_{\da}$ in the components $\cM, \ldots, \bar\cL$.

\subsection{Level $(\bq_+)^0$}

The first component $\cT(x,\q_+,w)$ in the expansion \p{4.33} is the so-called \emph{ chiral} stress-tensor multiplet studied extensively in \cite{Chicherin:2014uca}, in the twistor framework.\footnote{In reality,  the stress-energy tensor $T^{\a\da, \b\db}(x)$ is not contained in $\cT$ but is a component of the semi-superfield $\cS^{(\da\db)}= \ldots + (\q^2)_{(\a\b)}T^{\a\da, \b\db} $ (see \p{37} and \p{4.33}). }  Here we repeat the construction, in the LHC framework.  

We want to find a gauge-invariant operator of dimension $2$ which satisfies the chiral half of the R-analyticity conditions \p{5.3}, namely, 
\begin{align}\label{5.11}
\pa^{\a}_{+ a'} \cT=0\ \Rightarrow \ \cT=\cT(x, \q_+,w) \,.
\end{align}
The other half of \p{5.3} involves the RH-projected derivative $\bar \pa^{a}_{+\da}= w^{a}_{+A}\bar \pa_\da^A $, but it is trivial in our chiral setup, $\bar \pa_\da^A \cT=0$. Both properties, gauge invariance and R-analyticity, can be achieved by acting on the  Lagrangian $L_{\rm Z}$ \p{lint} with   the R-analyticity projection operator or equivalently the integral over the anti-analytic odd variables $\q_-$,
\begin{align}\label{439}
&\cT(x,\q_+,w) = \frac{1}{\omega} (\pa_{+})^4 L_{\rm Z} \equiv \frac{1}{\omega} \int d^4\q_- \, L_{\rm Z}\nt
& =  (\pa_{+})^4\ \tr\sum^\infty_{n=2}{(-1)^n  g^{n-2}\over n} \int du_1\ldots du_n\; {A^{++}(x, \q^+_1,u_1) \ldots A^{++}(x, \q^+_n,u_n) \over
(u^+_1u^+_2) \ldots (u^+_nu^+_1)}       \ .
\end{align}
Here we have displayed explicitly the gauge coupling constant $g$. 
{Note that the  constant $\omega$ drops out from $\cT$, so if we wish, we can set $\omega =0$ and still have a non-trivial $\cT$.}
The R-analyticity of $\cT$  is manifest because $ (\pa_{+})^5=0$. Its gauge invariance is based on the property  \p{3.32}. {Indeed,}
\begin{align} \notag
\delta_{\Lambda} \cT(x,\q_+,w)  = \frac{1}{\omega} (\pa_{+})^4 \int d u \Delta L = \frac{1}{4\omega} \int d u\ (\pa^-_+)^2  (\pa^+_+)^2 \Delta L = 0\,,
\end{align}
where we have split the derivatives  $(\pa_+)^4 = \frac{1}{4}(\pa^-_+)^2 (\pa^+_+)^2$, under the LH integral,  by projecting with LHs $u^{\pm}$. 
The two derivatives $(\pa^+_+)^2$ annihilate the gauge variation $\Delta L$. In fact, the gauge invariance of $\cT$ suffices for showing that of the action $S_{\rm Z}$ (see~\p{N4}). Indeed, the Grassmann measure $d^8 \q$ in $S_{\rm Z}$ can be alternatively split up using RHs, $(\pa)^8 = (\pa_-)^4 (\pa_+)^4$, therefore 
\begin{align} \label{TtoS}
S_{\rm Z} = \int d^4 x d^8 \q \ L_{\rm Z} = \omega\int d^4x (\pa_-)^4 \cT = \omega\int d^4x d^4\q_+\, \cT\,.
\end{align}
{Eq.~\p{TtoS} is at the origin of the Lagrangian insertion formula \p{ins} that relates the Born-level correlators and the integrands of the loop corrections.}

The operator $\cT$ is represented schematically in the first picture in Fig.~\ref{TMfig}. There each line corresponds to a superfield $A^{++}$ in the expansion \p{439}.

An alternative form of the operator $\cT$ is given by the standard expression for the stress-tensor multiplet (`supercurrent') in $\cN=4$ harmonic (R-analytic) superspace \cite{Hartwell:1994rp}:
\begin{align}\label{5.14}
T = -\frac{1}{8} {\rm tr} (W_{++})^2\,.
\end{align}
Here $W_{++}(x, \q^{a}_+,\bq_{+a'},w) = \bar w^A_{+a'} \ep^{a'b'} \bar w^B_{+b'} W_{AB}$ and $W_{AB}$ is the SYM curvature defined by the full (non-chiral) constraint \p{3.7}. In our chiral description it is truncated to $W_{++}(x,\q_+,w) + O(\bq_+)$. We then have  (see \p{3.12})
\begin{align}\label{TW}
\cT = -\frac{1}{8} {\rm tr} (W_{++})^2\quad {\rm with} \quad W_{++}= \bar w^A_{+a'} \ep^{a'b'} \bar w^B_{+b'} \pa^+_A \pa^+_B A^{--} \equiv (\pa^+_+)^2 A^{--}\,.
\end{align}
The equivalence of the two forms \p{439} and \p{TW} is shown in Appendix \ref{proofTW}. We remark that the last relation implies $\pa^+_{+a'} \cT = 0$, i.e. the shortening condition \p{5.11} projected with the LH $u^+_\a$. Using the LH independence of $\cT$ (see Appendix~\ref{proofTW}), we can remove this projection and recover the full constraint  \p{5.11}.

Let us now examine the component field content of $\cT$. The operator is made  from the semi-superfield $A^{++}$ only, therefore it  is appropriate  to use the WZ gauge  \p{23}. To simplify the discussion we restrict it to the Abelian (linearized or $g=0$) case and we drop the fermions, 
\begin{align}\notag
A^{++} =  (\q^{+})^{2\ AB} \phi_{AB} +  3(\q^+)^4 u^{-\a} u^{-\b} G_{\a\b}\,.
\end{align}
From \p{439} we obtain
\begin{align}
\cT & =  - \frac{1}{2} (\pa_+)^4 \int \frac{du_1 du_2}{(u^+_1 u^+_2)^2}\ A^{++}(x,\q \cdot u_1^+,u_1) A^{++}(x,\q \cdot u_2^+,u_2) \nt
& = {-\frac{1}{2}} (\phi_{++})^2 
+ {\frac{1}{2}} (\q^{\a}_+ \cdot \q^{\b}_+) \ \phi_{++} G_{\a\b} - {\frac{1}{2}} (\q_+)^4 G_{\a\b} G^{\a\b} \notag
\end{align}
with $(\q^{\a}_+ \cdot \q^{\b}_+) \equiv \q^{\a a}_+ \ep_{a b} \q^{\b b}_+$. This calculation uses the definitions \p{contr} and the  identitiy
\begin{align} \notag
(\pa_+)^4 \ \q^{\a A} \q^{\b B} \q^{\gamma C} \q^{\delta D} = \ep^{\a \delta} \ep^{\b \gamma} (\bar w^A_+ \cdot \bar w^C_+) (\bar w^B_+ \cdot \bar w^D_+) - \ep^{\a \gamma} \ep^{\b \delta} (\bar w^A_+ \cdot \bar w^D_+) (\bar w^B_+ \cdot \bar w^C_+)\,
\end{align}
with $ (\bar w^A_+ \cdot \bar w^B_+) \equiv  \bar w^A_{+a'} \ep^{a'b'} \bar w^B_{+b'} $.
The bottom component is the half-BPS operator \p{5.2}. At the level of $(\q_{+})^4$ in $\cT$ we are supposed to find the linearized {chiral on-shell Lagrangian} (see the first line in \p{4.48}). There the field $G_{\a\b}$ is just a Lagrange multiplier, a priori having nothing to do with the self-dual YM curvature $F_{\a\b}$. The two are identified only after adding the Chern-Simons piece of the action (see~\p{CS}) and solving the algebraic equation for $G_{\a\b}$, more precisely $F_{\a\b} = \omega G_{\a\b}$. 

\subsection{Level $\bq^{\da}_{+a'}$}

Let us now examine the first $\bQ$ superpartner of $\cT$, the fermionic semi-superfield $\cM$ {from the expansion \p{4.33}}. It is obtained as the $\bQ-$variation of $\cT$ defined in \p{439}. Taking into account \p{Lvar}, \p{4.28'} and \p{4.27}, we find
\begin{align}\label{5.12}
\bQ^A_\da \cT  = -(\pa_+)^4 \tr \int  du\;  \bar{\mathbb{Q}}^A_\da A^{++}  A^{--} = - (\pa_+)^4 \tr \int  du \;  \q^{+A} (A^+_\da + \pa^-_\da A^{++}) A^{--}\,.
\end{align}
On the other hand, from \p{4.33} we deduce $ \bQ^A_\da \cT  = -  \bar w^A_{+ a'}  \cM_\da^{a'}$,
hence\footnote{The other projection of the variation, $ w^a_{+A} \bQ^A_\da \cT $ is trivial due to the half-BPS nature of the stress-tensor multiplet. For explanations see Appendix~\ref{BPS}.}
\begin{align}\label{435}
 \cM_\da^{a'} = - w_{-A}^{a'}   \bQ^A_\da \cT = (\pa_+)^4\tr \int  du \;   \q^{+ a'} _-\, (A^+_\da +\pa^-_\da A^{++})  A^{--}\,.
\end{align}
The new operator is { automatically} gauge invariant as a consequence of the commutativity of the $\bQ$ supersymmetry and the gauge transformations~\p{gcom}. We emphasize that the prepotential $A^+_\da$ appears for the first time in the $\bQ$ variation of $\cT$, it is not necessary for the description of the chiral sector of the stress-tensor multiplet. 

\begin{figure}
\begin{center}
\includegraphics[height = 3.5 cm]{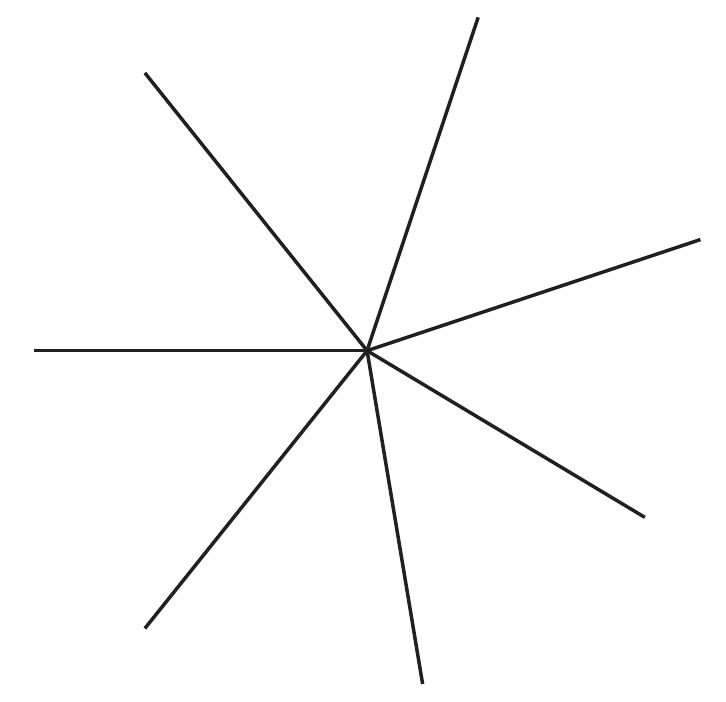} \qquad \qquad \includegraphics[height = 3.5 cm]{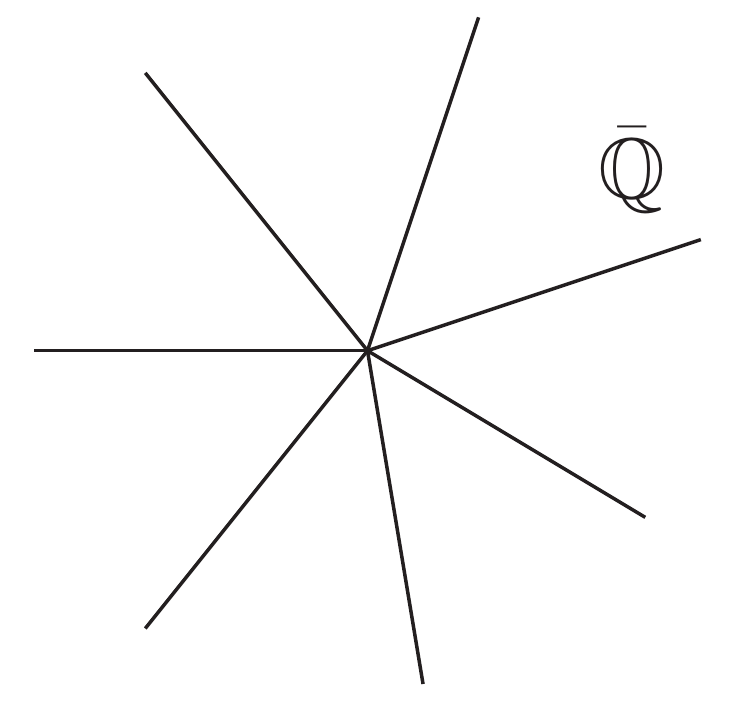} 
\end{center}
\caption{Schematic representation of the operators $\mathcal{T}$  \p{439} and $\cM$  \p{5.12}, respectively. 
The diagram corresponds to a term in the infinite expansion with lines representing the superfields $A^{++}$. $\cM$ is obtained from $\cT$  by a $\bar{\mathbb{Q}}$-variation (see~\p{4.27}) of one $A^{++}$.} \label{TMfig}
\end{figure}

The operator $\cM$ is depicted in the second picture in Fig.~\ref{TMfig}. Roughly speaking, it is obtained from $\cT$ by replacing one $A^{++}$ in the expansion \p{439} by $\bar{\mathbb{Q}}_{-} A^{++}$.

Let us examine the expression for the superpartner in the WZ gauge \p{23}, \p{25}. Restricting to the first component in $\cM$, we need the term $\sim (\q)^4$ under the harmonic integral. In the WZ gauge {and in the Abelian ($g=0$) case}
\begin{align}\label{512}
& A^{--} = (\q^{-})^{2 AB} \phi_{AB}+ \frac1{4}\ep_{ABCD}\q^{-B} \q^{-C} (\q^{+D} \psi^{-A} - \frac1{3} \q^{-D} \psi^{+A}) \nt
& \qquad  + (\q^-)^4 G^{++} - \q^{+A} (\q^-)^3_A G^{+-} + \ep_{ABCD} (\q^+)^{2 AB} (\q^-)^{2 CD} G^{--} \nt
& A^+_\da + \pa^-_\da A^{++} =\cA^+_\da + \q^{+A} \bar\psi_{\da A}  + O(\q^2)\,.  
\end{align}  
{Substituting this in \p{5.12} we obtain} $\bQ^A_\da \cT  = - \bar w^A_{+a'}\ep^{a'b'} \bar\psi_{+\da  b'} \phi_{++}$.
{Indeed, together with the explicit $\q^{+ A}$ in \p{5.12} the quartic term $(\q)^4$ under the integral contains $\phi \bar \psi$ and $\cA\psi^A $.
The second product does not survive the projector $(\pa_+)^4$ because the field $\psi^A$ alone cannot carry so many $\bar w_+$.}
Finally, the projection with $w^a_{+A}$ gives zero (R-analyticity), while that with $w^{a'}_{-A}$ in \p{435} gives the $\bQ-$superpartner $ \cM_\da^{a'} = \bar\psi^{ a'}_{+\da} \phi_{++}$ of the bottom component ${-\frac{1}{2}}\phi_{++} \phi_{++}$ of $\cT$, as shown in the second line of \p{4.48}.

\subsection{Level $ (\bq_+^2)_{(a'b')} $}

Now we implement the second $\bQ$-variation of ${\cT}$.
We start with the operator
\begin{align}\label{4.57}
\cR^{(a'b')} = -\frac14 \ep^{\da\db} \bQ^{(a'}_{-\da} \bQ^{b')}_{-\db}{\cT}\,.
\end{align}
As before, this operator is gauge invariant due to the mechanism \p{gcom}. It consists  of two types of terms, 
\begin{align}\label{4.58}
\cR^{(a'b')} = \cR^{(a'b')}_0+\cR^{(a'b')}_{\rm Z}\,. 
\end{align}
In the first type the two $\bQ$ act on different $A^{++}$  from ${\cT}$ (see~\p{439}), in the second type both $\bQ$ act on the same $A^{++}$.

\begin{figure}
\begin{center}
\includegraphics[height = 3.5 cm]{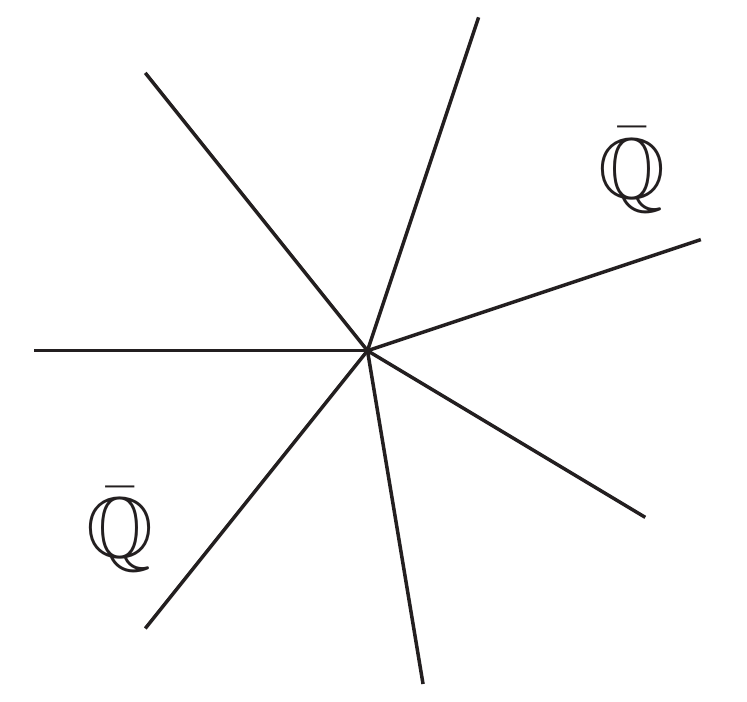} \qquad \includegraphics[height = 3.5 cm]{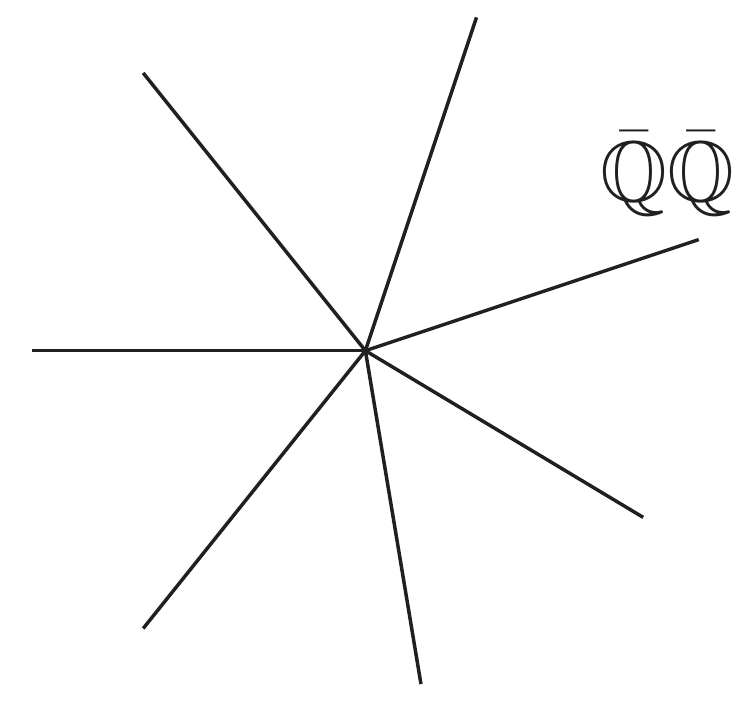} \qquad \includegraphics[height = 3.5 cm]{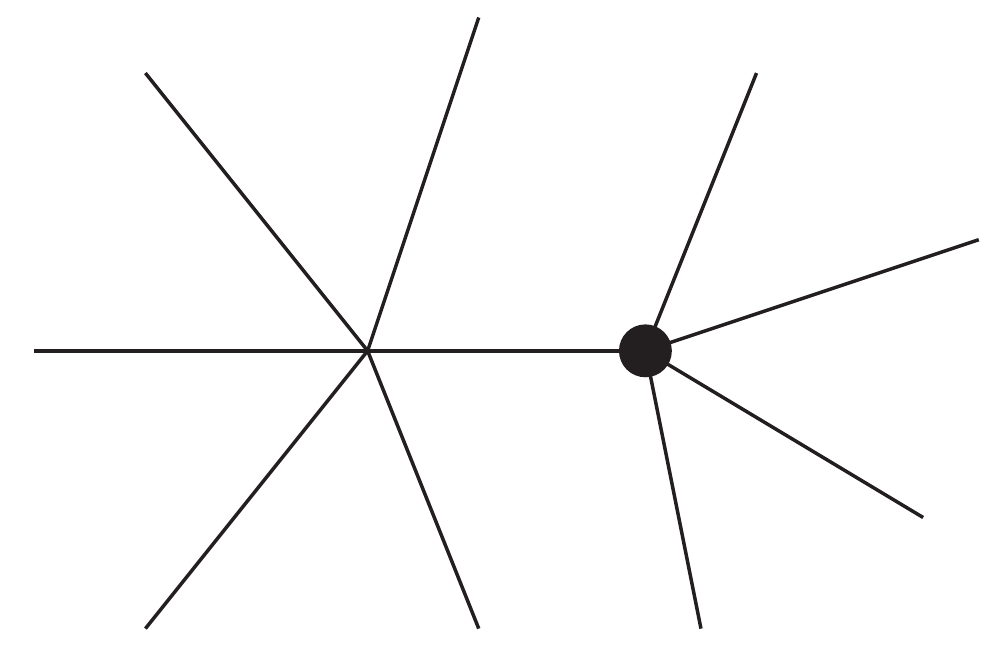}
\end{center}
\caption{Diagrams depicting the operators $\cR_0$ \p{O'a'b'} and $\cS^{'}$ \p{478}; $\cS^{''}$ \p{Odadb}; $\cR_{\rm Z}$ \p{monstr0}. 
They are obtained from the second picture in Fig.~\ref{TMfig} acting by $\bar{\mathbb{Q}}$ and $\bar Q_{\rm Z}$ (see~\p{4.27}).
The blob in the last picture corresponds to the L-analyticity projector $(\pa^+)^4$ in  \p{monstr0}.}
\label{RSfig}
\end{figure}

The first term $\cR^{(a'b')}_0$ is obtained with the help of $\bar {\mathbb{Q}}$ defined in \p{4.27}:  
\begin{align}
\cR^{(a'b')}_0  = -\frac14
(\pa_+)^4\tr  \sum_{n=2}^{\infty}\sum_{j = 1}^{n-1} (-1)^n g^{n-2} \int \frac{d u_1 \cdots d u_n}{(u_1^+ u_2^+)\cdots (u_n^+ u^+_1)}\  \ep^{\da\db}\, 
 A^{++}(u_1) \cdots A^{++}(u_{j-1})   \nt
\times\    \q^{+ (a'}_{j-}\left[\pa^-_{\da} A^{++} + A^+_{\da}\right](u_j) 
 A^{++}(u_{j+1}) \cdots A^{++}(u_{n-1}) \ \q^{+ b')}_{n-} \left[\pa^-_{\db} A^{++} + A^+_{\db}\right](u_n) \label{O'a'b'}\,,
\end{align}
due to the cyclic property of the trace.  After the symmetrization $(a'b')$, the two terms with $\bar{\mathbb{Q}}_{-} A^{++}$ are automatically antisymmetric in $\da,\db$, as follows from relabeling the points under the trace. 
The operator $\cR_0$ is depicted in the first picture in Fig.~\ref{RSfig}.

In the second term $\cR^{(a'b')}_{\rm Z}$ both ${\bQ}$ act on the same $A^{++}$. In this combination the double variation $\bar{\mathbb{Q}} \bar{\mathbb{Q}}$ does not appear.
Indeed, a simple calculation leads to
\begin{align} \label{cQcQ}
\bar {\mathbb{Q}}^{a'}_{-\da} \bar {\mathbb{Q}}^{b'}_{-\db} A^{++} = - \ep^{a'b'} (\q^+_-)^2 \left[ 
\pa^{-}_{(\da} A^{+}_{\db)} + {\textstyle\frac{1}{2}} \pa^-_{\da} \pa^-_{\db} A^{++} \right] 
\end{align}
where $(\q^+_-)^2=\q^{+ a'}_- \ep_{a'b'}  \q^{+ b'}_- $.
This expression vanishes upon the symmetrization $(a'b')$. Therefore we have 
\begin{align} \label{3.34}
\ep^{\da\db} \bQ^{(a'}_{-\da} \bQ^{b')}_{-\db} A^{++} = \ep^{\da\db} (\bar Q_{\rm Z})^{(a'}_{-\da} \bar {\mathbb{Q}}^{b')}_{-\db} A^{++}\,.
\end{align}
Using \p{4.27}, \p{5.12} and \p{3.34} we obtain 
\begin{align}
\cR^{(a'b')}_{\rm Z}  &= \frac14 (\pa_+)^4\tr \int d u\;
A^{--} \, \ep^{\da\db} \bQ^{(a'}_{-\da} \bQ^{b')}_{-\db} A^{++} \nt &=  \frac{\omega}{2}  (\pa_+)^4\tr \int d u\;
A^{--}  \, (\pa^{+})^4 \left[ \q^{+ (a'}_{-}\q^{- b')}_{-} A^{--} \right]\,. \label{monstr0} 
\end{align}
The operator $\cR_Z$ is depicted in the third picture in Fig.~\ref{RSfig}.
Taking into account \p{contr} we rewrite \p{monstr0} as follows
\begin{align}
& \frac{\omega}{8}  (\pa_+)^4\tr \int d u\; (\pa^+_+)^2 A^{--}   \, (\pa^+_-)^2 \left[ \q^{+ (a'}_{-}\q^{- b')}_{-} A^{--} \right]     \nt
&= \frac{\omega}{8} (\pa_+)^4 \int d u\; \q^{+ (a'}_{-}\q^{- b')}_{-} \;    \tr (W_{++} \, W_{--} )\,.  \notag
\end{align}
In the last line we have used $W_{++}= (\pa^+_+)^2 A^{--}$ from \p{TW} and its analog $W_{--} = (\pa^+_-)^2 A^{--}$. Also,  the corollary of \p{314} $\pa^{++}  \tr (W_{++} \, W_{--} ) =0$ implies that this LH-chargeless quantity is independent of $u^\pm$, so the harmonic integral becomes trivial (recall~\p{6}) and we finally obtain 
\begin{align}\label{monstr} 
\cR^{(a'b')}_{\rm Z}  = \frac{\omega}{16} (\pa_+)^4  \left[ \q^{\a(a'}_{-}  \q^{b')}_{-\a}\; \tr ( W_{++}\, W_{--}) \right].
\end{align}
This operator is gauge invariant by itself, which implies that its complement $\cR^{(a'b')}_0$ also is. 

{Note that $\cR_{\rm Z}$ is proportional to $\omega$, thus its occurrence  is similar in a sense to generating a loop correction  (see the discussion in Section \ref{s41}). 
We will not need it in the calculation of the Born-level correlators.
Another important property} of the operator $\cR^{(a'b')}_{\rm Z}$, closely related to the previous one,  is that it vanishes at the linearized (free or Abelian or $g=0$)  level, i.e.,  
\begin{align}\label{5.25}
\cR^{(a'b')}_{\rm Z} = O(g)\,.  
\end{align}
Since  the operator is gauge invariant, this can be shown in the WZ gauge. We find 
\begin{align}\notag
W_{--}= -2 \phi_{--} + \q^{\a a'}_- \psi_{-a' \a} + (\q^{\a}_-\cdot \q^\b_-) G_{\a\b} + O(g)
\end{align}
with $\phi_{--} = -\frac12 (w_{-A} \cdot w_{-B})\ \phi^{AB}$, $\psi^{a'}_{-\a} = w^{a'}_{-A} \psi^{A}_{\a}$ and $(\q^{\a}_-\cdot \q^\b_-) = \q^{\a a'}_- \ep_{a' b'} \q^{\b b'}_-$. Notice the absence of LHs, as follows from the linearized property $\pa^{++}W_{--}= 0$. Further, the derivatives $\pa_+$ in \p{monstr} do not act on $W_{++}$, so we need to consider $(\pa_+)^4 \left[ \q^{(a'}_{-}\cdot  \q^{b')}_{-}\;   W_{--} \right] = (\pa_+)^4 \left[ \q^{(a'}_{-}\cdot  \q^{b')}_{-}\;   \q^{\a}_-\cdot \q^\b_- G_{\a\b} + O(g)\right] = O(g)$.  

Let us find the non-Abelian component $O(g)$ at the bottom of $\cR_{\rm Z}$.
Restricting $A^{++}$ in the WZ gauge \p{23} to the scalars only, we obtain by means of \p{448}    
\begin{align}\notag
A^{--} = (\q^-)^{2 AB} \phi_{AB} - \frac12 g\ [\phi^{CE},\phi_{CD}] (\q^-)^3_{E} \q^{+D}  + O(\theta,g^2)\,.
\end{align}
From here we find the projection $W_{++}$ of the curvature according to \p{TW},
\begin{align}\notag
W_{++} = -2 \phi_{++} - g(w_{+E} \cdot w_{+H}) \ [\phi^{CE},\phi_{CD}] \q^{-H} \q^{+D}  + O(\theta,g^2)\,,
\end{align}
and the projection $W_{--}$ is obtained by replacing $w_{+A} \to w_{-A}$ in the previous formula, i.e. by exchanging lower level $+$ and $-$. Note that $W_{++}$ explicitly  
depends on the LHs through the projected odd variables $\q^\pm$ and carries zero LH-charge. It is only covariantly harmonic independent, see \p{314}.
Acting with the R-analyticity projector $(\pa_+)^4$ in \p{monstr} we find
\begin{align} \label{3.43}
\cR^{(a'b')}_{\rm Z} = -\frac{\omega g}{4}\ \tr\ \phi_{++} [\phi^{C(a'}_{-}, \phi_{+C d'}] \ep^{b')d'}\,,
\end{align}
with $\phi^{C a'}_{-} = \phi^{CD} w^{a'}_{-D}$ and $\phi_{+C a'} = \phi_{CD} \bar w^{D}_{+a'}$. 
This corresponds to the second term  in $\cR^{(a'b')}$ from \p{4.48}. The first term $\bar\psi\bar\psi$ arises from  $\cR^{(a'b')}_0 = -\frac14 \bar \psi^{(a'}_{+\da} \bar\psi^{b')\da}_{+}$. Indeed, restricting to the bilinear term in \p{O'a'b'}, we see that the necessary $(\q)^4$ come from the second line in \p{512}, while the harmonic integral is trivial.

\subsection{Level $ (\bq_+^2)_{(\da\db)} $}

Now we consider the second $\bQ$-variation of ${\cT}$ of the type
\begin{align}\label{4.72}
\cS_{(\da\db)} = \frac14 \ep_{a'b'} \bQ^{a'}_{-(\da} \bQ^{b'}_{-\db )}{\cT}\,.
\end{align}
{It is obtained as the first variation of \p{5.12}. 
This variation acts either upon $A^{--}$ or $ A^{+}_{\db} + \pa^-_{\db} A^{++}$. 
In other words, the two variations from \p{4.72} act either on two different $A^{++}$ in $\cT$ (see  \p{439}) 
or on the same $A^{++}$.} 
Consequently, the operator $\cS_{(\da\db)}$ consist of 
terms of two types, 
\begin{align}\label{4.76}
\cS_{(\da\db)} = \cS^{'}_{(\da\db)} + \cS^{''}_{(\da\db)} \,.
\end{align}
The expression for $\cS^{'}$ is similar to $\cR_0$ \p{O'a'b'},
\begin{align}
\cS^{'}_{(\da\db)} &= \frac14
 (\pa_+)^4\tr  \sum_{n=2}^{\infty}\sum_{j = 1}^{n-1} (-1)^n g^{n-2}
 \int \frac{d u_1 \cdots d u_n}{(u_1^+ u_2^+)\cdots (u_n^+ u^+_1)} \  \ep_{a'b'}  \;
 A^{++}(u_1) \cdots A^{++}(u_{j-1})  \nt
& \times\   \q^{+ a'}_{j-}\left[ \pa^-_{(\da} A^{++} + A^+_{(\da} \right](u_j) 
 A^{++}(u_{j+1}) \cdots A^{++}(u_{n-1}) \ \q^{+ b'}_{n-} \left[ \pa^-_{\db)} A^{++} + A^+_{\db)} \right](u_n)\,.     \label{478}
\end{align}
The operator $\cS^{'}$ is shown in the first graph in Fig.~\ref{RSfig}.
 $\cS^{''}$ is obtained from $\ep_{a'b'} \bQ^{a'}_{-(\da} \bQ^{b'}_{-\db )} A^{++}$.
Due to the symmetrization $(\da\db)$, $\bar Q_{\rm Z}$  does not contribute to the variation of $A^{+}_{\da}$ (see~\p{4.27}). So, we can replace $\bQ$ by $\bar{\mathbb{Q}}$ in the combination
$
\ep_{a'b'} \bQ^{a'}_{-(\da} \bQ^{b'}_{-\db )} A^{++} = 
\ep_{a'b'} \bar {\mathbb{Q}}^{a'}_{-(\da} \bar {\mathbb{Q}}^{b'}_{-\db )} A^{++} \,,
$
already evaluated in \p{cQcQ}. All other terms  appear in a symmetrized form. Thus we have
\begin{align}
\cS^{''}_{(\da\db)} &= -\frac12 (\pa_+)^4\tr \int  du \; (\q^+_-)^2\, 
\left({\textstyle \frac{1}{2}} \pa^-_\da \pa^-_\db A^{++} + \pa^{-}_{(\da} A^+_{\db)}\right) A^{--}\,. \label{Odadb}
\end{align}
The operator $\cS^{''}$ is depicted in the second picture in Fig.~\ref{RSfig}.

The operator $\cS$ is gauge invariant, but its two constituents $\cS^{'}$ and $\cS^{''}$ are not. Unlike the operator $\cR_{\rm Z}$, here both operators \p{478} and \p{Odadb} have linearized (Abelian) terms. In the WZ gauge  \p{512} one can easily see that $\cS'$ starts with $\cS^{'}_{(\da\db)} =  \frac14 \bar \psi_{+a'(\da} \bar\psi_{+\db) b'} \ep^{a'b'}$. The calculation is analogous to the one for $\cR_0$. 

One can also easily find the bottom component of $\cS^{''}$ in the Abelian approximation. 
The term $\sim (\q)^4$ under the harmonic integral comes from the lowest components in 
$A^{--} = (\q^-)^{2 AB} \phi_{AB} + O(\q^3)$ (see~\p{512})
and ${\textstyle \frac{1}{2}} \pa^-_\da \pa^-_\db A^{++} + \pa^{-}_{(\da} A^+_{\db)} = \pa^{-}_{(\da} \cA^+_{\db)} + O(\q)$.
Substituting this in \p{Odadb} we find 
 $\cS^{''}_{(\da\db)} = - \frac12 \phi_{++} \tilde F_{(\da \db)}$ with $\tilde F_{(\da \db)} = \pa^{\a}_{(\da} \cA^+_{\a\db)}\,$. Together the bottom components  of $\cS^{'}+\cS^{''}$ match the expression for $\cS$ in \p{4.48}.

\subsection{Levels $(\bq_+)^3$ and  $(\bq_+)^4$}

At the two top levels in the expansion \p{4.33} we observe a similar pattern. We define these operators 
as the triple and quadruple $\bQ$-variations of $\cT$, respectively,
\begin{align} \label{NL}
& \cN_{\da}^{a'} = \frac16\left(\bQ^{b'}_{-\da} \bQ^{\db}_{-b'} \bQ^{a'}_{-\db} - \bQ^{a'}_{-\db} \bQ^{\db}_{-b'} \bQ^{b'}_{-\da} \right) \cT \,,\nt
& \bar\cL = \frac{1}{24}\left( \bQ^{\da}_{-a'} \bQ^{a'}_{-\db} \bQ^{\db}_{-b'} \bQ^{b'}_{-\da} 
- \bQ^{\da}_{-a'} \bQ^{b'}_{-\da} \bQ^{\db}_{-b'} \bQ^{a'}_{-\db} \right) \cT\,.
\end{align}

First of all, let us consider the parts $\cN_0$ and $\bar\cL_{0}$ of these operators obtained by acting with $\bar{\mathbb{Q}}$, i.e. we drop
the term $\bQ_{\rm Z}$ from \p{4.28'} and replace $\bQ \to \bar{\mathbb{Q}}$ in the previous formula. 
Let us recall that the Abelian subalgebra of the $\bar{\mathbb{Q}}$-variations closes  off shell, therefore $(\bar{\mathbb{Q}})^5\cT =0$ and $\bar{\mathbb{Q}}^{a'}_{-\da} \bar\cL_0 = 0$. The order of the operators $\bar{\mathbb{Q}}$ is irrelevant, so in this case the two expressions in \p{NL} coincide.

\begin{figure}[h]
\begin{center}
\includegraphics[height = 2.5 cm]{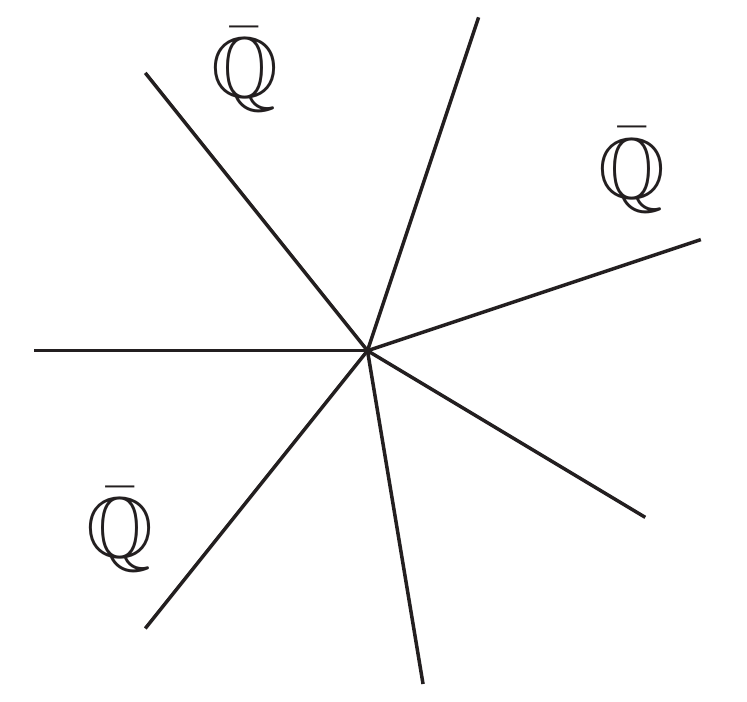} \quad \includegraphics[height = 2.5 cm]{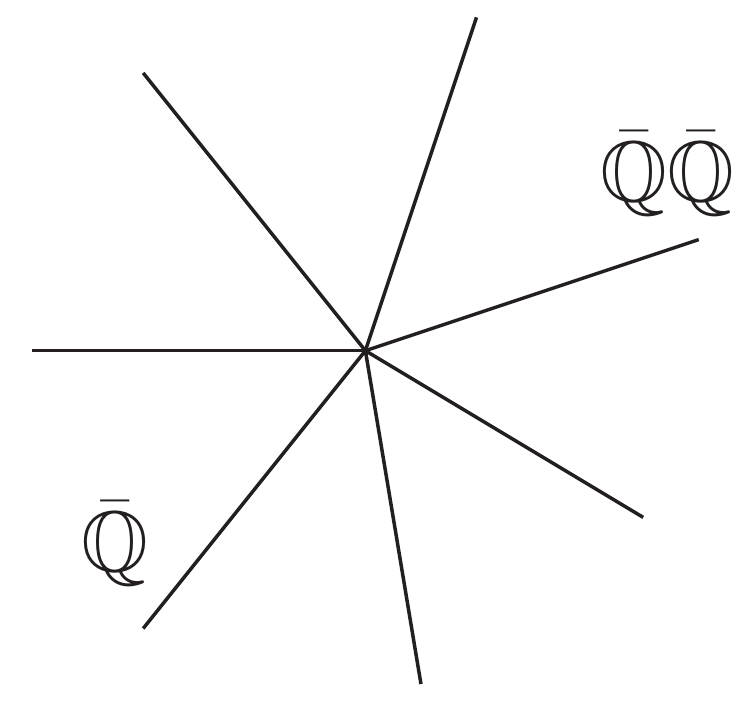} \quad \includegraphics[height = 2.5 cm]{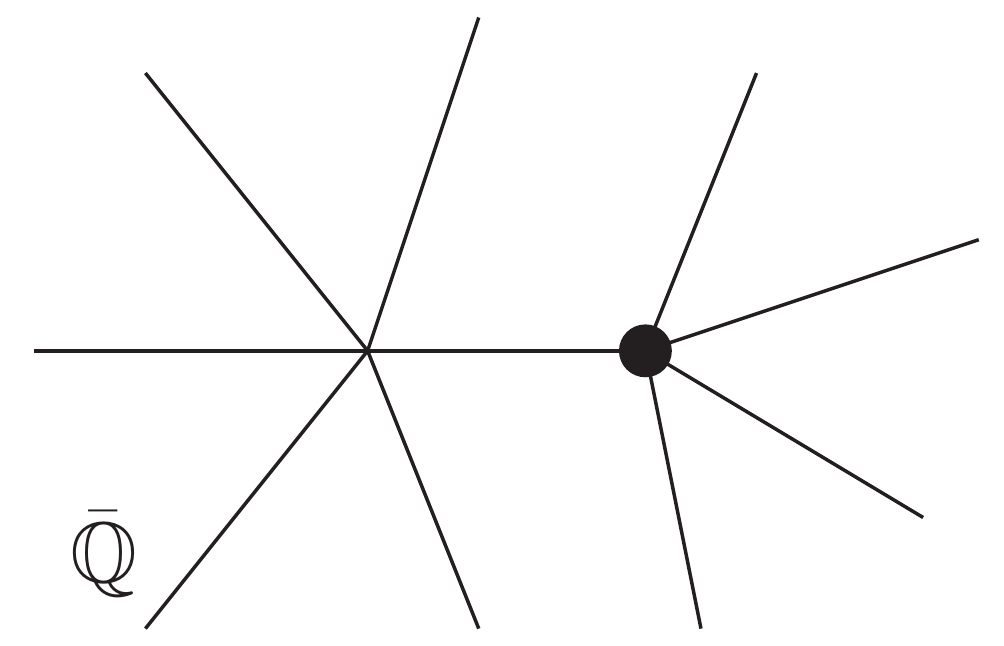}
\quad \includegraphics[height = 2.5 cm]{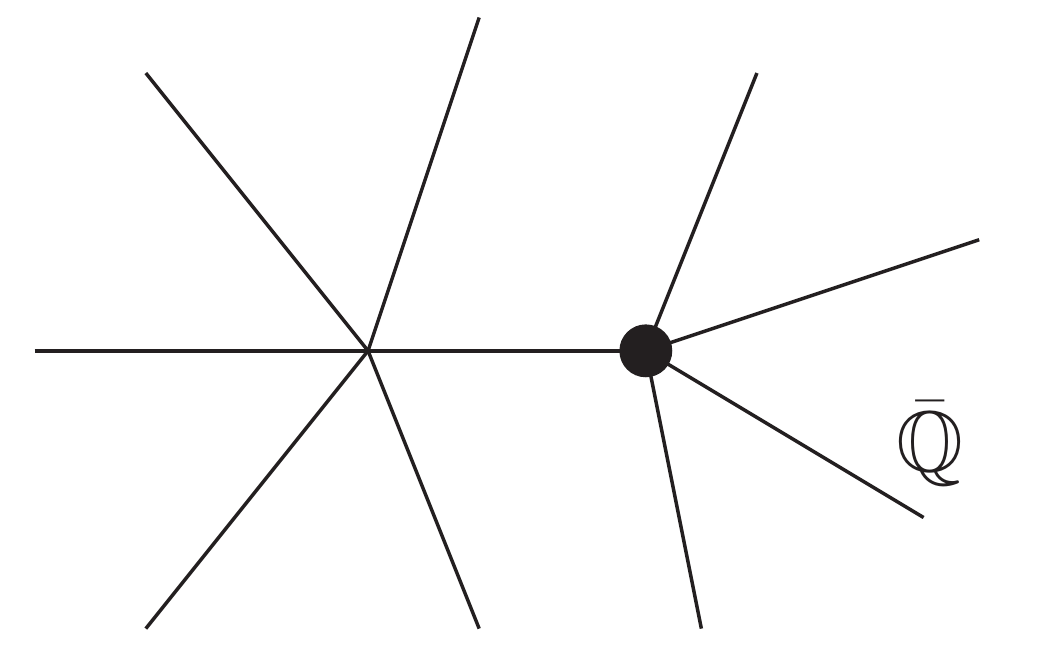}
\end{center}
\caption{The first two diagrams correspond to the operator $\cN_0$ and the last two to $\cN_{\rm Z}$ (see~\p{NLexp}).
They are obtained from the pictures in Fig.~\ref{RSfig} acting by $\bar{\mathbb{Q}}$ and $\bar Q_{\rm Z}$ (see~\p{4.27}).
The blob corresponds to the L-analyticity projector $(\pa^+)^4$.}
\label{Nfig}
\end{figure}

\begin{figure}[h]
\begin{center}
\includegraphics[height = 2.5 cm]{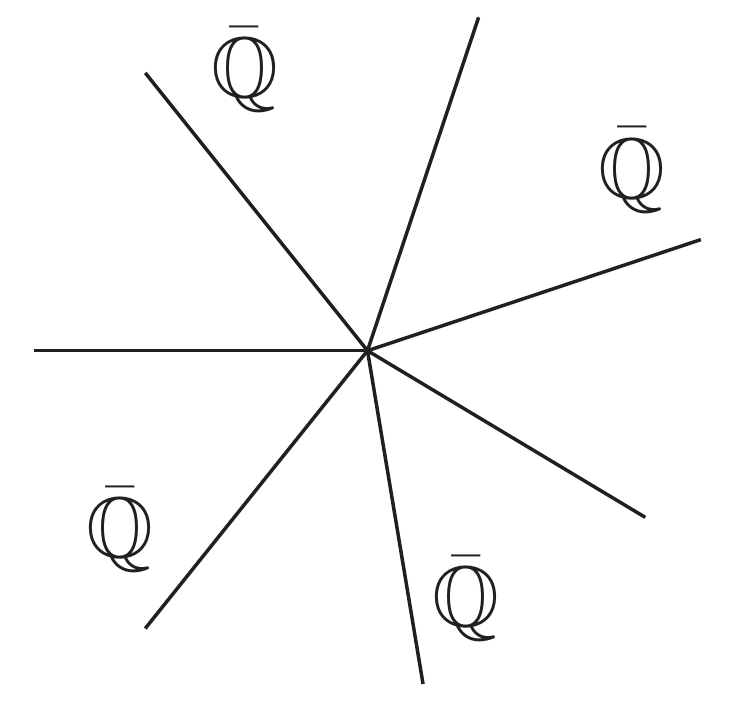} \quad \includegraphics[height = 2.5 cm]{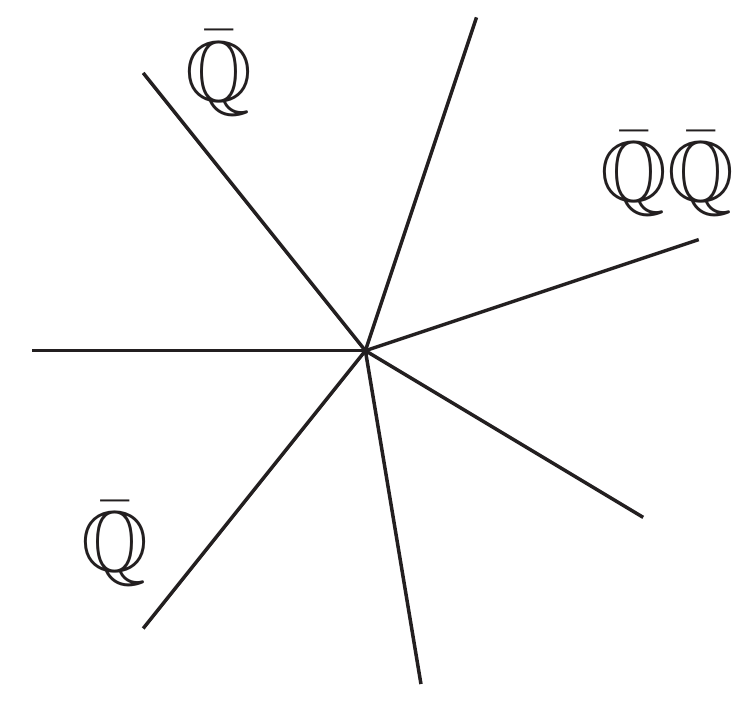} \quad \includegraphics[height = 2.5 cm]{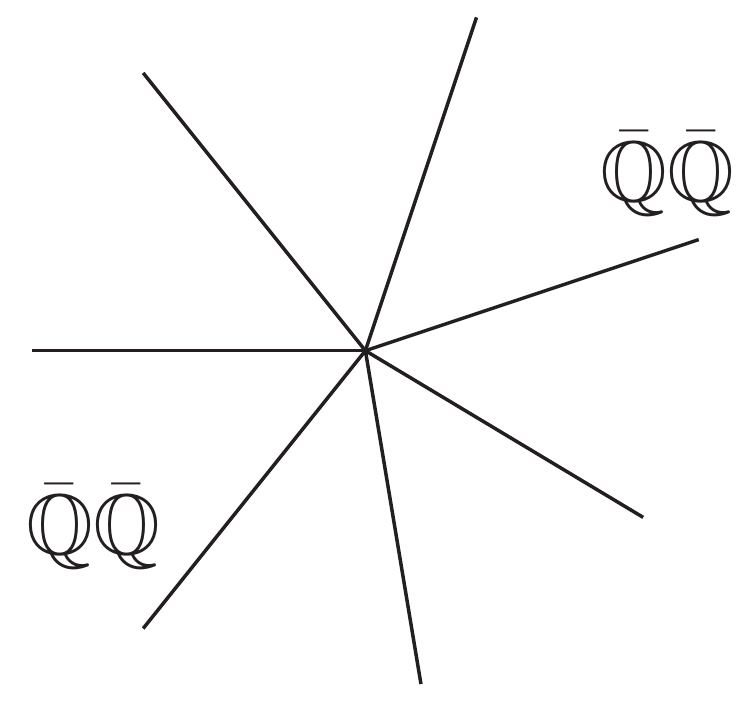} 

\vspace{0.5cm}

\includegraphics[height = 2.0 cm]{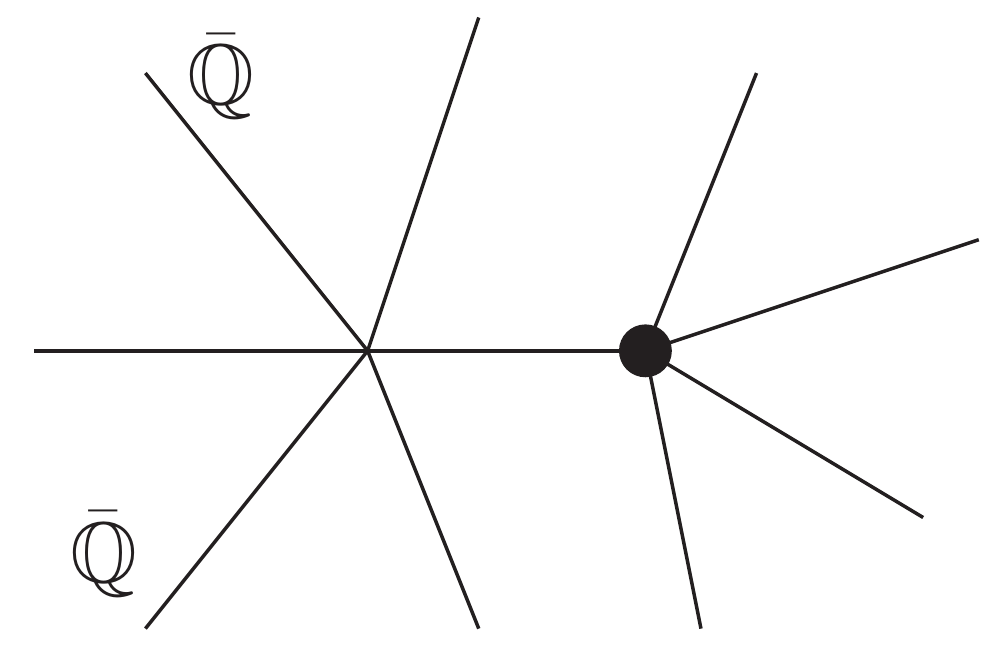} \quad 
\includegraphics[height = 2.0 cm]{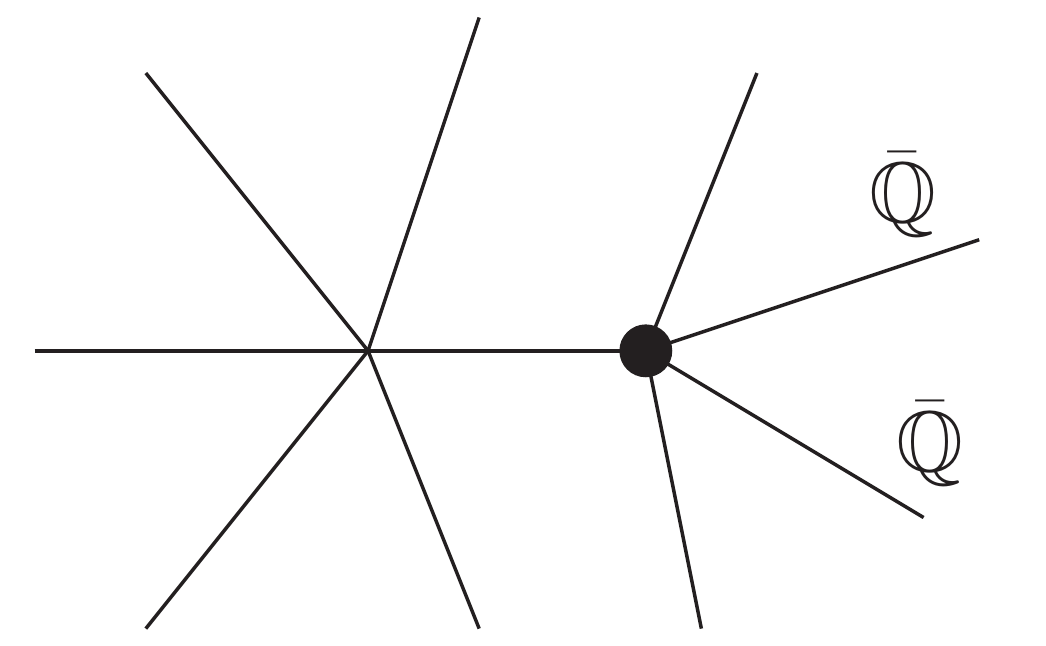} \quad \includegraphics[height = 2.0 cm]{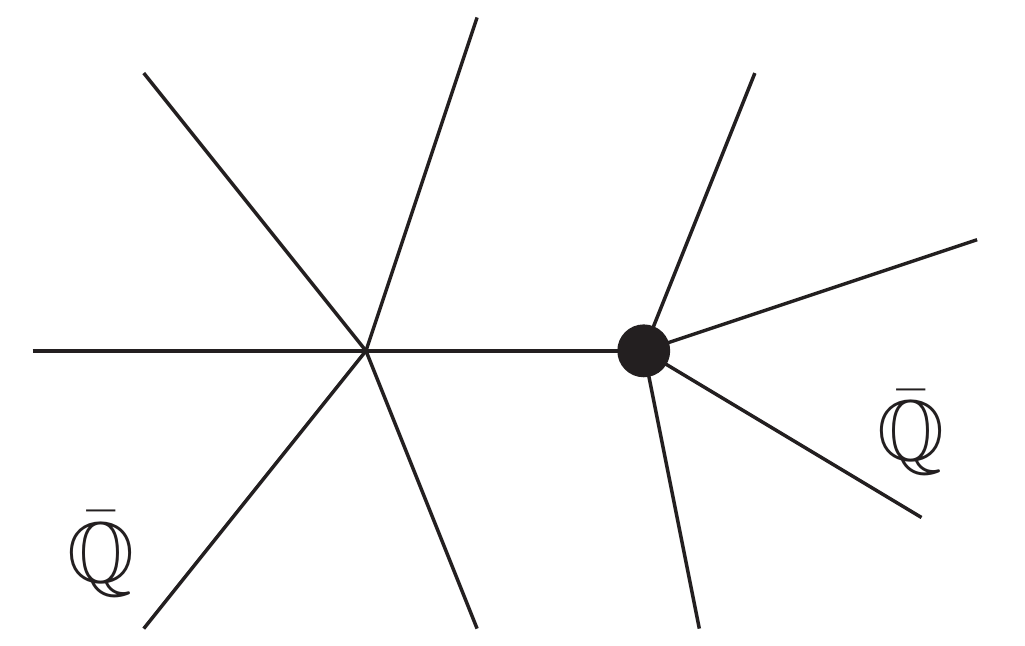} 

\vspace{0.5cm}

\includegraphics[height = 2.0 cm]{Qgr8.pdf} 
\quad \includegraphics[height = 2.0 cm]{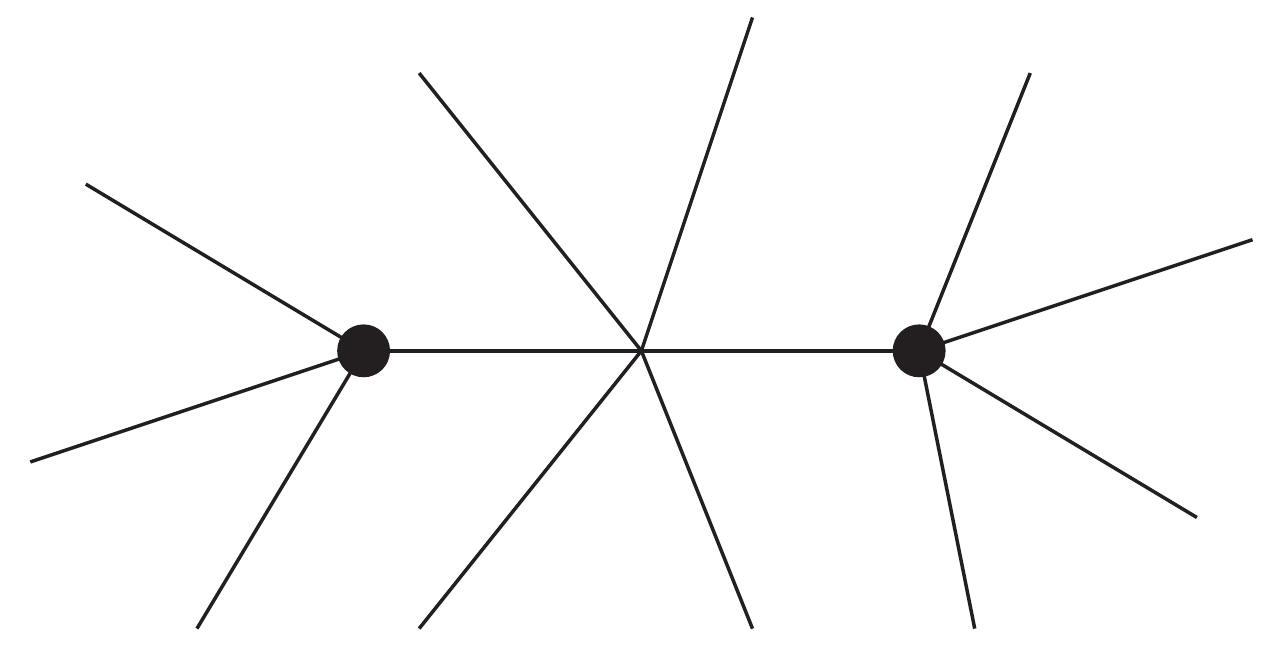} \quad \includegraphics[height = 2.0 cm]{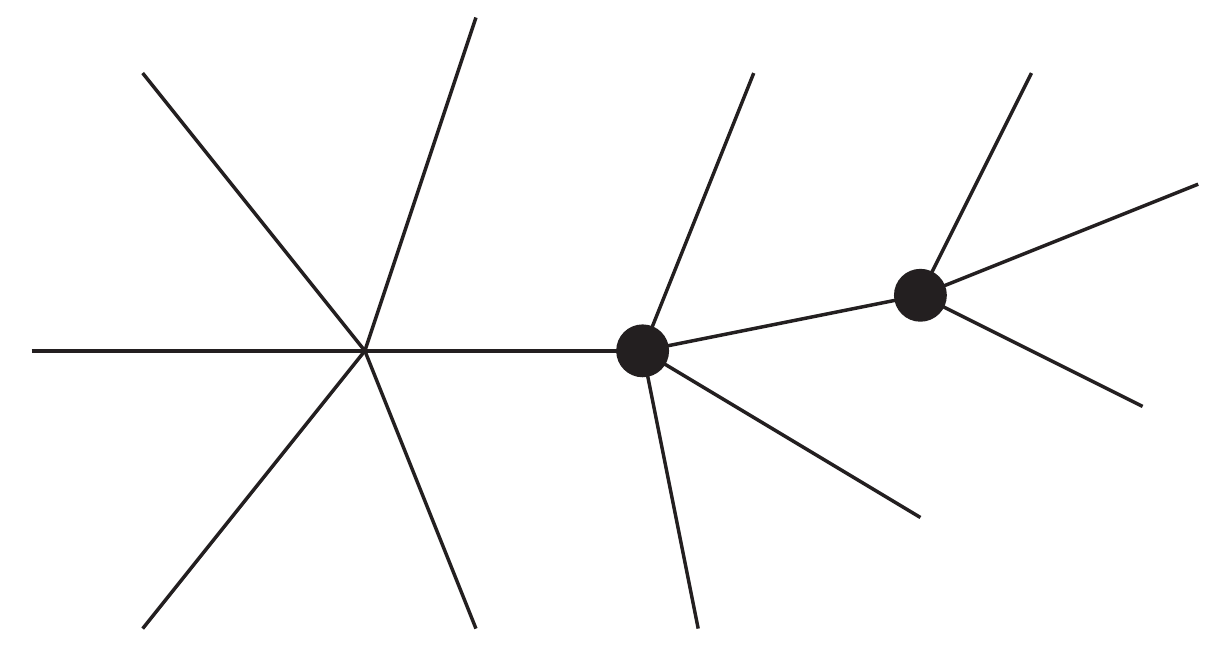}
\end{center}
\caption{The first three diagrams correspond to the operator $\bar{\cL}_0$; the next four to $\bar\cL_{\rm Z}$;
the last two  to $\bar\cL_{\rm Z Z}$ (see~\p{NLexp}).
They are obtained from the pictures in Fig.~\ref{Nfig} acting by $\bar{\mathbb{Q}}$ and $\bar Q_{\rm Z}$ (see~\p{4.27}).
Each blob corresponds to an L-analyticity projector $(\pa^+)^4$ defined with the appropriate harmonic $u^{+}$.
The 7th diagram here is a variation of the 2nd diagram in Fig.~\ref{Nfig}, so it is different from the 3rd diagram in Fig.~\ref{Nfig}.}
\label{Lbarfig}
\end{figure}

Further, $\bar{\mathbb{Q}}$ cannot hit a given $A^{++}$ in \p{439} more than twice,
\begin{align} \label{3.50}
\bar{\mathbb{Q}}^{a'}_{-\da} \bar{\mathbb{Q}}^{b'}_{-\db} \bar{\mathbb{Q}}^{c'}_{-\dot\gamma} A^{++} = 0\,.
\end{align}
This is due to the presence of $\q^+_{-}$ in $\bar{\mathbb{Q}}_{-}$ (see~\p{4.27}).
The operator $\cN_0$ consists of terms of two types depicted in the first two pictures in Fig.~\ref{Nfig}.
Analogously, $\bar\cL_0$ consists of terms of three types depicted in the first three pictures in Fig.~\ref{Lbarfig}.
They are the higher-order analogs of $\cM$ \p{435}, $\cR_0$ \p{O'a'b'} and $\cS$ \p{4.76}. 
In the WZ gauge \p{23}, \p{25} they produce the bilinear terms in the bottom components of the superfields $\cN$ and $\bar\cL$
shown in the last two lines in \p{4.48}. Namely, $\tilde F \bar\psi$ in $\cN_0$ comes from the second 
diagram in Fig.~\ref{Nfig}, and $\tilde F \tilde F$ in $\bar\cL_0$ (see  \p{5.10}) comes from the third diagram in Fig.~\ref{Lbarfig}. 
We will need only the $\cN_0$ and $\bar\cL_0$ pieces of \p{NL} for the calculation of the Born-level correlators.

Next, let us turn on the $\bQ_{\rm Z}$ part of $\bQ$ \p{4.28'}. It is proportional to the constant $\omega$ measuring the 
deviation from the self-dual sector. In this way we find the higher-order analogs of $\cR_{\rm Z}$ \p{monstr0}.
The operators \p{NL} can be decomposed as follows, 
\begin{align} \label{NLexp}
\cN = \cN_0 + \cN_{\rm Z} \;\;,\;\;
\bar\cL = \bar\cL_{0} + \bar\cL_{\rm Z} + \bar\cL_{\rm Z Z}\,,
\end{align}
where the subscripts Z and ZZ refer to the number of $\bQ_{\rm Z}$-variations involved.
Consequently, $\cN_{\rm Z},\, \bar\cL_{\rm Z} \sim \omega$ and  $\bar\cL_{\rm ZZ} \sim \omega^2$.

Let us recall that the $\bQ$-variations form a closed algebra (up to a gauge transformation which is irrelevant since we consider gauge-invariant operators) only on shell. Therefore, we are able to define $\cN_{\rm Z}$, $\bar\cL_{\rm Z}$ and $\bar\cL_{\rm ZZ}$ only 
modulo the field equations \p{3.6}, \p{3.14'}. The two terms in \p{NL} corresponding to different orderings of $\bar{Q}$ are equal on shell. Acting with the fifth $\bQ$ on $\cT$ we will not automatically get zero, but a non-trivial operator which vanishes only on shell.

There are two types of contributions to $\cN_{\rm Z}$  depicted in the last two pictures in Fig.~\ref{Nfig}.
They are obtained in two ways: (i) by the $\bar{\mathbb{Q}}$-variation of the last picture in Fig.~\ref{RSfig} which represents $\cR_{\rm Z}$ \p{monstr0}; (ii) by the $\bQ_{\rm Z}$-variation of the first picture in Fig.~\ref{RSfig} which represents $\cR_{0}$ \p{O'a'b'}.

At the top level of the expansion \p{4.33} the situation is more involved. The operator $\bar\cL_{Z}$ gets contributions from the 4th to 7th diagrams in Fig.~\ref{Lbarfig}. They are obtained in two ways: (i) acting by $\bar{\mathbb{Q}}$ on $\cN_{\rm Z}$ depicted in the last two pictures in Fig.~\ref{Nfig}; (ii) acting by $\bQ_{\rm Z}$ on $\cN_{0}$ depicted in the first two pictures in Fig.~\ref{Nfig}. 
Then we see a new possibility: We can act  on $\cN_{\rm Z}$ by $\bQ_{\rm}$, since it contains the connection $A^{+}_{\da}$.
In this way we obtain the last two diagrams in Fig.~\ref{Lbarfig}. They correspond to the $\bar\cL_{\rm ZZ}$-part of $\bar\cL$ \p{NLexp}. 

In the WZ gauge these operators arising from the $\bQ_{\rm Z}$-variations are responsible for the non-Abelian completions of the bottom components of the superfields in the last two lines in \p{4.48}.
The terms $\omega g[\phi,\phi]\bar\psi$ and  $\omega g\bar\psi[\phi,\bar\psi]$ come from $\cN_{\rm Z}$ and $\bar\cL_{\rm Z}$, respectively.
The term $\omega^2 g^2[\phi,\phi][\phi,\phi]$ originates from $\bar\cL_{\rm ZZ}$.  

The operators $\cN_{\rm Z}$, $\bar\cL_{\rm Z}$ and $\bar\cL_{\rm Z Z}$ are purely non-Abelian. Moreover, $\bar\cL_{\rm Z Z}$
appears only at the second order in the gauge coupling $g$, 
\begin{align} \label{nonAb}
\cN_{\rm Z} = O(g)\quad , \quad \bar\cL_{\rm Z} = O(g) \quad , \quad \bar\cL_{\rm Z Z} = O(g^2)\,. 
\end{align}
Each of the operators $\cN_{0}$, $\cN_{\rm Z}$, $\bar\cL_{0}$, $\bar\cL_{\rm Z}$ and $\bar\cL_{\rm Z Z}$ in \p{NLexp} is gauge invariant. 
It would be interesting to find their expressions in terms of the gauge curvature $W_{AB}$, like  \p{5.14} for $\cT$ and  \p{monstr} for $\cR_{\rm Z}$. 

The explicit expressions for the $\cN_0$ and $\bar\cL_0$ parts (see~\p{NLexp}) of $\cN$ and $\bar\cL$ are clear from Figs.~\ref{Nfig},\ref{Lbarfig}. However, we show only schematically  the non-Abelian completions \p{nonAb} in Figs.~\ref{Nfig},\ref{Lbarfig}, since they are rather involved and are not needed for our purposes here. They play no role in our study of the Born-level correlation functions of the different components \p{4.33} of the stress-tensor supermultiplet $T$ in Section \ref{s8}. Roughly speaking, the $\bQ_{\rm Z}$-variation introduces an additional L-analyticity projector $(\pa^+)^4$ in the game (see~\p{4.27}), which perturbs the match between the Grassmann degree of the Born-level correlation function and its order in the coupling $g$. We explain this in more detail in Section \ref{sanMHV}.

\section{Born-level correlation functions of the stress-tensor multiplet (chiral sector)} \label{s7}

\subsection{General structure of the correlation functions of the stress-tensor multiplet}\label{s41}

In this and the following section we study the $n-$point correlation functions of the stress-tensor multiplet \p{4.33}. The general structure of their Grassmann expansion is as follows:
\begin{align}\label{7.1}
\vev{T(1)\ldots T(n)} = &\sum_{p>0,r\geq0 } (\q_+)^{4p+r} (\bq_+)^{r} G^{\rm nMHV}_{p,r}(x,w)\nt
&+\sum_{r\geq0 }(\q_+)^{r} (\bq_+)^{r} G^{\rm MHV}_{r}(x,w)\nt
+& \sum_{p>0,r\geq0 } (\q_+)^{r}  (\bq_+)^{4p+r} G^{\rm \overline{nMHV}}_{p,r}(x,w)\,.
\end{align}
It is determined by the requirement of invariance under the  $\mathbb{Z}_4$ center of the R-symmetry group $SU(4)$, which acts on the odd variables as follows: $\q \to \zeta \q$, $\bq \to \zeta^{-1} \bq$ with $\zeta^4=1$. The expansion parameters satisfy the inequalities  
\begin{align}\notag
4p\leq 4n-16\,, \qquad r \leq 4n\,, \qquad 4p+r \leq 4n\,.
\end{align}
Indeed,  $\cN=4$ superconformal symmetry has $16+16$ odd generators which can be used to gauge away  16 $\q$'s and 16 $\bq$'s. This is why the maximal power of $\q_+$ in the purely chiral sector (at $\bq_+=0$) is $4n-16$ and not $4n$.

We classify the correlators $G(x,w)$ according to their Grassmann degree $4p$, i.e., the difference between the number of $\q_+$'s and $\bq_+$'s.  It is known that the (singular) light-like limit of the chiral sector of the correlators corresponds to scattering superamplitudes in $\cN=4$ SYM (see \cite{Alday:2010zy}--\cite{Eden:2011ku}). We therefore apply the amplitude terminology by dubbing the correlators $G_{r}$ with $p=0$ `MHV-like', $G_{p,r}$ with $p=1$ are called `NMHV-like' or more generally,  `non-MHV-like' for any $p>0$. Alternatively, the MHV-like correlators may be called non-nilpotent because they have a component at $\q=\bq=0$. The non-MHV-like correlators are nilpotent (they vanish  at $\q=\bq=0$). The third line in \p{7.1} corresponds to the conjugate of the first.\footnote{The stress-tensor multiplet $T(x,\q_+,\bq_+,w)$ is real in the sense of the special conjugation on harmonic superspace, see \cite{Galperin:1984av} for the case $\cN=2$ and \cite{Hartwell:1994rp} for $\cN=4$.} {Our aim in this paper is the calculation of the first two lines in \p{7.1}, using the chiral semi-superfield approach elaborated in the previous Sections, i.e. the correlators for which the number of $\q_+$ is greater or equal to the number of $\bq_{+}$}. Our chiral formulation of $\cN=4$ SYM becomes less efficient for the calculation of (mostly) anti-chiral objects (see Section \ref{sanMHV} for more details), but this is not a problem since the whole information about the nilpotent correlators is contained in the first line in \p{7.1}. The MHV-like correlators in the second line are self-conjugate.

We further classify the purely chiral correlators $G^{\rm nMHV}_{p,0}$ according to the level in the gauge coupling constant $g^{2\ell}$ with $\ell \geq 0$. The minimal possible level for a given type of correlator is known as the Born level. For MHV-like correlators it is $\ell=0$, i.e. it corresponds to the free theory. Such correlators are made from free propagators {in the conventional formulation of the theory}. For non-MHV-like correlators the Born level is $\ell = p$. The easiest way to see this is to recall that the top component in the chiral stress-tensor multiplet $\cT$ in \p{4.48} is $(\q_+)^4 L$ where $L$ is the (chiral on-shell) Lagrangian of $\cN=4$ SYM. Thus, among the components of $G^{\rm nMHV}_{p,0}$ we can find the correlator
\begin{align}\label{LOcor}
\vev{L(1) \ldots L(p) \cO(p+1) \ldots \cO(n)}\,,
\end{align}
where $\cO(x,w)$ is the bottom component of $\cT$ defined in \p{5.2}. This corresponds to $p$ Lagrangian insertions in the $(n-p)-$point correlator  of half-BPS operators $\cO$. Each Lagrangian insertion is equivalent to a derivative of the generating functional with respect to the coupling $g^2$, hence the lowest possible level $g^{2p}$. If the level in the coupling is higher than the minimal one, $\ell > p$, we call such correlators `loop corrections'. A characteristic feature of the Born level correlators is that they are rational functions of the space-time coordinates $x_i$, whereas the loop corrections involve genuine Feynman integrals of the $(\ell-p)-$loop type.\footnote{Exceptions from this rule are possible, see \cite{Korchemsky:2015ssa}. There it is shown that the {connected} MHV-like correlator $\vev{L(1) \bar L(2) \cO(3) \cO(4)}$, which corresponds to $G^{\rm MHV}_{r}$ with $n=4$, $r=4$,  starts at level $g^2$ and contains a one-loop integral. Its disconnected part is still rational, however.}

In this Section we are interested in the purely chiral sector  of the correlator \p{7.1} obtained by setting all $\bq=0$, i.e. the correlators $G^{\rm nMHV}_{p,0}$ and $G^{\rm MHV}_{0}$  in \p{7.1}. {It corresponds to dropping all semi-superfields but $\cT$ in the expansion \p{4.33}. Thus, here we deal with the correlators}
\begin{align} \label{chirCor}
\vev{\cT(1)\ldots \cT(n)}\,.
\end{align}
The question how to restore the dependence on $\bq$ for such correlators, i.e. how to obtain the sector $G^{\rm nMHV}_{p,r}$ and $G^{\rm MHV}_{r}$ in \p{7.1}, is discussed in Section \ref{s8}. As mentioned above, we do not consider $G^{\rm \overline{nMHV}}_{p,r}$ {in detail} for two reasons: (i) our chiral formalism is not convenient for their study; (ii) they can be obtained from $G^{\rm nMHV}_{p,r}$ by conjugation, or, in other words, using the equivalent  antichiral formalism.

In this Section we show several examples of  Born-level  calculations of  chiral correlators. We apply the Feynman rules in LHC superspace outlined at the end of Section \ref{s2} (for more details see~\cite{PartI}). The MHV-like case in Section \ref{s7.1}  corresponds to the free theory and is thus very simple. Nevertheless, it illustrates very well the techniques we use to evaluate the Feynman graphs. 

The non-MHV case in Section \ref{s7.2} shows the main advantage of the LHC formulation of $\cN=4$ SYM. As explained there, to obtain the Born level of a correlator of Grassmann degree $4p$ we ought to collect the necessary power $g^{2p}$. It can come either from  the expansion of $\cT$ in \p{439} in terms of the dynamical superfield $A^{++}$ at the external vertices of the correlator, or from the internal (interaction) vertices in a given Feynman graph. If a power of $g^2$ comes from an internal vertex from the {$S_{\rm Z}$ part of the} action \p{N4}, this automatically lowers the Grassmann degree of the Feynman graph below the required $(\q_+)^{4p}$. Such interaction terms appear in the loop corrections to correlators of lower Grassmann degree, but not at Born level for the given degree $4p$. As a consequence, the Born-level Feynman graphs are made from external multilinear vertices, coming from the expansion of $\cT$,  connected with free propagators \p{prop1}. {Thus the chiral Born-level correlators are of order $\omega^0$ in the normalization constant in \p{lint}}. At no step we need to do Feynman integrals $\int d^4 x$. This is very different from the usual component field Feynman graphs at Born level. There, to obtain a correlator of Grassmann degree $p>0$, one needs to employ genuine interaction vertices for the gluon, gluinos and scalars. This requires the evaluation of certain (not always simple) Feynman integrals. Although the end result is a rational function of {the coordinates} $x_i$, $i=1,\ldots,n$, the intermediate steps may be rather intricate. Examples of such component correlators were given in \cite{Chicherin:2014uca}, another one is shown in Appendix~\ref{apE}. Even if the computation of loop corrections is not the subject of this paper, we give an example of a correlator at one-loop level (involving an interaction vertex) in Appendix~\ref{loop}. Alternatively, we can view it as the calculation of the five-point, $n=5$, Born-level correlator \p{LOcor} with $p = 1$. This LHC supergraph calculation reproduces the partial non-renormalization theorem \cite{Eden:2000bk} at one-loop level.

The content of this section is closely related to the calculation of chiral nilpotent correlators in \cite{Chicherin:2014uca} whithin the alternative twistor formalism. The new original results on the non-chiral sector, for which the LHC formalism has been developed,  appear in Section \ref{s8}.

\subsection{MHV-like correlators}\label{s7.1}

\subsubsection{ There points: $\vev{\cT \cT \cT}$}

We start by calculating the three-point MHV-like correlator at order $O(g^0)$ \footnote{The three- and four-point correlators \p{7.1} must have $p=0$, i.e. they are necessarily MHV-like.  This follows from the absence of nilpotent superconformal R-analytic invariants with less than five points.}
\begin{align}
\label{5.32}
\langle {\cT}(1) {\cT}(2) {\cT}(3)\rangle_{g^0} \ \ = \ \ \begin{array}{c}\includegraphics[height = 2.5 cm]{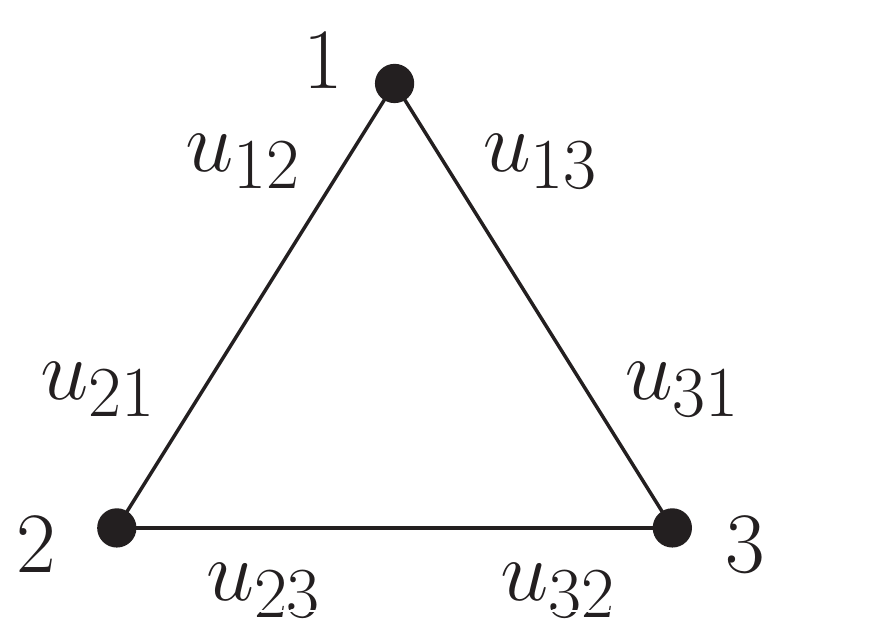}\end{array}
\end{align}
of the chiral operators $\cT$ defined in \p{439}. This correlator involves only bivalent external vertices from the non-polynomial expansion \p{439} of $\cT = {\frac{1}{\omega}}(\pa^+)^4 L_{\rm Z}$. Indeed, recall that the  $n$-valent vertex is proportional to $g^{n-2}$. 
Each vertex brings in the R-analyticity projector $(\pa_+)^4 = \int d^4\q_-$ and each propagator contains $\delta^4(\q^+)$ (see~\p{prop1}), therefore the correlator is of Grassmann degree $(\q)^0$, i.e. it is not nilpotent. 
The bivalent `interaction' vertices induced by the bilinear part of $S_{\rm Z}$ \p{N4} are not allowed. 
Indeed, they bring in eight Grassmann integrations $\int d^8\q$, which would kill any diagram
that consists only of bivalent vertices. There is a single contributing Feynman graph shown in \p{5.32},  which gives (we drop the color factor $N_c^2-1$)
\begin{align}
& - \int d^4 \q_{1-} \, d^4 \q_{2-} \, d^4 \q_{3-} 
\int \frac{d u_{12} \, d u_{21} \, d u_{13} \, d u_{31} \, d u_{23} \, d u_{32}}
{(u^{+}_{12} u^{+}_{13})^2 (u^{+}_{21} u^{+}_{23})^2 (u^{+}_{31} u^{+}_{32})^2}\nt
& \qquad \times\langle A^{++}(1) A^{++}(2)\rangle\ \langle A^{++}(1) A^{++}(3)\rangle\ \langle A^{++}(2) A^{++}(3)\rangle  
 \,. \label{4.6}
\end{align}
Here $u_{ij}$ denotes the harmonic at the bivalent vertex $i$ connected with the vertex $j$ and the denominator originates from \p{439}. 
Note that to each external point $i$ corresponds its own set of RHs, e.g. the R-analytic $\cT(i)$ depends on $\q_{i+}$ defined by \p{4.39} with its own $w_{i+}$, 
$\q_{i,+\a}^a = (w_i)^a_{+A}\ \q^{A}_{i, \a}$. 
Substituting the propagator $\langle A^{++}A^{++} \rangle$ 
from \p{prop1} in \p{4.6} we obtain
\begin{align}
&{- \frac{1}{\pi^3} \ } \int d^4 \q_{1-} \, d^4 \q_{2-} \, d^4 \q_{3-} 
\int \frac{d u_{12} \, d u_{21} \, d u_{13} \, d u_{31} \, d u_{23} \, d u_{32}}
{(u^{+}_{12} u^{+}_{13})^2 (u^{+}_{21} u^{+}_{23})^2 (u^{+}_{31} u^{+}_{32})^2} \nt 
&\times \left[ \delta(u_{12},u_{21}) \delta^2(x^{\+ +}_{12}) \delta^4(\q^+_{12}) \right]\
\left[\delta(u_{13},u_{31}) \delta^2(x^{\+ +}_{13})  \delta^4(\q^+_{13})\right]\  
\left[\delta(u_{23},u_{32}) \delta^2(x^{\+ +}_{23}) \delta^4(\q^+_{23})\right] \,. \notag
\end{align}
Here we apply the shorthand notations $(\q^+_{ij})^A\equiv (\q_i - \q_j)^{\a A} (u^+_{ij})_{\a}$, $x^{\+ +}_{ij} 
\equiv  \xi^{\+}_{\da} (x_{i}-x_j)^{\da\a} (u^{+}_{ij})_{\a} $ (recall \p{chan} and \p{419}).
Then we integrate out the harmonics $u_{ij}$ with $i>j$ by means of the harmonic delta functions \p{4.7.4},
\begin{align}\label{5.35}
&{- \frac{1}{\pi^3}\  } \int d^4 \q_{1-} \, d^4 \q_{2-} \, d^4 \q_{3-} 
\int \frac{d u_{12} \, d u_{13}\, d u_{23}}
{(u^{+}_{12} u^{+}_{13})^2 (u^{+}_{12} u^{+}_{23})^2 (u^{+}_{13} u^{+}_{23})^2}\ 
 \notag\\ & \   \times
\delta^2(x^{\+ +}_{12}) \delta^4(\q^+_{12})\
\delta^2(x^{\+ +}_{13}) \delta^4(\q^+_{13}) \ \delta^2(x^{\+ +}_{23}) \delta^4(\q^+_{23}) \,.
\end{align}

Next we calculate the Grassmann integrals as in  \cite{Chicherin:2014uca}.
We decompose each $\q^A$ into a pair of $\q_{+}^{a}$, $\q_{-}^{a'}$ with the help of the RHs (see \p{4.39}),
\begin{align}\notag
\q^A = \bar w^A_{-a} \q^a_+ + \bar w^A_{+a'} \q^{a'}_-\,,
\end{align} and 
correspondingly we represent each $\delta^4$ as a pair of $\delta^2$,
\begin{align}\label{5.36}
\delta^4(\q^+_{ij}) = y_{ij}^2\ \delta^2((\q_{i-}+B_{ij})\cdot u^{+}_{ij})\,
\delta^2((\q_{j-}+B_{ji})\cdot u^{+}_{ij})\,,
\end{align}
where\footnote{The quantity $A_{ij}$ from \cite{Chicherin:2014uca} 
is related to $B_{ij}$ as follows $A_{ij} = (u^{+}_{ij})_{\a} B^{\a}_{ij}$.}
\begin{align}\label{7.10}
 B^{a' \a}_{ij} = \left[\q^{\a a}_{i+} (w_{ij})_a{}^b - \q^{\a b}_{j+}\right] (w^{-1}_{ij})_b{}^{a'}
\end{align}
and 
\begin{align}\notag
 &(w_{ij})_a{}^b = \bar w_{i,-a}^A w_{j,+A}^b\,, \qquad (w_{ij})_{a'}{}^b= \bar w_{i,+a'}^A w_{j,+A}^b\,, \qquad  y_{ij}^2=\det  (w_{ij})_{a'}{}^b\,.
\end{align}
Notice that the argument of each $\delta^2$ in \p{5.36} is invariant under $Q$ supersymmetry (see  \p{compl}), $Q_A (\q_{i-}+B_{ij})^{b'}= [w_{i,-A}^{a'} (w_{ij})_{a'}{}^b +w_{i,+A}^a (w_{ij})_a{}^b - w^{b}_{j,+A}] (w^{-1}_{ij})_b{}^{b'}=0$.

It is often convenient to use a particular parametrization of the $SU(4)$ harmonic coset,
\begin{align}\label{yww}
w_{+B}^{a} = (\delta_b^a,y_{b'}^a)\,,\qquad w_{-B}^{a'} = (0,\delta_{b'}^{a'}) \,,\qquad 
\bar w_{-a}^B = (\delta_a^b,0)\,,\qquad \bar w_{+a'}^B = (-y_{a'}^b, \delta_{a'}^{b'})\,,
\end{align}
which amounts to choosing a gauge for the local coset denominator.\footnote{More precisely,  one uses the isomorphism $SU(4)/[SU(2)\times SU(2)'\times U(1)] \sim GL(4,C)/\cP$ where $\cP$ is the Borel subgroup of upper triangular matrices. The gauge fixes $\cP$ locally, so that the coset is parametrized by the four complex variables $y_{a'}^b$.} {Then}
\begin{align}\notag
 (w_{ij})_a{}^b = \delta_a^b\,, \qquad (w_{ij})_{a'}{}^b = (y_j - y_i)_{a'}{}^b \equiv  (y_{ji})_{a'}{}^b 
\end{align}
and \p{7.10} simplifies considerably
\begin{align}\label{B}
 B^{a' \a}_{ij} = B^{a' \a}_{ji}= -(\q_{ij+})^{\a b}\left(y^{-1}_{ij}\right)_b{}^{a'}  \; , \qquad \q_{ij+} \equiv \q_{i+}-\q_{j+}\,.
\end{align}

From \p{5.35} and \p{5.36} we see that, e.g.,  $\q_{1-}$ appears twice, in the two propagators $1\to2$ and $1\to3$ joining at the bivalent vertex at point $1$. Using the Grassmann delta functions $\delta^2((\q_{1-}+B_{12})\cdot u^{+}_{12})$ and $\delta^2((\q_{1-}+B_{13})\cdot u^{+}_{13})$ we can integrate out $\q_{1-}$. This produces an LH factor $(u^{+}_{12} u^{+}_{13})^2$ which cancels the corresponding factor from the denominator. Repeating the procedure at each point we find
\begin{align} \label{4.15}
{- \frac{1}{\pi^3}\ } y_{12}^2 y_{13}^2 y_{23}^2 \int d u_{12} \, d u_{13}\, d u_{23}\,
\delta^2(x^{\+ +}_{12})\delta^2(x^{\+ +}_{13})
\delta^2(x^{\+ +}_{23}) \,.
\end{align}
The remaining harmonic integrals are easy to calculate. In \cite{PartI} we discussed the harmonic integrals of the complex delta function and proved the formula
\begin{align}\label{intd}
\int du\, \delta^2( x^{\+ +})\ f(u^+,u^-) =\frac1{\pi x^2} f\bigl(\ x^{\+}/\sqrt{x^2}\ ,\ x^{\-}/\sqrt{x^2}\ \bigr)\,.
\end{align}
Thus we have the final result
\begin{align}\label{5.41}
\langle {\cT}(1) {\cT}(2) {\cT}(3)\rangle_{g^0}  = {- \frac{1}{\pi^6} \, } \frac{y_{12}^2 y_{13}^2 y_{23}^2}{x_{12}^2 x_{13}^2 x_{23}^2}\,.
\end{align}

This correlator  coincides with the three-point function of the half-BPS operators $\cO$ defined in \p{5.2}, 
$\vev{\cO(1) \cO(2) \cO(3)}_{g^0}$, made from free propagators $\vev{\phi_{++} \phi_{++}} =y^2/(2 \pi^2 x^2)$.

Clearly, the same calculation applies to the $n-$point correlator $\langle {\cT}(1) \ldots {\cT}(n)\rangle_{g^0} $, also made only from 
bivalent external vertices. As explained earlier, this correlator may become nilpotent (i.e., $\sim (\q_+)^{4p}$) starting at the coupling level $g^{2p}$.

\subsubsection{ Two points: $\vev{\cT \cT }$} 

The two-point MHV-like correlator
\begin{align}\label{5.42}
&\langle {\cT}(1) {\cT}(2)\rangle_{g^0} = \begin{array}{c}\includegraphics[height = 2.0 cm]{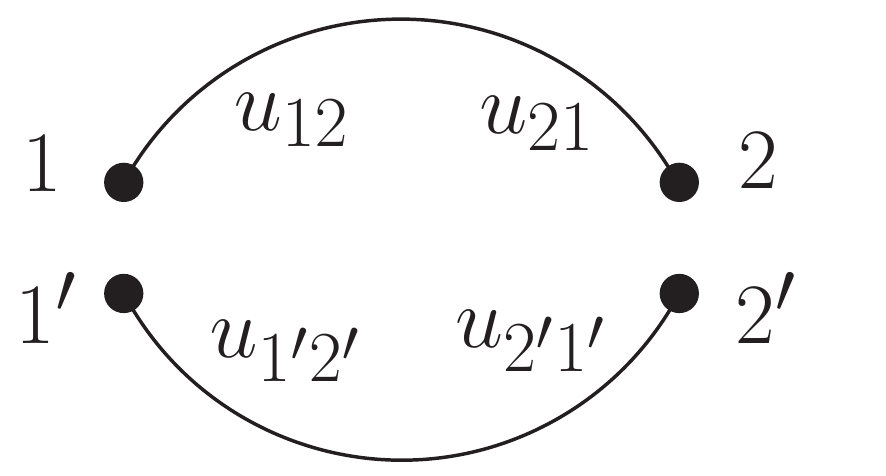}\end{array} \nt &= {\frac{1}{2}} \int  
 \frac{d u\ d^4 \q_{1-} \, d^4 \q_{2-}}
{(u^{+}_{12} u^{+}_{1'2'})^2 (u^{+}_{21} u^{+}_{2'1'})^2 } \langle A^{++}(1) A^{++}(2)\rangle\ \langle A^{++}(1') A^{++}(2')\rangle  \nt
& = {\frac{1}{2 \pi^2}} \int \frac{d u\ d^4 \q_{1-} \, d^4 \q_{2-}}
{(u^{+}_{12} u^{+}_{1'2'})^4 }  \delta^2(x^{\+}_{12} u^+_{12})\delta^2(x^{\+}_{12} u^{+}_{1'2'})\ \delta^4(\q_{12} u^+_{12}) \delta^4(\q_{12} u^{+}_{1'2'})
\end{align}
is somewhat degenerate and requires care in treating the singularities (on the picture we have split the points in the LH space).  
The two propagators $\vev{A^{++} A^{++}}$ connect the same points, which creates a potential problem with the denominator $(u^{+}_{12} u^{+}_{1'2'})^4 $. Indeed, the delta functions $\delta(x^{\+ +})$ identify the two harmonics, $u^{+}_{12} \sim  u^{+}_{1'2'} \sim x^{\+}_{12}$. The resolution is to first do the Grassmann integration using \p{5.36}. This results in a factor of $y^4_{12} (u^{+}_{12} u^{+}_{1'2'})^4$ which cancels out the harmonic denominator. Then the harmonic integrals can be safely done by means of \p{intd} producing a factor of $1/x^4_{12}$ and we find the familiar result
\begin{align}\notag
\langle {\cT}(1) {\cT}(2)\rangle_{g^0} = {\frac{1}{2 \pi^4}} \frac{y^4_{12}}{x^4_{12}}\,.
\end{align}

\subsection{Non-MHV-like correlators (chiral case)}\label{s7.2}

 In the MHV-like example \p{5.35} we encountered only bivalent vertices. Consequently, the number of Grassmann integrations $\int d^4\q_-$ matches the number of Grassmann delta functions from the propagators and the final result is not nilpotent. To obtain a non-MHV-like (i.e., nilpotent) correlator we need vertices of higher valency in the expansion of $\cT$. Consider, for instance, a 3-valent vertex at point 1 connected via propagators $\vev{A^{++}(1) A^{++}(j)}$ with points $j=2,3,4$:\footnote{Here we essentially repeat the calculation of \cite{Chicherin:2014uca} adapting it to our formalism.}
 \begin{align}\label{6.13}
\begin{array}{c}\includegraphics[height = 2.5 cm]{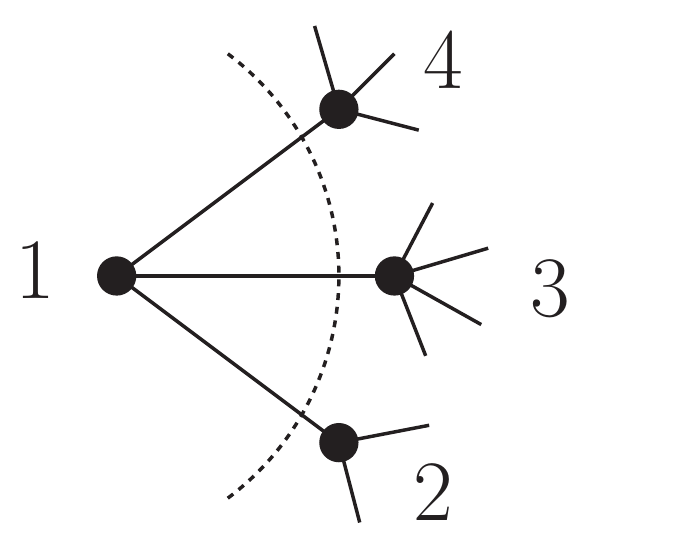}\end{array} =
{- \frac{1}{\pi^3} } \int \frac{d u\ d^4 \q_{1-} }
{(u^{+}_{12} u^{+}_{13})(u^{+}_{13} u^{+}_{14})(u^{+}_{14} u^{+}_{12})}  \prod_{j=2}^4\delta^2(x^{\+ +}_{1j} )\ \delta^4(\q^+_{1j})\,. 
\end{align}
The harmonic delta functions from the propagators have already been used to integrate out half of the LHs. Decomposing each Grassmann $\delta^4$ as in \p{5.36}, we see that $\q_{1-}$ appears in three different $\delta^2$:
\begin{align}
&\int d^4 \q_{1-} \prod_{j=2}^4 \delta^4(\q^+_{1j} ) \ \to \  \int d^4 \q_{1-}\prod_{j=2}^4 y^2_{1j} \delta^2((\q_{1-} + B_{1j})\cdot u^+_{1j})\nt 
& = y^2_{12}y^2_{13}y^2_{14}\  \delta^2\Big((u^{+}_{13} u^{+}_{14})(u^+_{12}B_{12}) + (u^{+}_{14} u^{+}_{12})(u^+_{13}B_{13}) + (u^{+}_{12} u^{+}_{13})(u^+_{14}B_{14}) \Big)\,. \notag
\end{align}
The other half of the Grassmann delta functions is to be used at the vertices where the other ends of the propagators are attached. 

The harmonic integrations in \p{6.13} are done by means of \p{intd}. Note that the normalization factors $\sqrt{x^2_{ij}}$ from \p{intd} drop out due to the invariance of the integrand under local rescalings of $u^+_{ij}$. The result is 
\begin{align}
{\frac{1}{\pi^6}} \frac{y^2_{12}y^2_{13}y^2_{14}}{x^2_{12}x^2_{13}x^2_{14}} \ R(1;2,3,4)\,,     \label{550'}
\end{align}
where
\begin{align}\label{Rinv}
R(1;2,3,4) = -\frac{\delta^2\Big((x^{\+}_{13} x^{\+}_{14})(x^{\+}_{12}B_{12}) + (x^{\+}_{14} x^{\+}_{12})(x^{\+}_{13}B_{13}) + (x^{\+}_{12} x^{\+}_{13})(x^{\+}_{14}B_{14}) \Big)}{(x^{\+}_{12} x^{\+}_{13})(x^{\+}_{13} x^{\+}_{14})(x^{\+}_{14} x^{\+}_{12})}\,.
\end{align}
Here $(x^{\+}_{ij} x^{\+}_{ik}) = \xi^{\+}_{\da} x^{\da \a}_{ij} x_{ik}^{\db}{}_{\a}  \xi^{\+}_{\db}$
and $(x^{\+}_{ij} B^{a'}_{ij}) = - \xi^{\+}_{\da} x^{\da \a}_{ij} (y^{-1}_{ij})^{a'}_{b} (\q_{+i}-\q_{+j})^b_{\a}$ (recall \p{B}).
This result is equivalent to the R-vertex in \cite{Chicherin:2014uca}, upon the identification $x^{\+}_{ij} \to   x^2_{ij}\sigma_{ij}$.   We remark that the expression in \p{Rinv} is invariant under $Q$ supersymmetry, $\delta \q^A_\a = \ep^A_\a$. Indeed, $\delta B^{a' \a}_{ij}= \left(y^{-1}_{ij}\right)^{a'b}( w_{+i} -  w_{+j})_{b B}\ep^{\a B} = \ep^{\a a'}$, where we have used the parametrization  \p{yww}. The variation of the argument of $\delta^2$ in  \p{Rinv} vanishes due to the Schouten identity.

The rest of the calculation of a complete Feynman graph with the above chiral structure follows exactly Ref.~\cite{Chicherin:2014uca}. Any non-MHV-like chiral correlator at Born level can be constructed from the elementary building block \p{Rinv} using the \emph{ effective Feynman rules}:

 (i) a propagator $\frac{1}{\pi^2}\,y^2_{ij}/x_{ij}^2$ corresponds to a line connecting vertices $i$ and $j$; 
 
 (ii) a $p$-valent vertex $i$ connected with vertices $j_1, \ldots,j_p$ factorizes into a product of 3-valent vertices 
\begin{align} \label{Rmult}
R(i;j_1,\ldots,j_p) = \prod_{k = 2}^{p-1} R(i;j_1,j_k,j_{k+1}) \,.
\end{align}

An important question is how the final result (the sum of all Feynman graphs) becomes independent of the gauge-fixing parameter $\xi^{\+}$ \p{417}. The argument given in \cite{Chicherin:2014uca} amounts to showing that the sum of graphs is free from spurious singularities. It is adapted (and extended to the non-chiral case) to our formalism in Appendix~\ref{nolambda}. A much more difficult problem is to eliminate $\xi^{\+}$ explicitly from the sum of Feynman graphs without breaking Lorentz invariance. Appendix~\ref{loop} contains an example of such a calculation where the LHC supergraph formalism is applied to the simplest non-trivial non-MHV-like correlator, the five-point NMHV-like correlator.

In conclusion of this short review we would like to emphasize the following point about the structure of the chirl correlators at Born level. A $p$-valent vertex from $\cT$ is proportional to $g^{p-2}$ and its contribution is of  Grassmann degree $(\q_+)^{4p}$. Thus, we see that for Feynman diagrams made out of external vertices from $\cT$ without interaction vertices from $S_{\rm Z}$, the order in the coupling and the Grassmann degree match. This is exactly what we expect for Born-level correlators. If we would include an interaction vertex from $S_{\rm Z}$, then according to \p{TtoS} this is equivalent to integrating out one of the external points. This would take us out of the class of the Born-level correlators. So, the Born-level correlators are given only by free Feynman diagrams. No $x$-space  integrations are needed. The whole information about the interactions is encoded in the operators $\cT$.

\section{Restoring the $\bq$ dependence at Born level} \label{s8}

In this section we work out the full $\q$ and $\bq$ dependence of the purely chiral correlators \p{chirCor} from the previous section. The latter were obtained by switching off $\bq$, i.e. by restricting to the operator $\cT$ in \p{4.33}. What we need to do to restore the dependence on $\bq$ is a $\bQ$ supersymmetry transformation at each point $i$ with parameters $\bq_{i+}$ (see~\p{4.39}). In other words, we need to replace $\cT$ by its $\bQ-$descendants from  \p{4.33} and calculate a whole new family of correlators. Together they form the full correlator \p{7.1}. Our result is surprisingly  simple: The dependence on $\bq$ is restored by making simple shifts in the space-time variables $x_{ij}$, see \p{545}. More precisely, the effective Feynman rules of the previous section, modified by the shifts, are valid for the non-chiral correlators from the first and second lines in \p{7.1}.

To show how non-trivial this result is, think of the following possibility allowed by $Q$ and $\bar Q$ supersymmetry alone. From the $\q_+$ at three points $p,q,r$ we can construct the linear combination  $\Theta_{pqr}= B_{pq} - B_{pr}$ invariant under $Q$ supersymmetry (recall the discussion below \p{Rinv}). Similarly, from the $\bq_+$ at three points (not necessarily different from the first three) we can construct a linear invariant of $\bQ$ supersymmetry, $\bar\Theta_{lmn}$. We can then add $\Theta_{ijk} \bar\Theta_{lmn}$ to the hatted variable $\hat x_{ij}$ in \p{545} (after canceling the $SU(4)$ weight with harmonics) and still be consistent with the full supersymmetry. Superconformal symmetry (generators $S$ and $\bar S$) impose further constraints but it nevertheless remains possible to construct nilpotent invariants involving 5 or more points. The non-trivial claim that we make here is that such arbitrary invariants do not appear, everything is contained in the minimal modification $\hat x_{ij}$ in \p{545}.

\subsection{MHV-like correlators}

We start by considering a simple example, which already shows the main features of how $\bQ$ supersymmetry helps us to restore the dependence on $\bq$ in the correlators.

 The Grassmann expansion of the Born-level MHV-like correlators (second line in \p{7.1}) goes in terms of $\q\bq$, in order to preserve the symmetry under  the $\mathbb{Z}_4$ center of the R-symmetry group $SU(4)$. Therefore, restoring the $\bq$ dependence of a chiral MHV-like correlator, we will automatically also restore the full $\q\bq$ dependence. In fact, in this simple case the answer is well known and it is entirely fixed by $\cN=4$ supersymmetry. In the three-point case we have
\begin{align}\label{545}
\vev{T(1) T(2) T(3)}_{g^0} = { - \frac{1}{\pi^6}}  \frac{ y_{12}^2  y_{13}^2 y_{23}^2}{\hat x_{12}^2  \hat x_{13}^2 \hat x_{23}^2}\,, \quad \hat x_{ij} = x_{ij} + \q_{ij+} y^{-1}_{ij} \bq_{ij+}\,, \quad \q_{ij+} \equiv \q_{i+} - \q_{j+} \,. 
\end{align}
The generalization to the $n$-point case is straightforward.
From our point of view, to restore the $\bq_{+}$ dependence in \p{5.32} we need to act with the supersymmetry shifts 
\begin{align}\label{7.2}
e^{\bq_{1+} \cdot \bQ_1 + \bq_{2+} \cdot \bQ_2 + \bq_{3+} \cdot \bQ_3}  \langle {\cT}(1) {\cT}(2) {\cT}(3)\rangle_{g^0}\,. 
\end{align}
This should have the effect of putting hats on $x_{ij},\ i<j = 1,2,3$. 

Let us show how to obtain this result by means of LHC supergraphs. We need to calculate the correlators of the various operators from the expansion \p{4.33}.

\subsubsection{First-order $\bQ$ variation} \label{s711}

The linear term in the finite $\bQ_1$ transformation \p{7.2} corresponds to the three-point function
\begin{align}\label{531}
\bq_{1+a'}^\da\langle \cM^{a'}_{\dot\alpha}(1) {\cT}(2) {\cT}(3)\rangle_{g^0} 
\end{align}
of two chiral operators $\cT$ and the first-level $\bQ$ descendant $\cM^{a'}_{\dot\alpha}$ defined in \p{439}  and \p{435}, respectively. This correlator involves only bivalent external vertices and no interaction vertices $S_{\rm Z}$~\p{N4}. The extra $\q_-$ in the definition of $\cM$ makes the correlator \p{531} nilpotent, $\sim (\q)^1$. 
There are two contributing Feynman graphs (see Fig.~\ref{MHVfig}) which differ by the permutation $2\leftrightarrows 3$:
\begin{figure}[!h]
\centerline {
\includegraphics[height = 2.5 cm]{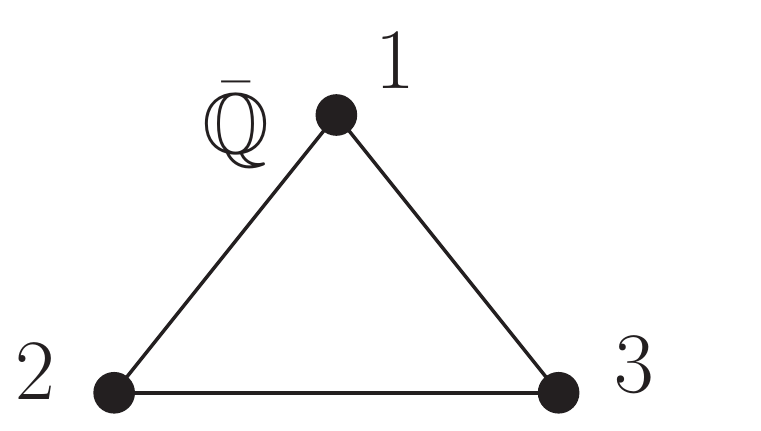} \qquad
    \includegraphics[height = 2.5 cm]{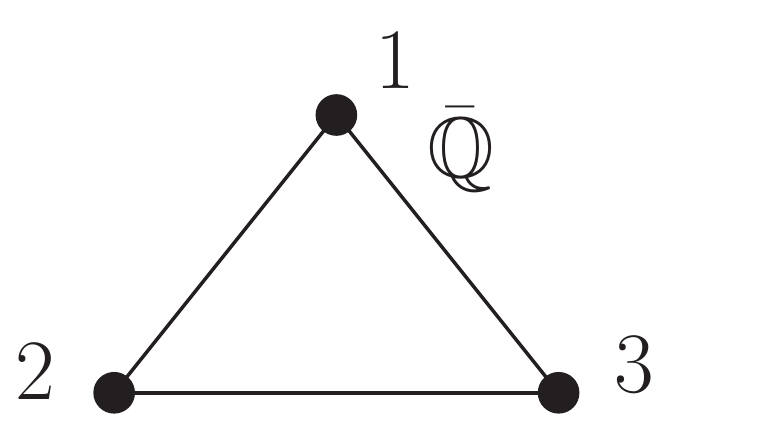} \qquad \includegraphics[height = 2.5 cm]{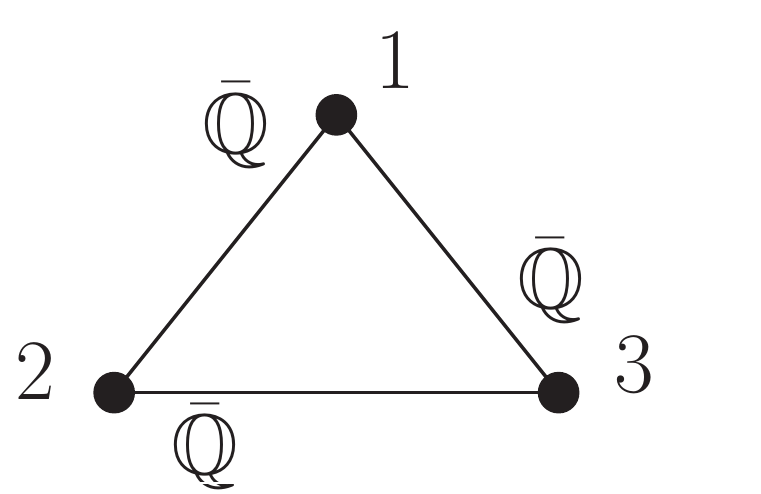} }
\caption{The first two LHC supergraphs contribute to $\langle \cM^{a'}_{\dot\alpha}(1) {\cT}(2) {\cT}(3)\rangle_{g^0}$.
The third  diagram contributes  to $\langle\cM^{a'}_{\dot\alpha}(1) \cM^{b'}_{\db}(2) \cM^{c'}_{\dot\gamma}(3)\rangle_{g^0}$.}
\label{MHVfig}		
\end{figure}
\\ In these pictures each $\bar{\mathbb{Q}}$ represents the transformation of 
the corresponding $A^{++}$ field, 
$\bar{\mathbb{Q}}^{a'}_{-\da} A^{++} = \q^{+a'}_{-} (A^{+}_{\da} + \pa^{-}_{\da} A^{++} )$ (recall~\p{4.27}). The first diagram contributes (compare with~\p{4.6})
\begin{align}\label{6.14}
& \int d^4 \q_{1-} \, d^4 \q_{2-} \, d^4 \q_{3-} 
\int \frac{d u_{12} \, d u_{21} \, d u_{13} \, d u_{31} \, d u_{23} \, d u_{32}}
{(u^{+}_{12} u^{+}_{13})^2 (u^{+}_{21} u^{+}_{23})^2 (u^{+}_{31} u^{+}_{32})^2}\ \langle A^{++}(1) A^{++}(3)\rangle \langle A^{++}(2) A^{++}(3)\rangle\nt
&  \qquad
\times (\q^{a'}_{1-} \cdot u^{+}_{12}) \left[ \langle A^{+}_{\da}(1) A^{++}(2)\rangle 
+ u^{-\a}_{12}  \pa_{\a\da}\langle A^{++}(1) A^{++}(2)\rangle \right]\,.
\end{align}
Substituting the propagators \p{prop1} and \p{prop2} we obtain
\begin{align}\label{6.15}
&{\frac{1}{\pi^3}} \int d^4 \q_{1-} \, d^4 \q_{2-} \, d^4 \q_{3-} 
\int \frac{d u_{12} \, d u_{21} \, d u_{13} \, d u_{31} \, d u_{23} \, d u_{32}}
{(u^{+}_{12} u^{+}_{13})^2 (u^{+}_{21} u^{+}_{23})^2 (u^{+}_{31} u^{+}_{32})^2}\nt 
&\qquad \times \left[\delta(u_{13},u_{31}) \delta^2(x^{\+ +}_{13})  \delta^4(\q^+_{13})\right] 
\left[\delta(u_{23},u_{32}) \delta^2(x^{\+ +}_{23}) \delta^4(\q^+_{23})\right] \nt
&\qquad \times  (\q^{a'}_{1-}\cdot u^{+}_{12})\left[ \frac{\xi^{\-}_{\da}}{x^{\- +}_{12}} 
+  u^{-\a}_{12}\frac{\pa}{\pa x_{12}^{\da\a}} \right] 
\delta(u_{12},u_{21}) \delta^2(x^{\+ +}_{12}) \delta^4(\q^+_{12})\,. 
\end{align}
Then we integrate out the half of the harmonics $u_{ji}$ with $j>i$ (see~\p{4.7.4}) and  the  odd variables $\q_-$ using \p{5.36}, which gives
\begin{align}\label{6.16}
&{-\frac{1}{\pi^3}} y_{12}^2 y_{13}^2 y_{23}^2\int du
\   \delta^2(x^{\+ +}_{13})   \delta^2(x^{\+ +}_{23})   \ (B^{\b a'}_{12} u^{+}_{12\,\b})\left[ \frac{\xi^{\-}_{\da}}{x^{\- +}_{12}} 
+  u^{-\a}_{12}\frac{\pa}{\pa x_{12}^{\da\a}} \right] 
 \delta^2(x^{\+ +}_{12})  \,. 
\end{align}
This expression almost coincides with its chiral counterpart \p{4.15}, except for the differential operator acting on $\delta^2(x^{\+ +}_{12})$.
We rewrite the last factor in \p{6.16} by means of the identity (here $u$ stands for the harmonic  $u_{12}$)
\begin{align}\label{6.17}
 u^{+}_{\b} \left[ \frac{\xi^{\-}_{\da} }{x^{\- +}}+  u^{-\a}\frac{\pa}{\pa x_{}^{\da\a}}\right]\delta^2(x^{\+ +})  
 =- \frac{\pa}{\pa x^{\da\b}} \delta^2(x^{\+ +})  +   \pa^{++}\left[ u^{-}_{\b} \frac{\xi^{\-}_{\da} }{x^{\- +}}\delta^2(x^{\+ +})\right]. 
\end{align}
It follows from the completeness relation for LHs $u^{+\a} u^-_{\b} - u^{-\a} u^+_{\b} = \delta^{\a}_{\b}$ (see~\p{2}), the property $\pa^{++} u^-_{\b} = u^+_{\b}$ (see  \p{4}) and the relation between the space-time and  the harmonic derivatives\footnote{When $\pa^{++}$  hits the singularity $1/x^{\- +}$, it gives a contact term $x^{\+ +}\delta^2(x^{\- +})$ (see Appendix~A.3 in \cite{PartI}) which vanishes due to $\delta^2(x^{\+ +})$. }
\begin{align} \notag
u^{+\a} \pa_{\da\a}\, f(x^{\- -}) =  {\xi^{\-}_{\da} }/{x^{\- +}}\, \pa^{++}\, f(x^{\- -})\,,
\end{align}
where $x^{\- -} = \xi^{\-}_{\da}x^{\da \a} u^{-}_{\a}$ (recall~\p{419}). In our case the role of $f$ is played by the
complex delta function $\delta(x^{\+ +},x^{\- -})$.

Since nothing else under the harmonic integral in \p{6.16} depends on $u_{12}$, the total harmonic derivative term in \p{6.17} vanishes.  
In this way the gauge fixing parameter  $\xi^{\-}$ drops out from \p{6.16} and we get
\begin{align}\label{6.21}
& {- \frac{1}{\pi^3}} y_{12}^2 y_{13}^2 y_{23}^2\int du
\   \delta^2(x^{\+ +}_{13})   \delta^2(x^{\+ +}_{23})   \ \left(-B^{\a a'}_{12}  \frac{\pa}{\pa x_{12}^{\da\a}}  \right)  
 \delta^2(x^{\+ +}_{12})  \,. 
\end{align}
 Finally, we do the harmonic integrations by means of \p{intd} giving the result \p{5.41} with an additional differential operator, 
\begin{align}\label{542}
{- \frac{1}{\pi^6}} (\q_{12+}y^{-1}_{12})^{a'\a} \frac{\pa}{\pa x_{12}^{\da\a}}  \left[ \frac{y_{12}^2 y_{13}^2 y_{23}^2}{x_{12}^2 x_{13}^2 x_{23}^2} \right]\,.  
\end{align}
Here and in the following we use a `directional' space-time derivative, i.e. we formally treat $x_{12}$, $x_{13}$ and $x_{23}$ as independent variables.
The second Feynman graph restores the permutation symmetry $2\leftrightarrow3$ with the final result 
\begin{align}\label{6.22}
{- \frac{1}{\pi^6}}\left[(\q_{12+}y^{-1}_{12})^{a'\a} \frac{\pa}{\pa x_{12}^{\da\a}} + (\q_{13+}y^{-1}_{13})^{a'\a} \frac{\pa}{\pa x_{13}^{\da\a}}  \right] \left[ \frac{y_{12}^2 y_{13}^2 y_{23}^2}{x_{12}^2 x_{13}^2 x_{23}^2} \right]\,.
\end{align} 
Completing \p{6.22} with $\bq_{1+}$ from the expansion \p{4.33}, we see full agreement with the expansion of $\hat x_{12}$ and $\hat x_{13}$ in \p{545}. 

The only new feature of this calculation, compared to the chiral case considered in Section \ref{s7.1}, 
is the identity \p{6.17} for the propagator connecting the fields $\bar{\mathbb{Q}}^{a'}_{+\da} A^{++}$ and $A^{++}$.
This trick is essentially local and it  applies to any Feynman diagram with a line connecting these two fields. 
For example, the last diagram in Fig.~\ref{MHVfig} contributes to 
$\langle\cM^{a'}_{\dot\alpha}(1) \cM^{b'}_{\db}(2) \cM^{c'}_{\dot\gamma}(3)\rangle_{g^0}$. 
Like \p{542}, it corresponds to the expansion of the supersymmetrized propagators in the correlator \p{545} 
with respect to $\hat{x}_{12}$, $\hat{x}_{23}$ and $\hat{x}_{13}$ 
to the first order in $\bar{\q}_{1+}$, $\bar{\q}_{2+}$ and $\bar{\q}_{3+}$, respectively, i.e. to the three-linear term $\bq_{1+}\bq_{2+}\bq_{3+}$ in the expansion of \p{545}.

\subsubsection{Second-order $\bQ$ variations} \label{712}

The next step in the restoration of the $\bq_{1+}$ dependence (see~\p{7.2}) requires doing the second $\bQ$ variation at point 1. This amounts to inserting the operators $\cR$ and $\cS$ from \p{4.33},
\begin{align}\label{711}
(\bq_{1+}^2)_{(a'b')} \vev{\cR^{(a'b')}(1){\cT}(2) {\cT}(3)}_{g^0}  + (\bq_{1+}^2)_{(\da\db)} \vev{\cS^{(\da\db)} (1){\cT}(2) {\cT}(3)}_{g^0}\,.
\end{align}

\begin{figure}[!h]
\centerline{
\includegraphics[height = 2.5 cm]{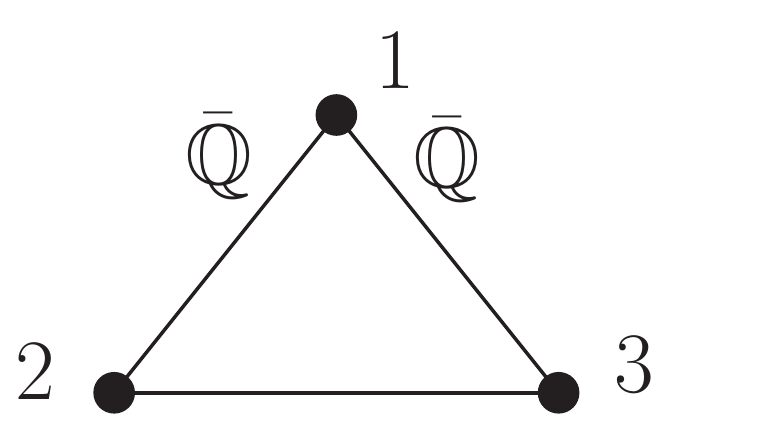} \quad \includegraphics[height = 2.5 cm]{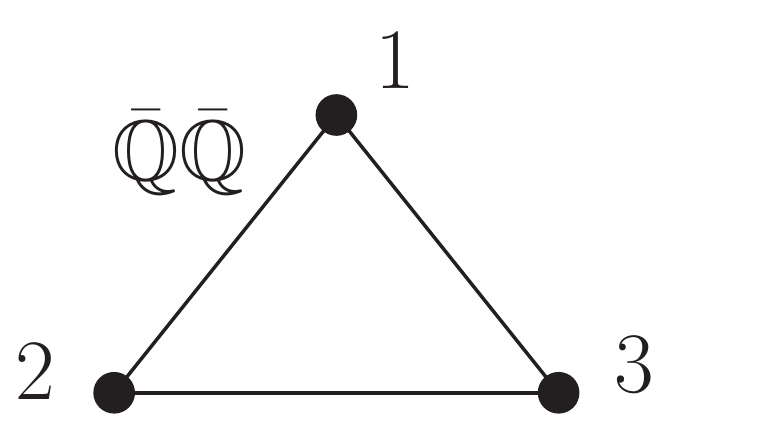} 
\quad \includegraphics[height = 2.5 cm]{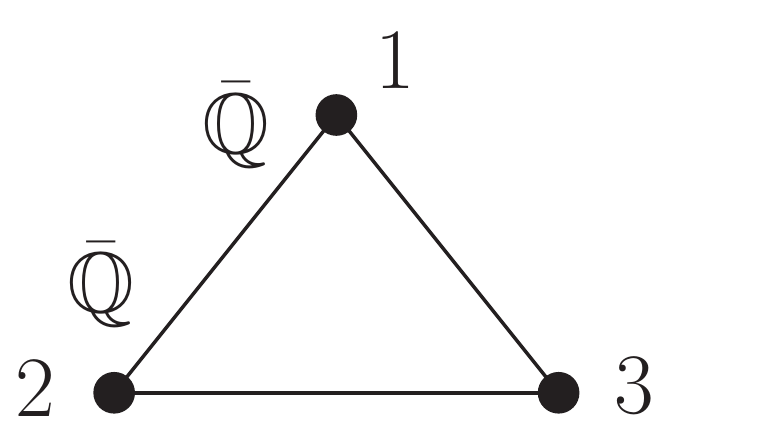}}
\caption{LHC supergraphs corresponding to: (i) $\langle \cR_{0}(1) {\cT}(2) {\cT}(3)\rangle_{g^0}$ 
(or $\langle \cS^{'}(1) {\cT}(2) {\cT}(3)\rangle_{g^0}$), (ii) $\langle \cS^{''}(1) {\cT}(2) {\cT}(3)\rangle_{g^0}$, (iii) $\langle \cM(1) {\cM}(2) {\cT}(3)\rangle_{g^0}$\,.}
\label{MHV2fig}
\end{figure}

Recall that $\cR$ and $\cS$ are sums of two terms (see~\p{4.58} and \p{4.76}).
Let us first consider the correlators with the insertions  $\cR_0$ and $\cS'$ defined in  \p{O'a'b'} and \p{478}, respectively. In these operators the two $\bar{\mathbb{Q}}$ variations are distributed over two superfields $A^{++}$. At order $g^0$  the operators $\cR_0$ and $\cS'$ are restricted to their bilinear terms with $\bar{\mathbb{Q}} A^{++}$. Both correlators correspond to the first picture in Fig.~\ref{MHV2fig}. The difference between the two operators comes only from the (anti)symmetrization of their indices.

We already know how to evaluate this diagram (compare to the third diagram in Fig.~\ref{MHVfig}).
The same manipulations as in Section \ref{s711}  give rise to two shifts of the type \p{542}, one for each line connecting $\bar{\mathbb{Q}}_{+} A^{++}$ and $A^{++}$,
\begin{align}\label{715}
- {\frac{1}{\pi^6}} \Bigl[(\q_{12+}y^{-1}_{12})^{a'\a} \frac{\pa}{\pa x_{12}^{\da\a}}\Bigr]  \Bigl[(\q_{13+}y^{-1}_{13})^{b'\b} \frac{\pa}{\pa x_{13}^{\b\db}}  \Bigr] \left[ \frac{y_{12}^2 y_{13}^2 y_{23}^2}{x_{12}^2 x_{13}^2 x_{23}^2} \right] ,
\end{align}
which is to be contracted with 
$-\bq^{\da}_{1+a'} \bq^{\db}_{1+b'} = -\frac12 \ep^{\da\db} (\bq_{1+}^2)_{(a'b')} + \frac12 \ep_{a'b'} (\bq_{1+}^2)^{(\da\db)}$. The first term originates from $\cR_{0}$ and the second from $\cS^{'}$.
The double derivative term \p{715} corresponds to the expansion of the supersymmetrized propagators in the correlator \p{545} with respect to $\hat{x}_{12}$ and $\hat{x}_{13}$
to the first order in $\bar{\q}_1$ for each of them.

At the quadratic level in $\bq_{1+}$ we also need to study the insertion of the operator $\cS^{''}_{(\da\db)}$ defined in \p{Odadb}. The difference with the previous case is that now both $\bar{\mathbb{Q}}$ variations act on the same superfield $A^{++}$ or equivalently on the same line of the Feynman graph (see the second picture in Fig.~\ref{MHV2fig}). After the integration over $u_{i1}$ with $i>1$ and $\q_{1-}$ we obtain the analog of \p{6.16}, but now with a second-order differential operator acting on $ \delta^2(x^{\+ +}_{12}) $: 
\begin{align} \label{2Var}
u^{+}_{\gamma}u^{+}_{\delta} \Bigl[ u^{-\a} \pa_{\a(\da}\xi^{\-}_{\db)} \frac{1}{x^{\- +}} 
+ \frac{1}{2} u^{-\a} u^{-\b} \pa_{\a\da} \pa_{\b\db} \Bigr] \delta^2(x^{\+ +})\,,
\end{align}
which is contracted with $B^{a' \gamma} \ep_{a'b'} B^{b' \delta}$. Here $u$ stands for the harmonic  $u_{12}$.
Repeating the manipulations from \p{6.17}, it is easy to recast this expression in the form
\begin{align} \label{2Var'}
\frac{1}{2} \pa_{(\gamma\da} \pa_{\delta)\db} \delta^2(x^{\+ +}) + \pa^{++}\left[u^-_{(\gamma} \Bigl( - \delta^\a_{\delta)} + \frac1{2} u^-_{\delta)}u^{+\a} \Bigr)  \pa_{\a(\da}  \xi^{\-}_{\db)} \frac{1}{x^{\- +}}   \right] \delta^2(x^{\+ +}) \,.
\end{align}
The second term is a total harmonic derivative, which vanishes under the harmonic integral $\int du_{12}$. We then carry out the remaining integrations in the Feynman graph (recall the step from \p{6.21} to \p{542}) and complete it with  $(\bq_1)^2$. The result is the second-order term in the expansion of the scalar propagator $y^2_{12}/\hat x^2_{12}$ from \p{545}, 
\begin{align}\label{722}
{-\frac{1}{\pi^6}} \left(-\frac12\right) \frac{1}{2} (\q_{12+}y^{-1}_{12})^{a'\a}\ep_{a'b'} (\q_{12+}y^{-1}_{12})^{b'\b} \frac{\pa^2}{\pa x_{12}^{\da\a}\pa  x_{12}^{\db\b}} \left[ \frac{y_{12}^2 y_{13}^2 y_{23}^2}{x_{12}^2 x_{13}^2 x_{23}^2} \right], 
\end{align} 
which is to be contracted with $(\bq_{1+})^{2(\da\db)}$. 
Putting together \p{715} and \p{722} (and the symmetric graphs with points 2 and 3 exchanged), we obtain the second-order term $(\bq_{1+})^2$ in the expansion of the exponential in \p{7.2}.

Finally, let us consider the bilinear terms in the $\bq_+$ expansion of the mixed type $\bq_{1+} \bq_{2+}$. In our sample correlator this corresponds to $\cM$ insertions (see~\p{435}) at points 1 and 2, see the third picture in Fig.~\ref{MHV2fig}.
The novelty is that a propagator is stretched between $\bar{\mathbb{Q}}_{-\da}^{a'} A^{++}$ and $\bar{\mathbb{Q}}_{-\db}^{b'} A^{++}$.
This diagram, compared to the chiral calculation, contains the factors $(\theta^{a'}_{1-} \,u^{+}_{12})$
and $(\theta^{b'}_{2-} \,u^{+}_{21})$ at the vertices 1 and 2, respectively, and the propagator  
\begin{align}
&\langle (A^{+}_{\da} + \pa^{-}_{\da} A^{++})(1) (A^{+}_{\db} + \pa^{-}_{\db} A^{++})(2) \rangle = \notag \\ 
&=
- 2\langle \bigl( \pa^{-}_{(\da} A^{+}_{\db)} +  {\textstyle \frac12} \pa^{-}_{\da} \pa^{-}_{\db} A^{++} \bigr)(1) 
A^{++}(2) \rangle\,, \label{sumprop}
\end{align}
where we use the absence of  the propagator $\vev{A^+_{\da}  A^+_\db}$ (see~\p{prop3}) and the antisymmetry of the propagator $\vev{A^+_{\da}  A^{++}}$ (see~\p{prop2}).
The integration over $\q_-$ turns $(\theta^{a'}_{1-} \,u^{+}_{12}) (\theta^{b'}_{2-} \,u^{+}_{12})$ into $(B^{a'}_{12} u^{+}_{12}) (B^{b'}_{12} u^{+}_{12})$, so we can effectively make the replacement (see~\p{cQcQ})
\begin{align} \notag
\langle \bigl( \bar{\mathbb{Q}}^{a'}_{-\da} A^{++} \bigr)(1) \bigl(\bar{\mathbb{Q}}^{b'}_{-\db} A^{++} \bigr)(2) \rangle \ \to \
\langle \bigl( \bar{\mathbb{Q}}^{a'}_{-\da} \bar{\mathbb{Q}}^{b'}_{-\db} A^{++} \bigr)(1) A^{++}(2) \rangle\,.
\end{align}  
Thus, the two first-order $\bar{\mathbb{Q}}$-variations at both ends of a propagator are equivalent to a $(\bar{\mathbb{Q}})^2$-variation at one end
that corresponds to the insertion of $\cS^{''}$ \p{Odadb}. 
The result is \p{722}, which is to be contracted with $2 (\bar{\theta}^{\da}_{1+} \cdot \bar{\theta}^{\db}_{2+})$. This is the bilinear term $\bq_{1+}\bq_{2+}$ in the expansion of the propagator $y^2_{12}/\hat x^2_{12}$ from \p{545}.

Remark that the expansion of the propagator $y^2_{12}/\hat x^2_{12}$ does not contain $(\bq_{+})^2_{(a'b')}$.
Indeed, it is accompanied by $\ep^{\da \db} \pa_{\a\da}\pa_{\b\db} (1/x^2) = -\frac14 \ep_{\a\b} \Box_x (1/x^2) \sim 0$, up to a contact term.
This symmetrization of indices corresponds to $\cR_{\rm Z}$ \p{monstr0} which does not contribute (see below).

\begin{figure}[!h]
\centerline{
\includegraphics[height = 2.5 cm]{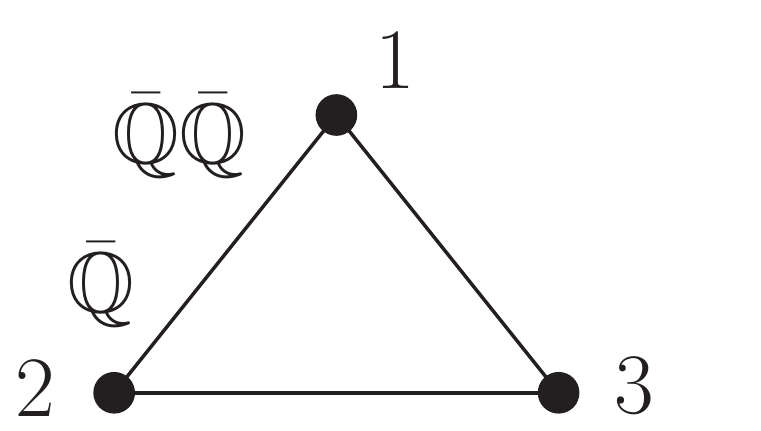} \quad \includegraphics[height = 2.5 cm]{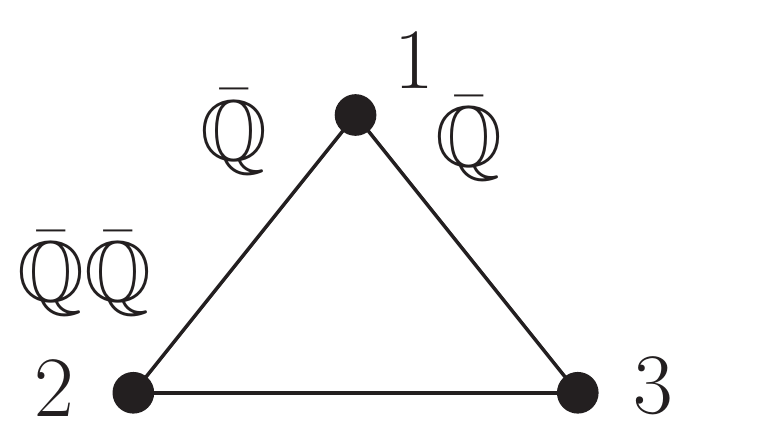} 
\quad \includegraphics[height = 2.5 cm]{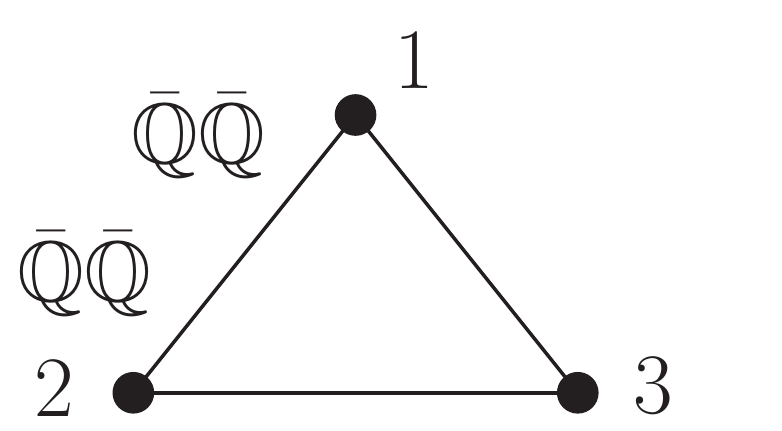} }
\caption{LHC supergraphs corresponding to: (i) $\langle \cS^{''}(1) {\cM}(2) {\cT}(3)\rangle_{g^0}$, (ii) 
$\langle \cS^{'}(1) {\cS^{''}}(2) {\cT}(3)\rangle_{g^0}$ (or $\langle \cR_{0}(1) {\cS^{''}}(2) {\cT}(3)\rangle_{g^0}$), (iii) 
$\langle \cS^{''}(1) {\cS^{''}}(2) {\cT}(3)\rangle_{g^0}$.}
\label{MHV3fig}
\end{figure}

The steps so far allowed us to obtain the desired result \p{545} to the second level in the $\bq_{+}$ expansion
with respect to each point. However, we have not yet considered the diagrams with insertions of $\cM, \cR_0, \cS$, 
where propagators are stretched between $\bar{\mathbb{Q}}_- \bar{\mathbb{Q}}_- A^{++}$ and $\bar{\mathbb{Q}}_- A^{++}$, 
and between a pair of $\bar{\mathbb{Q}}_- \bar{\mathbb{Q}}_- A^{++}$. Examples of such diagrams are depicted in Fig.~\ref{MHV3fig}. 
One can immediately see that they vanish. Indeed, after integrating out $\q_-$, both $\q_{1+}$ and $\q_{2+}$ turn into $B_{12}$,
but $(B_{12})^3 = 0$ (see~\p{B}).   

We have not yet considered the second term in \p{4.58},  the operator $\cR_{\rm Z}$ defined in \p{monstr0}. 
It corresponds to the $\bQ_{\rm Z}$-part of the $\bQ$ transformation~\p{4.28'}. It contains eight Grassmann derivatives (or integrals), contrary to ${\cT}$, $\cR_{0}$ and $\cS$ with four Grassmann derivatives (or integrals) each.
Consequently, an $\cR_{\rm Z}$ vertex lowers 
the Grassmann degree of the whole diagram, just like the vertices from the action $S_{\rm Z}$ \p{N4}. 
So in the MHV-like setting considered here such diagrams vanish.
Let us also recall the property \p{5.25} which forbids $\cR_{\rm Z}$ in the Born-level MHV-like correlators (order $O(g^0)$).

\subsubsection{Higher-order   $\bQ$ variations}

In our simple example \p{545} the hat expansion does not go beyond the second order for each propagator. 
Indeed, at the third order the hat shift produces the operator $(\q y^{-1} \bq)^3 \pa_x \Box_x (1/x^2) \sim 0$, up to a contact term. 
The analogous property is \p{3.50}, i.e. the triple action of the $\bar{\mathbb{Q}}$-variations on the same $A^{++}$ is trivial.
Therefore, the correlators with insertions of the operators $\cN_0$ and $\bar\cL$ (see Figs.~\ref{Nfig} and \ref{Lbarfig}) do not produce 
essentially new Feynman graphs. In order to restore the full dependence on $\bq$ we need to repeat the steps described in Sects.~\ref{s711} and \ref{712} until we reach the maximal level in $\bq$.
The operators $\cN_{\rm Z}$, $\bar\cL_{\rm Z}$ and $\bar\cL_{\rm Z Z}$ do not contribute due to \p{nonAb}.

Let us emphasize once more that the crucial simplification of the above calculation was due to the possibility to discard the $\bQ_{\rm Z}$-transformation \p{4.27}. It produces operators in the expansion \p{4.33} that do not contribute at Born level.

\subsection{Non-MHV-like correlators} \label{snMHV}

Here we show that the entire procedure used in the restoration of the $\bq_+$ dependence of the MHV-like correlator \p{545} applies, without any modification, to the Born-level non-MHV-like correlators (first line in \p{7.1}). The final result is surprisingly simple: We just need to replace all variables $x_{ij}$ in each R-vertex \p{Rinv} and each propagator by the corresponding hatted variables $\hat x_{ij}$ from \p{545}.

We illustrate the procedure on the $G^{\rm nMHV}_{1,1}$ correlator (NMHV-like) with an insertion of the operator 
$\cM^{a'}_{\dot\alpha}$ defined in \p{435}, 
\begin{align} \label{5.18}
\langle \cM^{a'}_{\dot\alpha}(1) {\cT}(2) \cdots {\cT}(n)\rangle_{g^2}\,.
\end{align}
In order to be more specific let us consider a part of a Feynman graph with a 3-valent vertex \p{6.13'}, where one of the propagators $\vev{A^{++} A^{++}}$ has been replaced by its first $\bar{\mathbb{Q}}$ variation at point 1. The rest of the diagram is not relevant.
As before, we apply the Feynman rules and do the $\q_{-}$ and half of the harmonic integrations. 
Compared to the analogous chiral calculation in Section \ref{s7.2}, an additional differential operator as in \p{6.16} is present.
As before in the MHV case, we apply the identity \p{6.17}. The second term in \p{6.17}, which is a total harmonic derivative, vanishes
upon integration by parts in $\int d u_{12}$. Indeed, the other factors in the diagram contain  $u_{12}^+$ but not $u_{12}^-$.\footnote{When integrating the total harmonic $\pa^{++}$ in \p{6.17} by parts we have to take care of the singularities in the denominator like in \p{6.13}. For example, $\pa^{++}_{u_{12}} (u^+_{12} u^+_{13})^{-1} = \delta(u_{12}, u_{13})$, but then the product $\delta^2(x^{\+ +}_{12}) \delta^2(x^{\+ +}_{13}) \delta(u_{12}, u_{13})$ implies that $x_{12}$ is collinear with $x_{13}$. We rule out such singular configurations.}
Therefore we can repeat the steps from \p{6.15} to \p{542} arriving at (see~\p{Rinv})
\begin{align}\label{6.13'}
\begin{array}{c}\includegraphics[height = 2.5 cm]{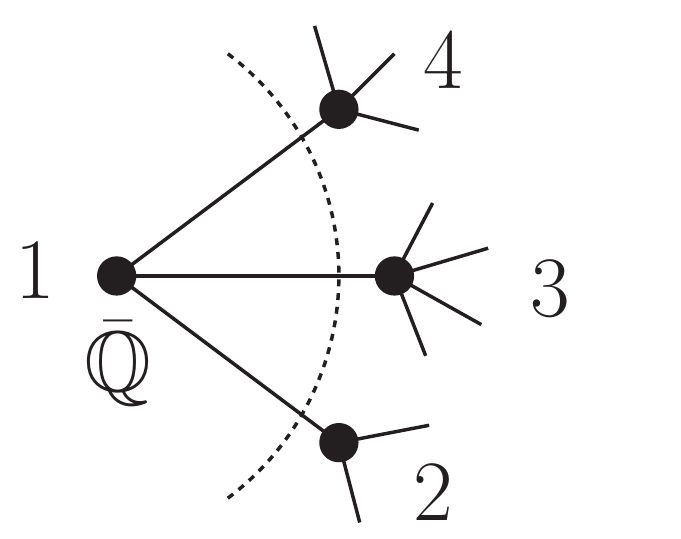}\end{array} =
(\q_{12+}y^{-1}_{12})^{a'\a} \frac{\pa}{\pa x_{12}^{\da\a}}  \left[ \frac{1}{\pi^6}\frac{y_{12}^2 y_{13}^2 y_{23}^2}{x_{12}^2 x_{13}^2 x_{23}^2}R(1;2,3,4) \right], 
\end{align}
where we formally treat $x_{12}$, $x_{13}$ and $x_{23}$ as independent variables, i.e. the derivative acts only on $x_{12}$.
Completing  with $\bq_{1+}$, we can interpret this result as the shift (at the linear level in the $\q\bq$ expansion)
\begin{align}\label{552}
{x}_{12}^{\da\a} \ \to \    x_{12}^{\da\a} 
+ \q_{12+}^{\a} y^{-1}_{12} \bq_{1+}^{\da}  \,.
\end{align}
The other diagrams contributing to \p{5.18}  reconstruct the $\bq_{1+}$-term in the shifts of all $x_{1i}$, $i=2,\ldots,n$.
Then, summing up all point permutations of the correlator \p{5.18} (completed with the corresponding $\bq_{+}$), 
we obtain the contribution of the  shift  \p{552}  (at the linear level) to the full hat shift \p{545} with $\bq_{12+}$. In the same way all $x_{ij}$ get hats: $x_{ij} \to \hat{x}_{ij}$.
Since the insertion of $\cM$ does not change the manipulations leading to the R-vertices,
the same result is valid at higher non-MHV-levels as well (the correlators $G^{\rm nMHV}_{p,1}$ in \p{7.1}).
They contain higher valence R-vertices \p{Rmult} in which $x_{ij}$  undergo hat shifts.

Going beyond the linear level in the $\q_+\bq_+$ expansion is not different from what we did in the MHV-like case in Section \ref{712}. At the quadratic level (the correlators $G^{\rm nMHV}_{p,2}$ in \p{7.1}) we need to consider: (i) a single insertion of the operators $\cR$ or $\cS$, (ii) a double insertion of $\cM$. Once again, we discard the part $\cR_{\rm Z}$ of $\cR$ because it takes us out of the Born-level approximation. The operators $\cR_0$ and $\cS'$ (and a pair of $\cM$) amount to a pair of propagators stretched between $\bar{\mathbb{Q}} A^{++}$ and $A^{++}$, while $\cS^{''}$ generates a propagator between $\bar{\mathbb{Q}}\bar{\mathbb{Q}} A^{++}$ and $A^{++}$. The latter is equivalent to a propagator between a pair of $\bar{\mathbb{Q}} A^{++}$ that is generated by the double $\cM$ insertion. The calculations from Section \ref{712} go through without modifications. We are still able to apply the identity \p{2Var} = \p{2Var'} and to drop the term with a total harmonic derivative. At this stage we reproduce the hat shifts at the quadratic level. The higher levels ($G^{\rm nMHV}_{p,r}$ with $r>2$) involve also the operators $\cN_0$ and $\bar\cL_0$, which do not introduce new structures in the graphs. We thus reconstruct the full hat shift $x_{ij} \to \hat{x}_{ij}$ in the effective superdiagrams (R-vertices and scalar propagators) for the chiral correlator \p{chirCor}.

However, there is a subtlety at the nMHV-level as compared to the MHV-like correlators, which we need to clarify.
As described earlier, going on with the $\bq_+$ expansion we will never apply more than two generators $\bar{\mathbb{Q}}$ to the same line. If we wish to reproduce this behavior from the expansion of the hatted variables $\hat x_{ij}$, it should terminate at the quadratic level $(\bq_{ij+})^2$. In the case of the free propagators in \p{545} this followed from the property $\Box (1/x^2)=0$ (up to a contact term), but why is it true for the R-vertices \p{Rinv}? The answer is simple: In the R-vertex we encounter only the projections $x^{\+}_{ij}$. Hence, when we put hats on such variables, we are dealing only with the projected odd variables $(\bq^{\+}_{ij})_{+a'}$. They cannot appear more than twice in the hat expansion. Consequently, the expansion of the `hatted' R-vertex does not produce $(\bq_{ij+})^2_{(a'b')}$. We have seen in Section \ref{712} that this holds for the propagator $y^2/\hat{x}^2$ as well. In other words, the R-vertex depends on half of the space-time coordinates and hence is annihilated by the directional d'Alembertian $\Box_{ij} = -4\pa_{ij}^{\-} \pa_{ij}^{\+}$. 

Finally, the mechanism of insertion of  $\cM, \cR_0, \cS, \cN_0$ and $\bar\cL_0$ described above cannot generate terms of the type $(\bq_{ij+})^2_{(a'b')}$. Is this true for the expansion of the hatted chiral correlator? As we have just seen, such terms do not arise from the expansion of the propagators $y^2/\hat{x}^2_{ij}$ or of the shifted R-vertices. The higher valence R-vertex \p{Rmult} can always be rewritten so that $\hat{x}_{ij}$ appears only in one of its R-vertex factors. So, we do not need to consider cross-terms coming from the $\bar\q_+$ expansion of the product of two R-vertices. The only possible source of $(\bq_{ij+})^2_{(a'b')}$ terms are  the cross-terms with a linear $\bq_+$ term from a propagator factor $y^2/\hat{x}^2$ and a linear term from the accompanying shifted R-vertex: 
\begin{align} \label{5.22}
x^{\da\a} B^{(a'}_{\a} \times B^{b')\b} \frac{\pa}{\pa x^{\da\b}} R \,.
\end{align}
Here $x$ stands for $x_{ij}$  and the derivative acts only on $x_{ij}$ in the expression of the R-vertex \p{Rinv}. The latter is independent of  $x^{\-}$, so $\pa_{\b\da} R = \xi^{\+}_{\da} \frac{\pa}{\pa x^{\+ \b}} R$. The identity $x^2 B^{\b} = x^{\- \b} (x^{\+} B) - x^{\+ \b} (x^{\-} B)$ and the invariance of the R-vertex \p{Rinv} under local rescaling $x_{ij} \to \rho\ x_{ij}$, i.e. $x^{\+ \b} \frac{\pa}{\pa x^{\+ \b}} R = 0$, imply 
\begin{align}\notag
x^2 B^{b'\b} \frac{\pa}{\pa x^{\da\b}} R = \xi^{\+}_{\da} (x^{\+} B^{b'}) x^{\- \b} \frac{\pa}{\pa x^{\+ \b}} R\,.
\end{align}
Therefore eq.~\p{5.22} is proportional to $(x^{\+} B^{(a'})(x^{\+} B^{b')}) = 0$.

In summary, we can say that the operator $(B^{(a'}_{ij}B^{b')}_{ij})\Box_{ij}$
annihilates the effective supergraphs representing the chiral correlator \p{chirCor}. To avoid any misunderstanding let us stress once more that this operator is to be applied to the effective Feynman graphs (not to the chiral correlator itself!) where we treat all variables $x_{ij}$ as independent. A similar argument shows that the expansion of the hatted variables cannot give rise to terms like $(\bq_{ij+})^3$ and $(\bq_{ij+})^4$. We conclude that the insertion of  $\cT,\cM,\cR_0, \cS,\cN_0$ and $\bar\cL_0$ completely reproduces the hat expansion of the effective supergraphs.

This reinterpretation of the modified Feynman rules  explains why  the non-MHV-like correlators with restored $\bq-$dependence remain gauge invariant, i.e., independent of $\xi^{\dt\pm}$. The hat-shift in the propagators, $\hat x_{ij} = x_{ij} + \q_{ij+} y^{-1}_{ij} \bq_{ij+}$,  is independent of $\xi$. After the replacement $x_{ij} \to \hat{x}_{ij}$ the R-vertex \p{Rinv} becomes
\begin{align}\notag
\hat{R}(1;2,3,4)= -\frac{\delta^2\Big((\hat x^{\+}_{13} \hat x^{\+}_{14})(\hat x^{\+}_{12} \q_{12+})(y_{12})^{-1} + {\rm cycle}(2,3,4) \Big)}{(\hat x^{\+}_{12} \hat x^{\+}_{13})(\hat x^{\+}_{13} \hat x^{\+}_{14})(\hat x^{\+}_{14} \hat x^{\+}_{12})} \ .
\end{align}
Formally, this is the same as the expression in \cite{Chicherin:2014uca}, in the gauge $Z_*=(0,\xi^{\+}_\da,0)$  and with the identification $\sigma_{ij}  = \hat x^{\+}_{ij}/x^{2}_{ij}$. Then we can apply the  argument from  \cite{Chicherin:2014uca} showing that the dependence on $\xi^{\+}$ cancels out in the sum of all Feynman graphs (for more details see Appendix~\ref{nolambda}).

An important remark about the above effective Feynman rules (putting hats on all variables $x_{ij}$) is that this should be done \emph{ before} any attempt to simplify the R-vertices like, for instance, the explicit elimination of the $\xi-$dependence. In particular, the simplification may make use of algebraic identities of the type $x_{ij} + x_{jk}=x_{ik}$. Clearly, this is not true for the hatted variables! 

Our hatting procedure works for all higher valence R-vertices \p{Rmult} needed in an $\mathrm{N^p MHV}$-like correlators in the Born approximation (at order $O(g^{2p})$). They are characterized by the difference between the total numbers of $\q$'s and $\bq$'s being equal to $4p$. Knowing the Born-level correlators, we can find  loop corrections to correlators with fewer points by means of the Lagrangian insertion formula
\begin{align} \label{ins}
\frac{\pa}{\pa g^2} \langle T(1) \ldots T(n-1) \rangle 
= \omega \int d^4 x_{n} d^4 \q_{n+}\ \langle T(1) \ldots T(n-1) T(x_{n},\q_{n+}, \bq_{n+} = 0)\rangle\,.
\end{align}
In view of \p{4.33} and \p{TtoS} this is equivalent to including the interaction vertices $S_{\rm Z}$ \p{N4} in the Feynman graphs.

In order to collect more evidence for the validity of our effective non-chiral Feynman rules, we performed an independent check. We calculated the component $\q_{1+}\bq_{2+}(\q_{3+})^4$ of the five- and six-point correlators $G^{\rm nMHV}_{1,1}$ \p{7.1}, 
using conventional Feynman rules, see~\p{G5} and \p{G6}, respectively. 
Then we compared them with the same component extracted from $\langle \cT(1)\cM(2)\cT(3) \ldots \cT(n) \rangle$ ($n=5,6$), calculated by means of the effective non-chiral supergraphs, 
and found full agreement. More details can be found in Appendix~\ref{apE}.

\subsection{Other correlators} \label{sanMHV}

When we reconstructed the Born-level non-chiral correlators $G^{\rm MHV}_{r}$ and $G^{\rm nMHV}_{p,r}$ in \p{7.1}, we only used part of the operators considered in Section \ref{s5}. For example, we never used $\cR_{\rm Z}$ \p{monstr0}. Now we wish to discuss the correlators  for which these operators are relevant. It is convenient to 
split the operators in three classes: (i) degree 4, (ii) degree 8, (iii) degree 12 operators,
according to the number of Grassmann derivatives in their expression in terms of the gauge connections $A^{++}$ and $A^{+}_{\da}$, see Table~\ref{Degtab}. For example, $\cT$ \p{439} involves the projector $(\pa_+)^4$, i.e. four Grassmann derivatives, whereas $\cR_{\rm Z}$ \p{monstr0} involves a pair of projectors $(\pa_+)^4$ and $(\pa^+)^4$, i.e. eight Grassmann derivatives.
We also treat the $S_{\rm Z}$ part of the action \p{N4} as a degree-8 operator because of its Grassmann measure $(\pa)^8$. Previously we only needed degree-4 operators, which are generated by $\bar{\mathbb{Q}}$-variations of $\cT$.
The degree-8 operators $\cR_{\rm Z}$, $\cN_{\rm Z}$ and $\bar\cL_{\rm Z}$ are generated by a single $\bQ_{\rm Z}$-variation, and the degree-12 operator $\bar\cL_{\rm ZZ}$ is generated by a double $\bQ_{\rm Z}$-variation. 
So, we can equivalently identify them by counting the powers of $\omega$.

\begin{table}[!h]
\begin{center}
\begin{tabular}{c|c|c}
degree-4 & degree-8 & degree-12 \\ \hline & & \\  [-2.5ex] $\omega^0$ & $\omega^1$ & $\omega^2$ \\  \hline  & & \\ [-2.0ex] 
$\cT,\,\cM,\,\cR_0,\,\cS,\,\cN_0,\,\bar\cL_0$ & $S_{\rm Z},\,\cR_{\rm Z},\,\cN_{\rm Z},\,\bar\cL_{\rm Z}$ & $\bar\cL_{\rm ZZ}$  
\end{tabular}
\end{center}
\caption{Classification of the operators from the expansion \p{4.33} of the stress-tensor supermultiplet $T(\q_+,\bq_{+})$ 
according to the number of Grassmann derivatives they bring  in Feynman graphs, see \p{439}, \p{435}, \p{O'a'b'}, \p{monstr0}, \p{478}, \p{Odadb}.}
\label{Degtab}
\end{table}

Let us now consider the component $(\bq^{2}_{1+})_{(a'b')}$ (completed by two $\q_{+}$-variables) of the MHV-like correlator $G^{\rm MHV}_{2}$ \p{7.1}. It does not appear in the $\bq$-expansion of the supersymmetrized free correlator (order $O(g^0)$), e.g. \p{545},  only the $(\bq^{2}_{1+})_{(\da\db)}$ component arises there at the quadratic level. To see such a component, we need to go to order $O(g^2)$ in the perturbative expansion. It is generated by the degree-8 operator $\cR_{\rm Z}$ \p{monstr0}. As an example, consider the seven-point correlator,
\begin{align} \label{525}
\langle \cR^{(a'b')}_{\rm Z}(1) \cT(2) \ldots \cT(7) \rangle_{g^2} 
= \begin{array}{c}\includegraphics[height = 2.5 cm]{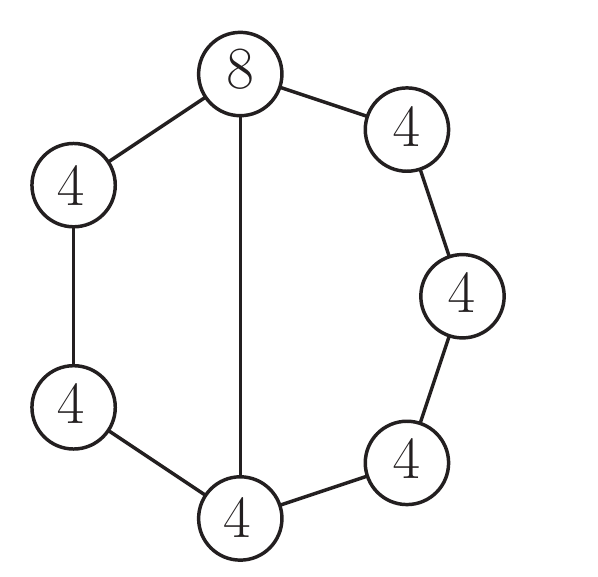}\end{array} + \quad  \text{permutations}
\end{align}
The numbers in the picture denote the degrees of the operators (see Table~\ref{Degtab}). This is obviously a rational function, since it does not involve space-time integrations. Counting the Grassmann degree of the diagrams, more precisely the difference between the number of $\q_+$ and $\bq_+$ variables, is easy. Each propagator contributes $(+4)$ units, each vertex contributes minus its degree ($(-4)$ or $(-8)$ units for the vertices in the above diagram). Thus for \p{525} we get zero, i.e. it has an equal number of $\q_+$ and $\bq_+$ (equal to two), as it should be for an MHV-like correlator.

Finally, we would like to comment on the `mostly antichiral' supercorrelators $G_{p,r}^{\overline{\mathrm{nMHV}}}$ from \p{7.1}.
Consider the  five-point Born-level supercorrelator $G_{1,r}^{\overline{\mathrm{nMHV}}}$ at order $g^2$, i.e. for which 
the number of $\bq$'s exceeds the number of $\q$'s by four. One of its components, 
the correlator (see \p{5.2}, \p{5.10})
\begin{align} \label{bLOs}
\langle \bar{{L}}(1) \,\cO(2) \cdots \,\cO(5)  \rangle
\end{align} 
belongs to this class. We can obtain it by conjugating $\langle L(1) \,\cO(2) \cdots \,\cO(5)  \rangle$
which we already know (see \p{Intr} and footnote \ref{f18}). 
However, the question is how to reproduce it, in principal, in the chiral formalism.

\begin{figure}[!h]
\centerline{
\includegraphics[height = 2.8 cm]{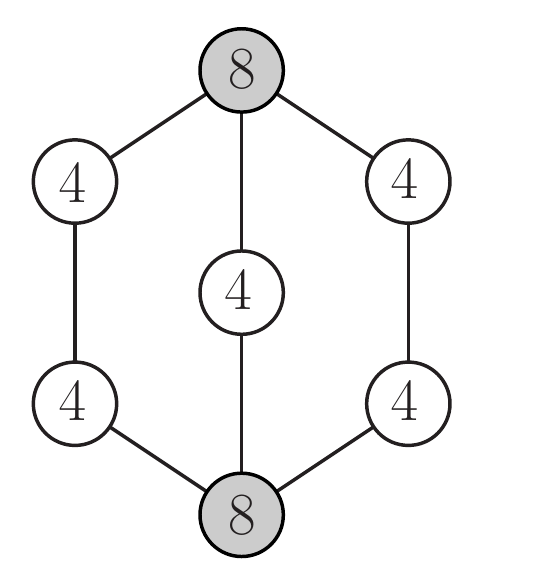} \quad \includegraphics[height = 2.8 cm]{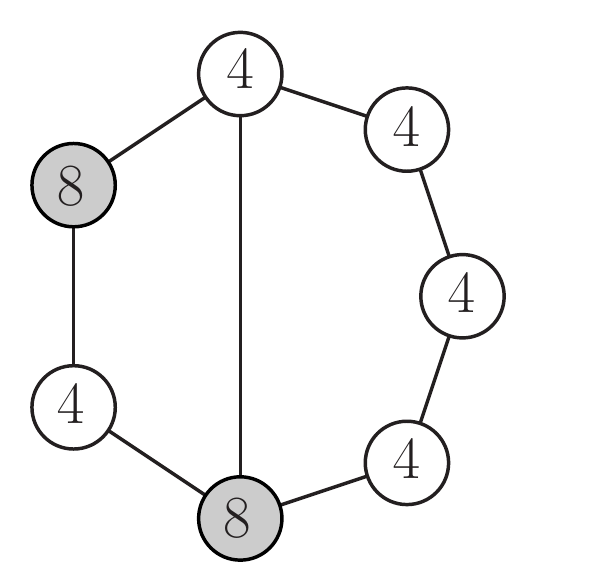} 
\quad \includegraphics[height = 2.8 cm]{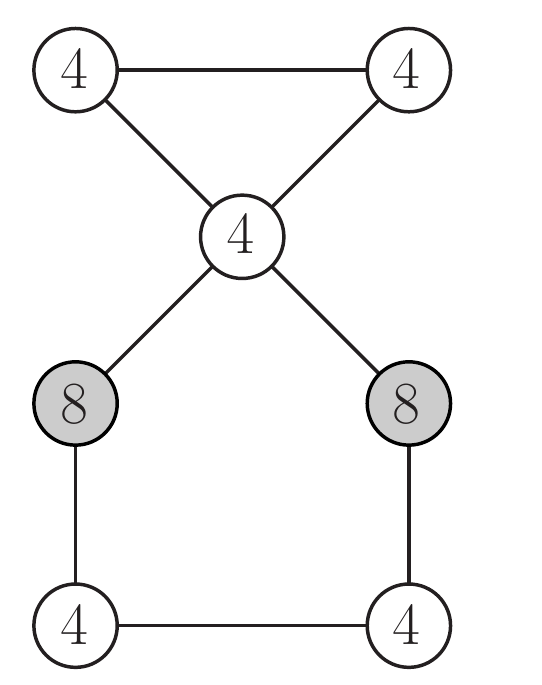} \quad \includegraphics[height = 2.8 cm]{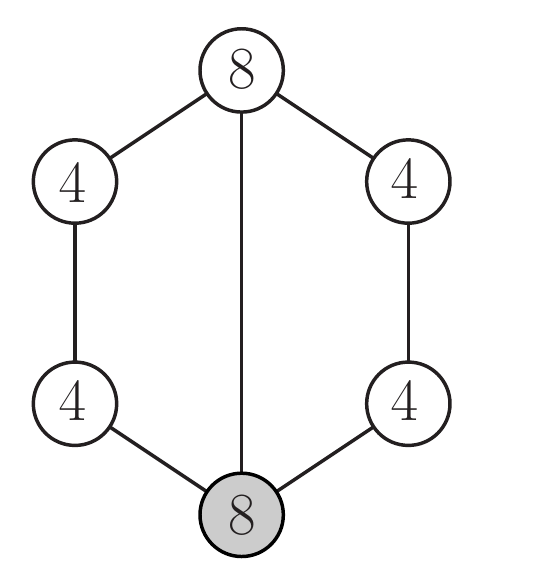} 
\quad \includegraphics[height = 2.5 cm]{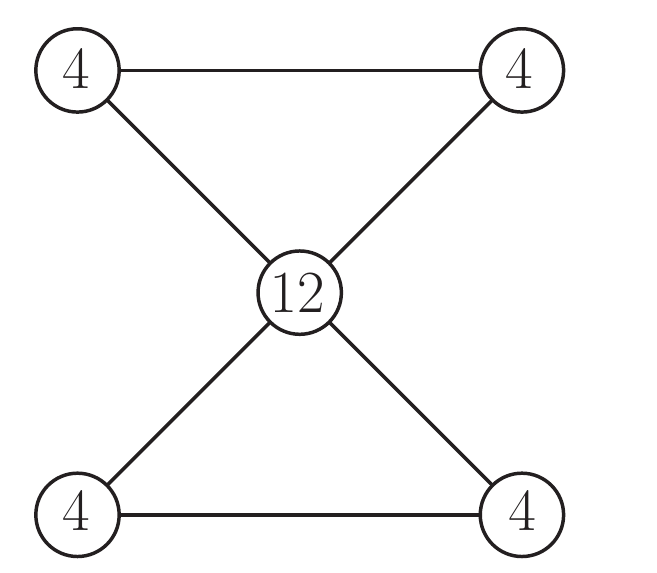}}
\caption{Graphs contributing to the correlator $\langle \bar{{L}}(1) \,\cO(2) \cdots \,\cO(5)  \rangle$.
They come in three types: (i) graphs with two interaction vertices $S_{\rm Z}$ denoted by gray blobs, 
(ii) graphs with one interaction vertex $S_{\rm Z}$ and one degree-8 operator, (iii) graphs without interaction vertices but with one degree-12 operator.}
\label{antifig}
\end{figure}

There are a number of contributing supergraphs, some of which are shown in Fig.~\ref{antifig}. They can be split in three classes according to the number of space-time integrations.  All of them are proportional to $\omega^2$. The diagrams involve non-trivial space-time integrations, although their sum is a rational function -- the Born-level correlator. Thus, the chiral semi-superfield approach to the correlators  $G_{p,r}^{\overline{\mathrm{nMHV}}}$ is analogous to the conventional Feynman graph calculation of the  components of the Born-level supercorrelator (an example of such a calculation is given in Appendix~\ref{apE}).

This variety of Feynman graphs and especially the presence of non-trivial space-time integrals shows that it would be rather hard to calculate $G_{p,r}^{\overline{\mathrm{nMHV}}}$  in the chiral framework. However, if we adopt the equivalent antichiral framework,   the  correlators $G_{p,r}^{\overline{\mathrm{nMHV}}}$ are given by free Feynman rules (just like the non-MHV-like correlators considered in Section \ref{snMHV}), without any space-time integrals. This means that the complicated expression for a mostly antichiral  correlator, computed in the chiral framework, should drastically simplify due to some intricate algebraic and integral identities.

{In conclusion, we would like to recall that the crucial simplification in our treatment of mostly chiral Born-level supercorrelators  was due to the possibility to discard the second, non-linear term in the $\bQ$ transformation \p{4.27} related to the interaction Lagrangian. We have shown that the linearized contribution of this term vanishes, while the non-linear terms are of higher order in the coupling $g$ and hence irrelevant at Born level. }

\section{Chiral and R-analytic bases in superspace}\label{s9}

In this section we show that the main result of Section \ref{s8}, namely that the $\bq$ dependence is restored simply by putting hats on $x_{ij}$, has a natural interpretation as a change of basis from chiral to R-analytic superspace. This is indeed the simplest way to render the chiral Feynman rules compatible with $\bQ$ supersymmetry. However, it is by no means the only way, so the question remains why our final answer for the non-chiral correlators is so simple.

\subsection{Chiral and R-analytic basis}\label{s9.1}

The dependence on $\bq$ for chiral or R-analytic functions needs to be considered in the appropriate basis in superspace. Let us illustrate this on the familiar example of chiral superspace. In the real basis of the full superspace the covariant derivative $\bar D$ has the form $\bar D= \pa/\pa\bq - \frac12 \q \pa/\pa x$. A chiral superfield satisfies the condition $\bar D \Phi(x,\q,\bq)=0$. In order to solve it, we define the chiral basis
\begin{align}\label{36}
x_{\rm ch} = x + \frac1{2} \q^A\bq_A \,,
\end{align}
in which $\bar D= \pa/\pa\bq$ and the chirality condition can be easily solved:
\begin{align}\notag
\bar D \Phi(x_{\rm ch},\q,\bq)=0\ \Rightarrow \ \Phi = \Phi(x_{\rm ch},\q)\,.
\end{align}
Similarly, an R-analytic superfield is defined by two conditions in terms of RH-projected covariant derivatives (recall~\p{5.3}):
\begin{align}\label{38}
\bar w^A_{+a'} D^\a_A \Phi(x,\q,\bq, u)= w^{a}_{+A} \bar D^A_\da \Phi(x,\q,\bq, u)=0\,.
\end{align}
To solve these constraints we define the R-analytic basis with coordinates $x^{\da\a}_{\rm an}$, $\q^a_+$, $\q^{a'}_{-}$, $\bq^{\da}_{+a'}$, $\bq^{\da}_{-a}$ \p{4.39}, \p{AnB}.
In this basis the covariant derivatives $\bar w^A_{+a'} D_A$ and $w^{a}_{+A} \bar D^A$ are short (see \p{DaB}), so the R-analyticity constraints \p{38} simply imply $\Phi=\Phi(x_{\rm an}, \q^{a}_+,\bq_{+a'},w)$.

Of particular interest for us is the  change of variables from the chiral basis \p{36} to the R-analytic one \p{AnB},
\begin{align}\label{3.39}
 x_{\rm ch} = x_{\rm an} +    \q^{a'}_- \bq_{+a'} \,,
\end{align} 
where the completeness condition for the RHs \p{compl} is taken into account.
The different $x$'s transform according to the type of basis, real, chiral or R-analytic:
\begin{align}\label{8.9}
&{\rm RB:}\quad  \delta x =  \frac1{2}(\q^A \bep_A -\ep^A \bq_A)\,, \qquad \delta \q= \ep\,, \qquad \delta\bq = \bep\nt
&{\rm CB:}\quad  \delta x_{\rm ch} =  \q \bep\,, \qquad \delta \q= \ep\,, \qquad \delta\bq = \bep\nt
&{\rm AB:}\quad  \delta x_{\rm an} =  \q^{a}_+ \bep_{-a} -\ep^{a'}_- \bq_{+a'}\,, \qquad \delta \q_\pm= \ep_\pm\,, \qquad \delta\bq_\pm = \bep_\pm
\end{align}
where $\ep_\pm = \ep^A  w_{\pm A}$ and $\bep_\pm = \bep_A \bar w^A_\pm$. 
Let us verify that the change of basis \p{3.39} reproduces the correct transformation of $x_{\rm ch}$ (recall~\p{compl}):
\begin{align}\notag
\delta x_{\rm ch} = (\q_+ \bep_- -\ep_- \bq_+) +(\ep_- \bq_+ + \q_- \bep_+) = \q^A(w^{a}_{+A} \bar w^B_{-a} + w^{a'}_{-A} \bar w^B_{+a'}) \bep_B = \q^A \bep_A\,.
\end{align}

Now, let us turn to our propagators $\vev{A^{++} A^{++}}$ and $\vev{A^+_\da A^{++}}$ (see~\p{prop1}, \p{prop2}) containing $\delta^2(x^{\+ +}) \delta^4(\q^+)$. Originally, the quantization was done in terms of `semi-superfields' $A(x,\q,u)$, no anti-chiral variables $\bq$ were used. Does this mean that our propagators are defined in the chiral basis? The answer is affirmative and follows from the way $Q$ supersymmetry is realized on them. Indeed, the propagators are invariant under simple shifts $Q \q = \ep$, without touching the space-time variables $x$. This corresponds to the chiral basis in \p{8.9}. The propagators satisfy the chirality condition
\begin{align}\label{41}
    \bar D^A \left[\delta^2(x_{\rm ch}^{\+ +}) \delta^4(\q^+) \right]=0 \quad  {\rm with} \ \bar D^A = \pa/\pa\bq_A\,.
\end{align}
Then it is natural to do the R-analyticity projection in \p{4.6}  with the covariant spinor derivatives,
\begin{align}\label{42}
(D_{+a'})^4 \left[ \delta^2(x_{\rm ch}^{\+ +}) \delta^4(\q^+) \ldots \delta^2(x_{\rm ch}^{\+ +}) \delta^4(\q^+)\right]\,.
\end{align}
This projection satisfies both constraints \p{38}, the first due to the projector in \p{42}, the second as a corollary of the chirality \p{41} (recall that $\{\bar D^{a}_+, D_{+a'}\}=0$). 

If we want the covariant derivatives in \p{42} to become partial derivatives and hence $(D_{+})^4 = \int d^4\q_-$, we need to  change the basis:
\begin{align}\label{43}
&(\pa/\pa\q^{a'}_-)^4 \left[ \delta^2(x_{\rm an}^{\++}) \delta^4(\q^+) \ldots \delta^2(x_{\rm an}^{\+ +}) \delta^4(\q^+)\right]\nt
&=\left(\frac{\pa}{\pa\q^{a'}_-} + \bq_{+a'}  \frac{\pa}{\pa x_{\rm ch}} \right)^4 \left[ \delta^2(x_{\rm ch}^{\+ +}) \delta^4(\q^+) \ldots \delta^2(x_{\rm ch}^{\+ +}) \delta^4(\q^+)\right]\nt
&=\left(D_{+a'}  \right)^4 \left[ \delta^2(x_{\rm ch}^{\+ +}) \delta^4(\q^+) \ldots \delta^2(x_{\rm ch}^{\+ +}) \delta^4(\q^+)\right]\,,
\end{align}
as stated in \p{42}. In the first line in \p{43} the $\bq$ dependence is present in the arguments $x_{\rm an}=x_{\rm ch}   - \q_-\bq_+$. We claim that this simple change of basis restores the $\bQ$ symmetry of any Feynman graph.

\subsection{$\bQ$ invariance of the R-vertices}

The $\bQ$ invariance of our Feynman graphs is not immediately obvious because the propagator  
$\langle A^{++} A^{++}\rangle \sim \delta^2(x^{\+ +}) \delta^4(\q^{+})$ is not invariant.\footnote{In the twistor formulation of Ref.~\cite{Adamo:2011cb} the chiral propagator $\delta^{4|4}(Z_*+\sigma_{ij} Z_i + \sigma_{ji} Z_j)$  is manifestly $\bQ$ invariant. In particular, with our choice $Z_*=(0,\xi,0)$ the propagator amounts to  $\delta^2(\sigma_{ij}  + \sigma_{ji})\,\delta^2(\xi + \sigma_{ij}x_{i}  + \sigma_{ji}x_{j}) \,\delta^4(\sigma_{ij}\q_i  + \sigma_{ji}\q_j)$. The variation $\bQ x_i = \q_i$ is then compensated by the Grassmann delta function. Roughly speaking, the analog of $\sigma$ for us is $u^+$, but the twistor formalism does not employ $u^-$. This leads to some differences in the way $\bQ$ supersymmetry is realized, as we explain in this subsection.}
Indeed, under the shift $x \to x + \bar{\ep}^{A} \q_{A}$ the propagator transforms as follows (here we use the expanded notation $\delta^2(z)=\delta(z,\bar z)$)
\begin{align}\notag
\langle A^{++} A^{++}\rangle\ &\rightarrow \  
\delta\big(x^{\+ +}\,  , \, x^{\- -}  + \bar{\ep}^{\dot -} \q ^{-}\big) \ \delta^4(\q^{+})\,,
\end{align}
and the projection $\bar{\ep}^{\dot -}$ remains. Nevertheless, as we show below, in the effective Feynman rules the full $\bQ$ supersymmetry is restored simply by making the change of basis \p{3.39}.

Let us consider the part of a Feynman graph where   vertex $i$ is connected with  vertices $j,k,l$. In particular, the propagator stretched between $i$ and $j$ contributes (after integrating out $\delta( u_{ij}, u_{ji})$; recall the calculation in Section \ref{s7.1})
\begin{align}\label{344}
\begin{array}{c}\includegraphics[height = 2.0 cm]{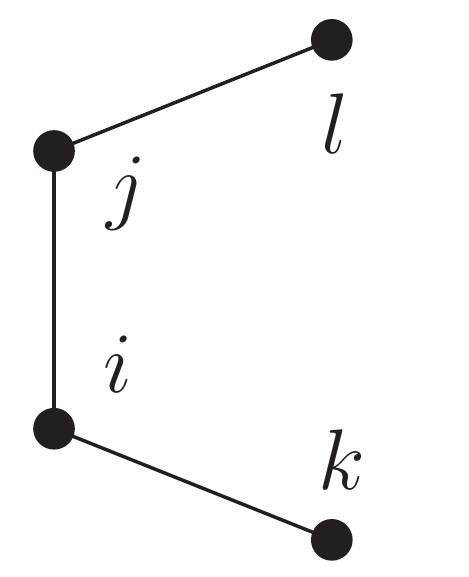}\end{array} \Rightarrow\
\int du\   
\delta\bigl(x^{\+ +}_{ij{\rm \, ch}},  x^{\- -}_{ij{\rm \, ch}}\bigr) y^2_{ij} 
\delta^2\left( u^{+}_{ij}\cdot (\q_{j-} + B_{ij}) \right) \delta^2\left( u^{+}_{ij}\cdot (\q_{i-} + B_{ij})\right). 
\end{align}
The Grassmann delta functions from this and the other propagators allow us to find $\q_-$ as functions of $\q_+$,
\begin{align}
\q_{i-} = \frac1{( u^{+}_{ij}\, u^{+}_{ik})}  
\left[u^{+}_{ij} (u^{+}_{ik} B_{ik}) - u^{+}_{ik} (u^{+}_{ij} B_{ij})\right]  \nt
\q_{j-} = \frac1{( u^{+}_{ji}\, u^{+}_{jl})}  
\left[u^{+}_{ji} (u^{+}_{jl} B_{jl}) - u^{+}_{jl} (u^{+}_{ji} B_{ji})\right] \label{thetaj-}\,.
\end{align}
 
After the change of basis \p{3.39} with $\q_-$ from \p{thetaj-},  the bosonic delta function in \p{344} has support on (we drop the subscript `an'; recall~\p{B})
\begin{align}\label{9.15}
0=x^{\+}_{ij{\rm \, ch}}    \cdot u^{+}_{ij} = (x^{\+}_{ij} +  \q_{i-} \bar \q^{\+ }_{i+} -  \q_{j-} \bar \q^{\+ }_{j+}) \cdot u^{+}_{ij} = (x^{\+}_{ij} - B_{ij} \bar\q^{\+ }_{ij+}) \cdot u^{+}_{ij} = \hat x^{\+}_{ij}\cdot u^{+}_{ij}\,,
\end{align}
and similarly for the conjugate variable $x^{\-}_{ij{\rm \, ch}}   \cdot u^{-}_{ij} =0$.  Notice that the projections of $\q_{i-}\,, \, \q_{j-}$ involving the other vertices $k,l$ have dropped out. Eq.~\p{9.15} allows us to solve for $u^{+}_{ij} = \rho_{ij}\, \hat{x}_{ij}^{\+ }$ where $\rho_{ij}$ is an arbitrary normalization factor. The rest of the Feynman diagram (propagators and LH denominators from $S_{\rm Z}$ \p{N4}, \p{439}) is homogeneous in $u^{+}_{ij}$ of degree zero, so this factor is irrelevant. 

Then we consider the second  condition following from the support of the bosonic delta function
\begin{align}\label{8.17}
0=x^{\-}_{ij{\rm \, ch}}    \cdot u^{-}_{ij} 
= (x^{\-}_{ij} +  \q_{i-} \bar \q^{\dot -}_{i+} -  \q_{j-} \bar \q^{\dot -}_{j+}) \cdot u^{-}_{ij}\,,
\end{align}
where $ \q_{i-}$ and $ \q_{j-}$ are now expressed in terms of $\hat{x}^{\+ } \sim u^+$ according to \p{9.15} and its analogs for the other propagators:
\begin{align}
\q_{i-} = \frac1{( \hat{x}^{\+ }_{ij}\,\hat{x}^{\+ }_{ik})}  
\left[\hat{x}^{\+ }_{ij} (\hat{x}^{\+ }_{ik} B_{ik}) - \hat{x}^{\+ }_{ik} (\hat{x}^{\+ }_{ij} B_{ij})\right]  
\nt
\q_{j-} = \frac1{( \hat{x}^{\+ }_{ji}\, \hat{x}^{\+ }_{jl})}  
\left[\hat{x}^{\+ }_{ji} (\hat{x}^{\+ }_{jl} B_{jl}) - \hat{x}^{\+ }_{jl} (\hat{x}^{\+ }_{ji} B_{ji})\right] \,.  \label{8.18}
\end{align}
 Again the normalization factors $\rho$ drop out due to homogeneity. Eq.~\p{8.17} enables us to find $u^-_{ij}$ up to normalization. The only place where $u^-$ appears is the Jacobian produced by the harmonic integral 
$\pi \int d u \, \delta(x^{\+} \cdot u^{+}, x^{\-} \cdot u^{-}) =1/(x^{\-}\cdot x^{\+})= 1/x^2 \equiv J^{-1}$ (see \p{intd}), or after the change of basis,
$$
J^{-1}= \hat{x}^{\+ }_{ij} \cdot \left( x^{\-}_{ij} +  \q_{i-} \bar \q^{\dot -}_{i+} -  \q_{j-} \bar \q^{\dot -}_{j+} \right)
$$
with $ \q_{i-}$ and $ \q_{j-}$ from   \p{8.18}.
We have $(\hat{x}^{\+ }_{ij}  \q_{i-}) = - ( \hat{x}^{\+ }_{ij} B_{ij})$, 
$(\hat{x}^{\+ }_{ij}  \q_{j-}) = - ( \hat{x}^{\+ }_{ij} B_{ij})$, so
$$
J^{-1}= ( \hat{x}^{\+ }_{ij}  x^{\-}_{ij} ) - (\hat{x}^{\+ }_{ij} B_{ij} \bar \q^{\dot -}_{ij+} )
= (\hat{x}^{\+ }_{ij} \hat{x}^{\dot -}_{ij}) = \hat{x}_{ij}^2 \,.
$$
This result, multiplied by the RH factor $y^2_{ij}$ from \p{344}, produces the expected  propagator factor, see  \p{545}. Together with the replacement $u^{+}_{ij}  \to  \hat{x}_{ij}^{\+ }$ this completes  the effective Feynman rules with manifest $\bQ$ invariance.

In conclusion, the simple form of the modified Feynman rules can be interpreted as the basis change needed to make compatible the initial formulation of the propagators as chiral objects with the R-analytic nature of the operators $\cT$, etc. This of course does not explain why the fully supersymmetric correlators do not involve more sophisticated nilpotent anti-chiral invariants. This questions requires further study. 

\section{Conclusions}

In this paper we use the LHC superspace approach to obtain effective Feynman rules 
leading to the multipoint Born-level correlation functions of the stress-tensor multiplet.
The chiral sector of the  multiplet is naturally formulated in terms of LHC superfields.
So, the derivation of the Feynman rules for the chiral correlators is straightforward. The description of the 
non-chiral part of the multiplet is more involved. In our chiral approach the multiplet 
splits up into several chiral layers corresponding to the expansion in terms of $\bq_+$. Nevertheless, at the level of the 
effective Feynman rules we succeeded to put the layers together in a manifestly supersymmetric expression for the 
non-chiral correlator. 

The same method is applicable to other supermultiplets as well.
Indeed, since we know the expression for the gauge super-curvature in terms of LHC superfields, 
$W_{AB} = \pa^+_A \pa^+_B A^{--}$ (see~\p{3.12}, \p{448}), we can construct many other multiplets.
For example, projecting $W_{AB}$ with RHs we find the (chiral truncation of the)  half-BPS multiplet of  weight $k$: 
$\cO^{(k)} = \tr \left( W_{++} \right)^k$. The (chiral truncation of the) Konishi multiplet is $K = \ep^{ABCD} W_{AB} W_{CD}$.
Clearly, the Feynman rules elaborated for $\cO^{(2)}$ (stress-tensor multiplet) are straightforward to generalized to $\cO^{(k)}$ and $K$.
The only complication is that they do not result in concise R-invariants like \p{Rinv}, i.e. the Grassmann integrations may be more laborious. 
Nevertheless, this is a purely algebraic task but not a fundamental obstacle. 
In \cite{Koster:2014fva} the twistor Feynman rules were applied to the calculation of the one-loop dilatation operator with manifest supersymmetry. Our formalism can be equally useful in such calculations of the dilatation operator, perhaps also at higher perturbative levels.

The non-chiral completions of the supermultiplets and the corresponding correlators can also be 
constructed. For the R-analytic  half-BPS $\cO^{(k)}$ the result is rather obvious. 
One just needs to take up the chiral Feynman rules and  replace $x_{ij}$ by $\hat x_{ij}$ (see~\p{545}). 
For other correlators the restoration of $\bq$ effectively leads to using the corresponding 
supersymmetrization $\hat x_{ij}$ of $x_{ij}$. If two Konishi operators appear at the vertices $i$ and $j$, we should use the supersymmetrization $x_{ij} \to \hat x_{ij} = x_{ij} + \frac12 \q_i \bq_{i} - \frac12 \q_j \bq_{j}$. 
If points $i$ and $j$ correspond to a half-BPS and a Konishi operators, respectively, then $x_{ij} \to \hat x_{ij} = x_{ij} + \q_{ij+} y^{-1}_{ij} \bq_{i+} - \frac12 \q_j \bq_{j}$. 

In conclusion, we emphasize once more the surprising simplicity of our main result. The reason why the non-chiral supersymmetrization of the multipoint Born-level correlators amounts to a simple shift of the space-time coordinates remains mysterious. This could be a manifestation of some hidden symmetry of the correlators, perhaps an off-shell analog of the famous Yangian symmetry of the superamplitudes \cite{Drummond:2008vq,Berkovits:2008ic,Beisert:2008iq,Drummond:2009fd}. Or it could follow from the analytic properties of the correlators, which would be incompatible with any other kind of nilpotent superconformal invariant. We hope to come back to these interesting questions in the future.

\section*{Acknowledgements}

 We profited from numerous discussions with T. Adamo, I. Bandos, B. Eden, P. Heslop, E. Ivanov,  L. Mason and D. Skinner.  We are particularly grateful to G. Korchemsky for making comments on the draft. We acknowledge partial support by the French National Agency for Research (ANR) under contract StrongInt (BLANC-SIMI-4-2011). The work of D.C. has been supported by the ``Investissements d'avenir, Labex ENIGMASS'' and partially supported by the RFBR grant 14-01-00341.


\newpage

\section*{Appendices} 

\appendix

\section{Proof of the alternative form of $\cT$}\label{proofTW}

Here we show the equivalence of the two forms \p{439} and \p{TW} of the chiral truncation $\cT$ of the stress-tensor multiplet (see~\p{4.33}). This proof follows closely the analogous equivalence of the traditional chiral form of the action of $\cN=2$ SYM and its harmonic superspace version given in \cite{Galperin:2001uw}. 

For simplicity we restrict ourselves to the Abelian (free case) where $L_{\rm Z}$ is given by the bilinear term  in \p{439}, 
$L_{\rm Z} = -\frac{\omega}{2} \int du\, A^{++} A^{--}$ (see \p{448}). We start with the form \p{439} and rewrite it as follows:
\begin{align}\notag
\cT &= {-\frac{1}{2}} (\pa_+)^4 \int du \ A^{++} A^{--} = {-\frac{1}{8}} \int du \ (\pa^+_+)^2 (\pa^-_+)^2 (A^{++} A^{--} ) \\ 
&=  {-\frac{1}{8}}\int du \  (\pa^-_+)^2 (A^{++} W_{++}) 
\end{align}
where $W_{++} = (\pa^+_+)^2 A^{--}$ (see \p{TW}). We then write $(\pa^-_+)^2 = \pa^{--} (\pa^-_+\pa^+_+)$ and integrate the harmonic derivative by parts. It annihilates the curvature $W_{++}$. To see this, we first remark that the linearized version of \p{314} implies  $\pa^{++} W_{++} =0$. Since $W_{++}$ has zero LH charge (scalar), the two properties $\pa^{++} W_{++} =0$ and $\pa^{--} W_{++} =0$ are equivalent to each other and to the fact that $W_{++}$ does not depend on the LHs. So, after the integration by parts $\pa^{--} $ can hit only $A^{++}$ and turn it into $\pa^{++} A^{--}$ with the help of \p{3.11}. We thus get (using $\pa^+_+ W_{++} = (\pa^+_+)^3 A^{--}=0$)
\begin{align}
\cT &= {\frac{1}{8}} \int du \ \pa^-_+\pa^+_+ (\pa^{++} A^{--} W_{++}) = {-\frac{1}{8}} \int du \ (\pa^+_+)^2 ( A^{--} W_{++}) = {-\frac{1}{8} }\int du \ (W_{++})^2\nt
& = {-\frac{1}{8}} (W_{++})^2\,.\notag
\end{align}
At the last step we used the fact that $W_{++}$ is independent of $u^\pm$, so $\int du\ 1=1$ (see~\p{6}). We have obtained the alternative form \p{TW}.

\section{Realization of the half-BPS condition on $\cT$}\label{BPS}

The stress-tensor multiplet \p{4.33} is half-BPS. This is translated into the shortening conditions \p{5.3}. The shortening becomes manifest in the R-analytic basis \p{AnB} where the spinor derivatives \p{DaB}, which define the R-analytic superfield, do not have terms with space-time derivatives (see Section \ref{s9.1}). The situation is the opposite with the supersymmetry generators in the same basis:
\begin{align}
&Q_{-\a a} = \pa/\pa \q^{\a a}_{+}\,, \qquad \bar Q^{a'}_{-\da} =\pa/\pa \bq^{\da}_{+a'}\nt
& Q_{+\a a'} = \pa/\pa \q^{\a a'}_{-} - \bq^{\da}_{+a'} \pa_{\a \da}\,, \qquad \bar Q^{a}_{+\da} = \pa/\pa \bq^{\da}_{-a} - \q^{\a a}_{+} \pa_{\a \da} \,. \notag
\end{align}
The generator $\bar Q^{a'}_{-}$ is what we used in Section \ref{s5} to relate the various semi-superfield components in \p{4.33} to each other. The generator $\bar Q^a_{+}$ should, according to the standard formulation, ``annihilate the half-BPS primary". In reality, due to the space-time derivative term in it, we expect a relation of the type
\begin{align}\label{c2}
\bar Q^a_{+\da} \cT = -\q^{\a a}_{+} \pa_{\a\da} \cT
\end{align}
for the bottom component in the $\bq_+$ expansion \p{4.33}. 
Let us show that this is the case.

We project relation \p{5.12} with the RH $w^a_{+A}$ and obtain 
\begin{align}\label{c3}
\bQ^a_{+\da} \cT  = - (\pa_+)^4 \tr \int  du \;  \q^{+a}_{+} (A^+_\da + \pa^-_\da A^{++}) A^{--} = {-\frac{1}{4}} \tr \int  du \;  \q^{+a}_{+} [(\nabla^-_+)^2 \nabla^{++} A^-_\da]\,  W_{++}\,.
\end{align}
Here we have split $(\pa_+)^4 = \frac14 (\nabla^-_+)^2 (\pa^+_+)^2$ under the trace, used the L-analyticity $\pa^+ A^+_\da=\pa^+ A^{++}=0$ and the definition \p{TW}. The super-curvature $W_{++}$ has the property \\$\nabla^-_+ W_{++}=0$, which follows from $\pa^+_+ W_{++}=0$ by differentiation with $\nabla^{--}$ and from $[\nabla^{--}, \pa^+_+] = \nabla^-_+$ and $\nabla^{--} W_{++}=0$ (see Appendix~\ref{proofTW}). We have also used the commutation relation $[\nabla^{++}, \nabla^-_\da]= \nabla^+_\da$ (the gauge covariantization of the obvious relation $[\pa^{++}, \pa^-_\da]= \pa^+_\da$), which  implies that $\nabla^{++} A^-_\da = A^+_\da + \pa^-_\da A^{++}$.

Next, we pull the harmonic derivative $\nabla^{++}$ out of the first factor in the second relation in \p{c3}, integrate it by parts and use $\nabla^{++} W_{++}=0$. On the way  $\nabla^{++}$ changes one of the spinor derivatives, $[\nabla^{++}, \nabla^-_+]= \pa^+_+$ and we obtain
$[\nabla^{++} ,(\nabla^-_+)^2] = (\pa^+_+ \nabla^-_+) + (\nabla^-_+ \pa^+_+ ) = 2 (\nabla^-_+ \pa^+_+ ) + \pa^+_+ A^{-}_{+}$. In \cite{PartI} we
have proved that $A^-_{A} = - \pa^{+}_{A} A^{--}$. Consequently, in view of \p{TW} we have $\pa^+_+ A^{-}_{+} =- W_{++}$. This
does not contribute due to the trace $\tr\ ( [W_{++},A^-_{\da}] W_{++} )= 0$, thus we have
\begin{align}\label{c4}
\bQ_{+a\da} \cT  = {\frac{1}{2}}\tr \int  du \;  \q^{+}_{+a} [(\nabla^-_+ \pa^+_+)  A^-_\da]\,  W_{++}\ .
\end{align}
To proceed, we need to show that
\begin{align}\label{c5}
(\nabla^-_+ \pa^+_+)  A^-_\da = - \nabla^-_\da W_{++} \,.
\end{align} 
This relation has overall LH charge $-1$, therefore, if we can show that it holds after differentiation with $\nabla^{++}$, it will also hold as it is (the operator $\pa^{++}$ is invertible on harmonic functions with negative charge). Thus, we act on \p{c5} with $\nabla^{++}$, use again $\nabla^{++} A^-_\da = A^+_\da + \pa^-_\da A^{++}$ and the L-analyticity of $A^+_\da, A^{++}$ to obtain
\begin{align}\label{c6}
(\nabla^-_+ \pa^+_+)  A^-_\da = - \nabla^-_\da W_{++}  \quad \stackrel{\nabla^{++}}{\longrightarrow} \quad (\pa^+_+)^2 A^-_\da =  - \nabla^+_\da W_{++} \,.
\end{align}
 The commutation relation $[\nabla^{--}, \nabla^+_\da]= \nabla^-_\da$ implies $A^-_\da = \pa^{--} A^+_\da - \nabla^+_\da A^{--}$. Hitting this with $(\pa^+_+)^2 $, the term with $\pa^{--} A^+_\da$ vanishes and we get the right-hand side of \p{c6}.
 
 Finally, with the help of \p{c5} the variation \p{c4} becomes
 \begin{align}\notag
\bQ^a_{+\da} \cT  = {- \frac{1}{4}}\int  du \;  \q^{+a}_{+} \pa^-_\da \tr(W_{++})^2 = 2 \int  du \;  \q^{+a}_{+} \pa^-_\da \cT = - \q^{\a a}_{+} \pa_{\a\da} \cT
\end{align}
because $\cT = {-\frac{1}{8}}\tr(W_{++})^2$ (see \p{TW}) is independent of the LHs and the integral picks the Lorentz singlet in $\q^{+a}_{+} \pa^-_\da$. So, we have confirmed the property \p{c2}.


\section{One-loop four-point chiral correlator}\label{loop}

In this section we calculate the one-loop correction to the four-point chiral MHV-like correlator   
$\langle \mathcal{T}(1) \mathcal{T}(2) \mathcal{T}(3) \mathcal{T}(4) \rangle$
by means of the Feynman rules in LHC superspace. 
This  result has been obtained long ago via Feynman graphs in $\cN=2$ harmonic superspace \cite{Eden:2000mv} and then generalized to $\cN=4$ harmonic (or R-analytic) superspace in \cite{Heslop:2002hp}. With four points there are no nilpotent invariants, so the purely chiral four-point correlator is restricted to its $\q=0$ component. It is given by a universal polynomial in the variables $x,y$, which captures the $SU(4)$  structure, multiplied by a function $\Phi(u,v)$ of the conformal cross-ratios containing the loop corrections. At one loop this function is given by the one-loop `box' (or `cross') integral:\footnote{The normalization factor is chosen to agree with the LHC Feynman graph calculation presented below. In fact, by comparing  \p{Intr} to the standard result of Ref.~\cite{Heslop:2002hp} we can fix the conventional value of $\omega$. }
\begin{align}
&\langle \mathcal{T}(1) \mathcal{T}(2) \mathcal{T}(3) \mathcal{T}(4) \rangle_{g^2} =-\frac{\omega g^2}{\pi^{12}} \Bigl[
\frac{y_{12}^2 y_{23}^2 y_{34}^2 y_{14}^2}{x_{12}^2 x_{23}^2 x_{34}^2 x_{14}^2}(x_{13}^2 x_{24}^2 - x_{12}^2 x_{34}^2 - x_{14}^2 x_{23}^2) \nt
& +
\frac{y_{12}^2 y_{13}^2 y_{24}^2 y_{34}^2}{x_{12}^2 x_{13}^2 x_{24}^2 x_{34}^2}(x_{14}^2 x_{23}^2 - x_{12}^2 x_{34}^2 - x_{13}^2 x_{24}^2) + \frac{y_{13}^2 y_{14}^2 y_{23}^2 y_{24}^2}{x_{13}^2 x_{14}^2 x_{23}^2 x_{24}^2}(x_{12}^2 x_{34}^2 - x_{14}^2 x_{23}^2 - x_{13}^2 x_{24}^2)\nt
& +
\frac{y_{12}^4 y_{34}^4}{x_{12}^2 x_{34}^2} + 
\frac{y_{13}^4 y_{24}^4}{x_{13}^2 x_{24}^2} + 
\frac{y_{14}^4 y_{23}^4}{x_{14}^2 x_{23}^2} \Biggr] \int\frac{d^4 x_5}{x_{15}^2 x_{25}^2 x_{35}^2 x_{45}^2}\,. \label{Intr}
\end{align}
This supersymmetric calculation will allow us to illustrate 
the LHC Feynman rules from Section \ref{s2} at work. Unlike the Born-level calculations presented in Section \ref{s7}, this one requires an interaction vertex and consequently integration over space-time.\footnote{In reality, this calculation is equivalent to taking the Born-level result for the NMHV-like correlator $\langle \mathcal{T}(1) \ldots \mathcal{T}(5) \rangle \sim g^2 (\q)^4$ and integrating it over the fifth space-time point. \label{f18}} 

There are several  graphs that can possibly contribute. They are indicated below up to permutations of the points $1$, $2$, $3$, $4$. The black blobs represent the external points, and the gray blobs are the interaction vertices $S_{\rm Z}$ \p{N4}. A space-time integral is assigned to each gray blob. In the following we will omit the factors $g^2$ and $\omega$ produced by each graph. 

\begin{figure}[!h]
\centerline{
\begin{tabular}{cccc}
\includegraphics[width = 3 cm]{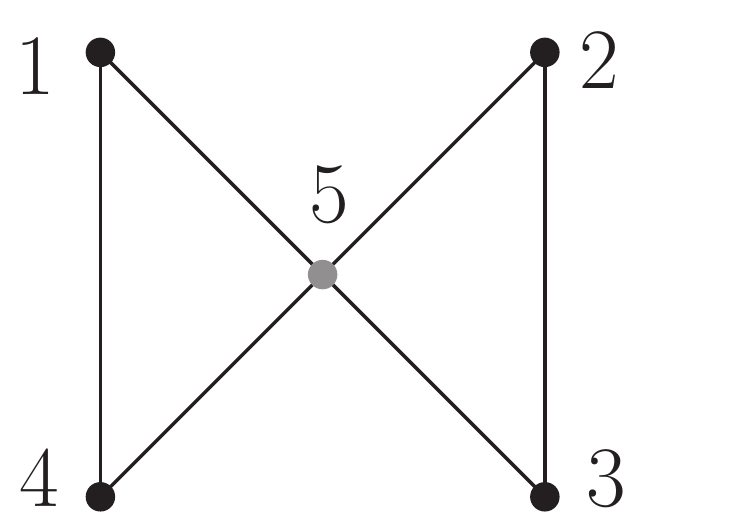} &
\includegraphics[width = 3 cm]{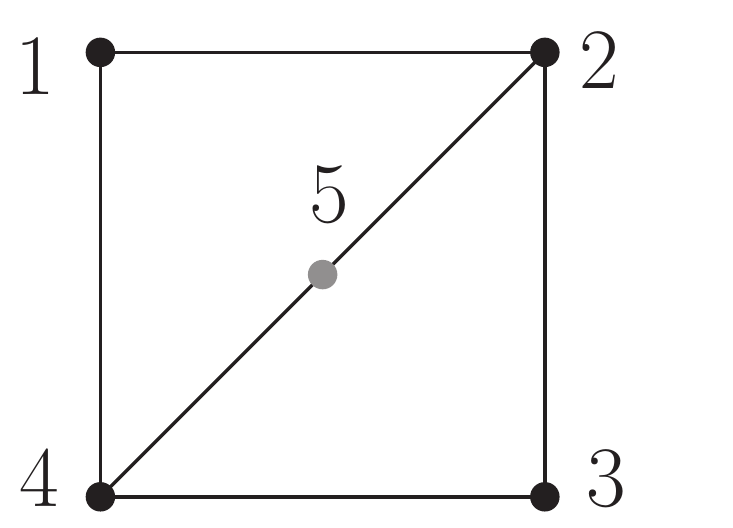} &
\includegraphics[width = 3 cm]{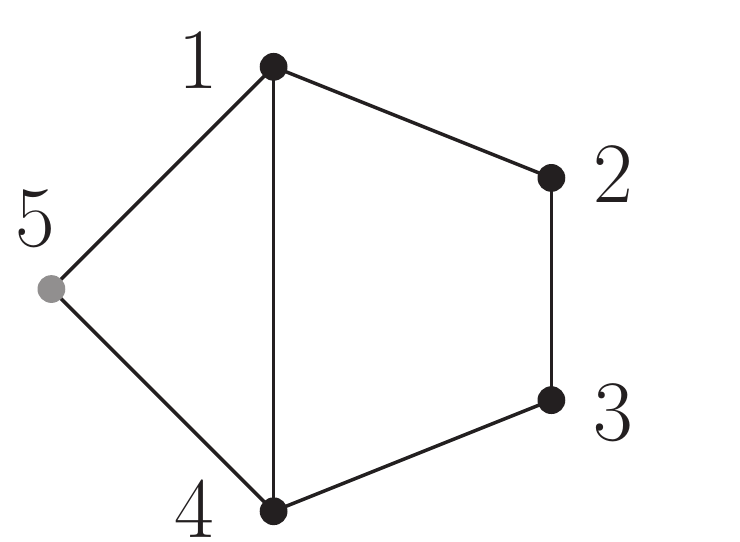} &
\includegraphics[width = 3 cm]{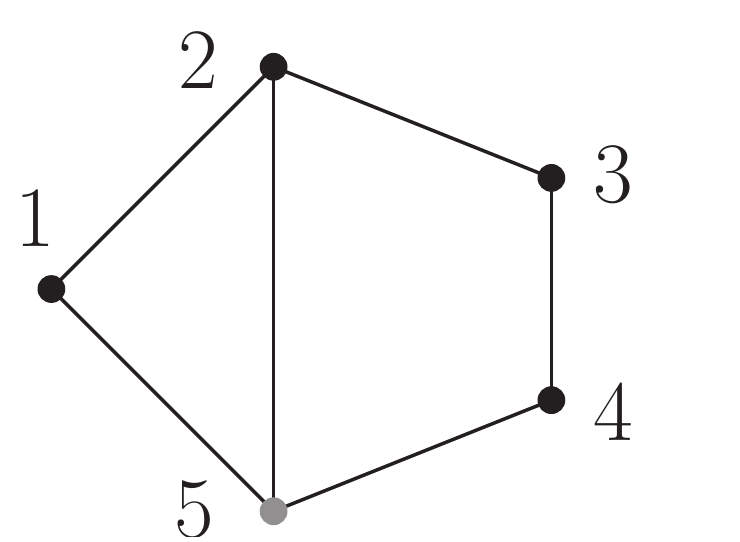} \\
Fig.~\ref{loop}.1 & Fig.~\ref{loop}.2 & Fig.~\ref{loop}.3 & Fig.~\ref{loop}.4 
\end{tabular}}

\vspace{0.3 cm}

\centerline{
\begin{tabular}{ccc}
\includegraphics[width = 3 cm]{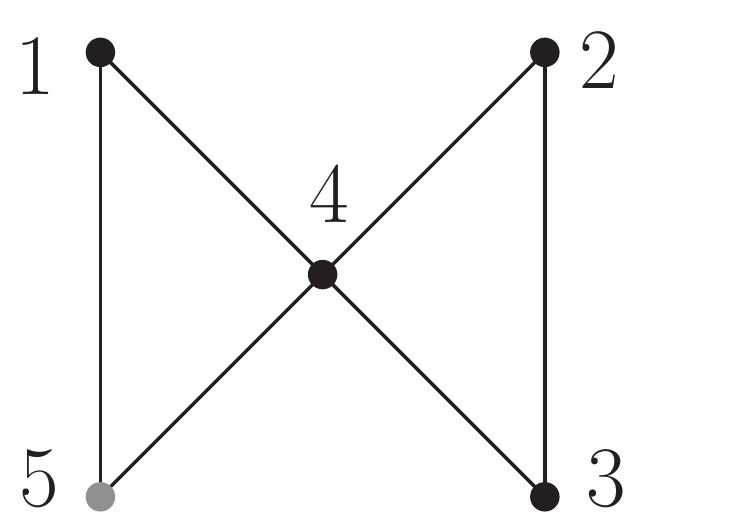} &
\includegraphics[width = 3 cm]{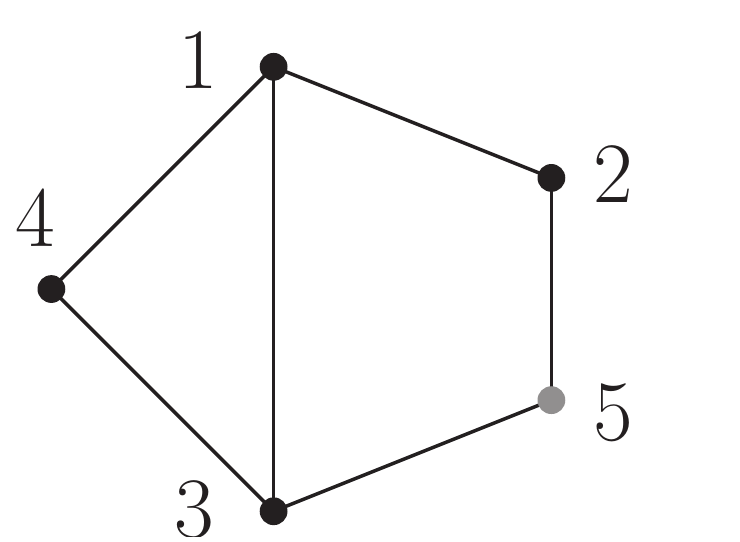} &
\includegraphics[width = 3 cm]{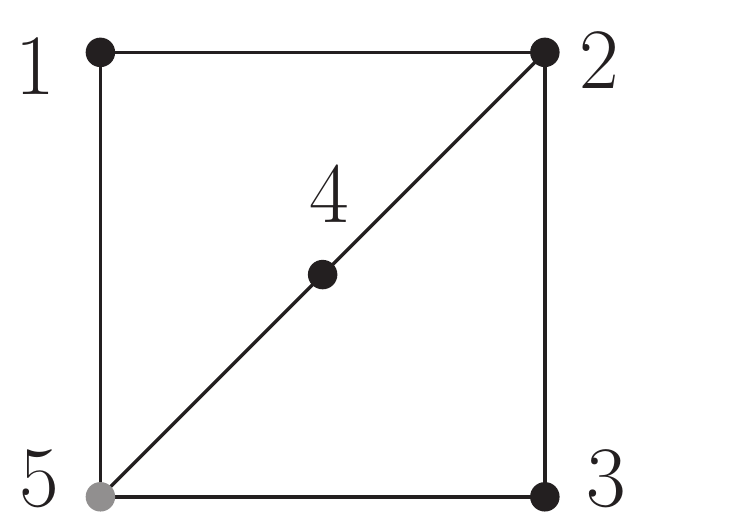} \\
Fig.~\ref{loop}.5 & Fig.~\ref{loop}.6 & Fig.~\ref{loop}.7 
\end{tabular}}
\end{figure}
The Grassmann integrations are done with the help of the identities (recall~\p{5.36}, \p{B}) 
\begin{align}
\delta^4(\theta_i - \theta_j) &= y_{ij}^2 \,\delta^2\left(\theta_{j-} - \theta_{ij+}y_{ij}^{-1}\right) 
\delta^2\left(\theta_{i-} - \theta_{ij+}y_{ij}^{-1} \right) ,\nt
\delta^4(\theta_i - \chi) &= \delta^2\left(\theta_{i+} - (\chi i_{+})\right) 
\delta^2\left(\theta_{i-} - \theta_{i+}y^{-1}_{ij} + (\chi j_{+})y^{-1}_{ij}\right), \label{4to2x2b}
\end{align}
where $\theta_{ij+} \equiv \theta_{i+} - \theta_{j+}$ and $(\chi i_{+}) \equiv \chi^A (w_i)_{+A}$  with $i_{+},j_{+}$ being shorthands for the RHs. Here we do not display the Lorentz spinor indices.
The second RH $j_{+}$ on the right-hand side of  \p{4to2x2b} is at our disposal  
since the left-hand side depends only on $i_{+}$.  We profit  from the relations $(x^{\+}_{ik} x^{\+}_{jk}) = (x^{\+}_{ij}x^{\+}_{ik})=(x^{\+}_{ij}x^{\+}_{jk})$ as well. 

After the Grassmann integration it is easy to see that
the diagrams in Figs.~\ref{loop}.5,6,7 do not contribute. The principal contribution comes from the graph in Fig.~\ref{loop}.1:
\begin{align}
&\mathcal{I}_1(1,2,3,4)  = \frac{1}{\pi^6} y_{14}^2 y_{23}^2 \int d^4 x_5 \int d u\,
\frac{ \delta(x^{\+ +}_{15})\delta(x^{\+ +}_{25})\delta(x^{\+ +}_{35})\delta(x^{\+ +}_{45}) }
{(u^+_{15}u^+_{25})(u^+_{25}u^+_{35})(u^+_{35}u^+_{45})(u^+_{45}u^+_{15})}\delta(x^{\+ +}_{14}) \delta(x^{\+ +}_{23})\notag\\ 
&\times \bigl[ y_{12}^2 y_{34}^2 (u_{25}^+ u_{45}^+) (u_{15}^+ u_{35}^+) (u_{25}^+ u_{35}^+) (u_{15}^+ u_{45}^+) 
+ y_{14}^2 y_{23}^2 (u_{25}^+ u_{45}^+) (u_{15}^+ u_{35}^+) (u_{15}^+ u_{25}^+) (u_{35}^+ u_{45}^+)\notag \\
& \qquad - y_{13}^2 y_{24}^2 (u_{25}^+ u_{35}^+) (u_{15}^+ u_{45}^+) (u_{15}^+ u_{25}^+) (u_{35}^+ u_{45}^+) \bigr] .\label{I1b}
\end{align}
The expression \p{I1b} and its permutations give part of the expected answer \p{Intr}.
Indeed, $\mathcal{I}_1(1,2,3,4) + \mathcal{I}_1(1,3,2,4)$ contributes in particular 
to the $y$-structure $y_{14}^4 y_{23}^4$ (recall~\p{intd}),
\begin{align}\notag
-\frac{1}{\pi^{12}}\frac{y_{14}^4 y_{23}^4}{x_{14}^2 x_{23}^2} \int  \frac{d^4 x_5}{x_{15}^2 x_{25}^2 x_{35}^2 x_{45}^2}\,.
\end{align}
Together with the other permutations of $\mathcal{I}_1$ we reproduce
the last line in \p{Intr}. 
The remaining graphs contribute only to the $y$-structure $y_{12}^2 y_{23}^2 y_{34}^2 y_{14}^2$ and its permutations. 

In order to complete the calculation we only need to identify the $y$-structure 
$y_{12}^2 y_{23}^2 y_{34}^2 y_{14}^2$ within the diagrams.
We find that this structure originates from 
$\mathcal{I}_1(1,2,3,4)$ and its permutations $\mathcal{I}_1(1,3,2,4)$, 
$\mathcal{I}_1(1,4,3,2)$, $\mathcal{I}_1(1,3,4,2)$, contributing the factor 
\begin{align} \label{C6}
\frac{x_{12}^2 x_{34}^2 + x_{14}^2 x_{23}^2}{x_{12}^2 x_{23}^2 x_{34}^2 x_{14}^2} 
- \frac{(x^{\+}_{15} x^{\+}_{35})(x^{\+}_{25} x^{\+}_{45})}{x_{12}^2 x_{23}^2 x_{34}^2 x_{14}^2} 
\left[ \frac{ x_{12}^2 x_{34}^2}{(x^{\+}_{15} x^{\+}_{25})( x^{\+}_{35} x^{\+}_{45})} 
+ \frac{ x_{14}^2 x_{23}^2 }{(x^{\+}_{15} x^{\+}_{45})(x^{\+}_{25}x^{\+}_{35})}\right]
\end{align}
(here $x^{\+  \a}_{ij} \equiv \xi^{\+}_{\da}x^{\da \a}_{ij}$). Once again, the first term in \p{C6} can be identified in \p{Intr}.
Thus we only need to show that the $x^{\+}$-dependent part combines
with the remaining topologies into (here we omit the cross-integral factor)
\begin{align} \label{rest}
-\frac{x_{13}^2 x_{24}^2}{x_{12}^2 x_{23}^2 x_{34}^2 x_{14}^2}\,. 
\end{align}
The diagrams in Figs.~\ref{loop}.2,3,4 contribute
\begin{align}
&\mathcal{I}_2(1,2,3,4) = \frac{1}{\pi^{12}}\frac{y_{12}^2 y_{23}^2 y_{34}^2 y_{14}^2}{x_{12}^2 x_{23}^2 x_{34}^2 x_{14}^2} \int  
\frac{d^4 x_5}{x_{25}^2 x_{45}^2} \frac{(x^{\+}_{25}x^{\+}_{45})^2 (x^{\+}_{12}x^{\+}_{23}) (x^{\+}_{14}x^{\+}_{34})}
{(x^{\+}_{14}x^{\+}_{45})(x^{\+}_{34}x^{\+}_{45})(x^{\+}_{23}x^{\+}_{25})(x^{\+}_{12}x^{\+}_{25})} \,,\nt
& \mathcal{I}_3(1,2,3,4) = -\frac{1}{\pi^{12}}\frac{y_{12}^2 y_{23}^2 y_{34}^2 y_{14}^2}{x_{12}^2 x_{23}^2 x_{34}^2 x_{14}^2} \int 
\frac{d^4 x_5 }{x_{15}^2 x_{45}^2} \frac{(x^{\+}_{14}x^{\+}_{34}) (x^{\+}_{12}x^{\+}_{14})}
{(x^{\+}_{12}x^{\+}_{15})(x^{\+}_{34}x^{\+}_{45})}\,,\nt
& \mathcal{I}_4(1,2,3,4) = -\frac{1}{\pi^{12}}\frac{y_{12}^2 y_{23}^2 y_{34}^2 y_{14}^2}{x_{12}^2 x_{23}^2 x_{34}^2 x_{14}^2} \int 
\frac{d^4 x_5 }{x_{15}^2 x_{25}^2 x_{45}^2} \frac{(x^{\+}_{25}x^{\+}_{45}) (x^{\+}_{12}x^{\+}_{23})}
{(x^{\+}_{23}x^{\+}_{25})(x^{\+}_{15}x^{\+}_{45})}\,. \notag
\end{align}
The permutations: (i) $\mathcal{I}_2(2,3,4,1)$; (ii) $\mathcal{I}_3(2,3,4,1)$, $\mathcal{I}_3(3,4,1,2)$, $\mathcal{I}_3(4,1,2,3)$; 
\\ (iii) $\mathcal{I}_4(2,3,4,1)$, $\mathcal{I}_4(3,4,1,2)$, $\mathcal{I}_4(4,1,2,3)$, $\mathcal{I}_4(1,4,3,2)$,
$\mathcal{I}_4(4,3,2,1)$, $\mathcal{I}_4(3,2,1,4)$, \\$\mathcal{I}_4(2,1,4,3)$, contain the same $y$-structure.
Summing up $\mathcal{I}_1$, $\mathcal{I}_2$, $\mathcal{I}_3$, $\mathcal{I}_4$ along with their permutations 
we find that the contribution to the $y$-structure $y_{12}^2 y_{23}^2 y_{34}^2 y_{14}^2$ agrees with \p{Intr}. 
The gauge-fixing parameter $\xi^{\+}$ drops out due to the following non-trivial identity (recall eq.~\p{rest})\footnote{A similar identity has been found by P. Heslop and R. Doobary in the twistor approach.}
\begin{align}
&x_{13}^2 x_{24}^2 = x_{12}^2 x_{34}^2 \frac{(x^{\+}_{15} x^{\+}_{35})(x^{\+}_{25} x^{\+}_{45})}{(x^{\+}_{15} x^{\+}_{25})(x^{\+}_{35} x^{\+}_{45})}
+ x_{14}^2 x_{23}^2 \frac{(x^{\+}_{15} x^{\+}_{35})(x^{\+}_{25} x^{\+}_{45})}{(x^{\+}_{15} x^{\+}_{45})(x^{\+}_{25} x^{\+}_{35})} \notag \\
&- x_{15}^2 x_{35}^2 
\frac{(x^{\+}_{25} x^{\+}_{45})^2(x^{\+}_{12} x^{\+}_{23})(x^{\+}_{14} x^{\+}_{34})}
{(x^{\+}_{14} x^{\+}_{45})(x^{\+}_{34} x^{\+}_{45})(x^{\+}_{23} x^{\+}_{25})(x^{\+}_{12} x^{\+}_{25})} 
- x_{25}^2 x_{45}^2 
\frac{(x^{\+}_{15} x^{\+}_{35})^2(x^{\+}_{12} x^{\+}_{14})(x^{\+}_{23} x^{\+}_{34})}
{(x^{\+}_{34} x^{\+}_{35})(x^{\+}_{23} x^{\+}_{35})(x^{\+}_{12} x^{\+}_{15})(x^{\+}_{14} x^{\+}_{15})} \notag \\
&+ x_{25}^2 x_{35}^2 \frac{(x^{\+}_{14} x^{\+}_{34})(x^{\+}_{12} x^{\+}_{14})}{(x^{\+}_{12} x^{\+}_{15})(x^{\+}_{34} x^{\+}_{45})}
- x_{35}^2 x_{45}^2 \frac{(x^{\+}_{12} x^{\+}_{14})(x^{\+}_{12} x^{\+}_{23})}{(x^{\+}_{23} x^{\+}_{25})(x^{\+}_{14} x^{\+}_{15})} 
+ x_{15}^2 x_{45}^2 \frac{(x^{\+}_{12} x^{\+}_{23})(x^{\+}_{23} x^{\+}_{34})}{(x^{\+}_{34} x^{\+}_{35})(x^{\+}_{12} x^{\+}_{25})} \notag \\
&- x_{15}^2 x_{25}^2 \frac{(x^{\+}_{23} x^{\+}_{34})(x^{\+}_{14} x^{\+}_{34})}{(x^{\+}_{14} x^{\+}_{45})(x^{\+}_{23} x^{\+}_{35})} 
+ x_{14}^2 x_{35}^2 \frac{(x^{\+}_{25} x^{\+}_{45})(x^{\+}_{12} x^{\+}_{23})}{(x^{\+}_{23} x^{\+}_{25})(x^{\+}_{15} x^{\+}_{45})}
+ x_{12}^2 x_{45}^2 \frac{(x^{\+}_{15} x^{\+}_{35})(x^{\+}_{23} x^{\+}_{34})}{(x^{\+}_{34} x^{\+}_{35})(x^{\+}_{15} x^{\+}_{25})}  \notag \\
&- x_{15}^2 x_{23}^2 \frac{(x^{\+}_{25} x^{\+}_{45})(x^{\+}_{14} x^{\+}_{34})}{(x^{\+}_{14} x^{\+}_{45})(x^{\+}_{25} x^{\+}_{35})}
- x_{25}^2 x_{34}^2 \frac{(x^{\+}_{15} x^{\+}_{35})(x^{\+}_{12} x^{\+}_{14})}{(x^{\+}_{12} x^{\+}_{15})(x^{\+}_{35} x^{\+}_{45})}  
- x_{12}^2 x_{35}^2 \frac{(x^{\+}_{25} x^{\+}_{45})(x^{\+}_{14} x^{\+}_{34})}{(x^{\+}_{34} x^{\+}_{45})(x^{\+}_{15} x^{\+}_{25})} \notag \\
&+ x_{14}^2 x_{25}^2 \frac{(x^{\+}_{15} x^{\+}_{35})(x^{\+}_{23} x^{\+}_{34})}{(x^{\+}_{23} x^{\+}_{35})(x^{\+}_{15} x^{\+}_{45})} 
+ x_{15}^2 x_{34}^2 \frac{(x^{\+}_{25} x^{\+}_{45})(x^{\+}_{12} x^{\+}_{23})}{(x^{\+}_{12} x^{\+}_{25})(x^{\+}_{35} x^{\+}_{45})}
- x_{23}^2 x_{45}^2 \frac{(x^{\+}_{15} x^{\+}_{35})(x^{\+}_{12} x^{\+}_{14})}{(x^{\+}_{14} x^{\+}_{15})(x^{\+}_{25} x^{\+}_{35})}\,.\notag
\end{align}


\section{Independence of the gauge-fixing parameter}\label{nolambda}


Here we adapt the argument given in \cite{Chicherin:2014uca} to our formalism and, in particular, to 
the non-chiral modification of the effective Feynman rules by the substitution $x_{ij} \to \hat{x}_{ij}$. We show  that the gauge-fixing spinor $\xi^{\+}$ drops out from  the sum of all Feynman graphs for the NMHV-like multipoint correlator. 

Let us start with the chiral case (all $\bq=0$). 
The argument of \cite{Chicherin:2014uca} amounts to proving the absence of spurious poles of the type
\begin{align}\label{d1}
(123) \equiv (x^{\+}_{12} x^{\+}_{13}) = [\xi^{\+}|\tilde  x_{12}x_{13} | \xi^{\+}] =0\,.
\end{align} 
Such poles occur in the individual R-vertices \p{Rinv} when  $\xi^{\+}$ becomes an eigenspinor of the matrix, $\tilde x_{12}x_{13} |\xi^{\+}] = \rho |\xi^{\+}]$. Since the R-vertices are homogeneous of degree zero under the rescaling of $\xi^{\+}$, the absence of poles in $\xi^{\+}$ implies that the sum of all Feynman graphs is independent of it.

The identity $x_{12}+x_{23}=x_{13}$ yields three equivalent conditions 
\begin{align} \label{e1'}
0=(123) = (x^{\+}_{12} x^{\+}_{13}) = (x^{\+}_{23} x^{\+}_{21}) = (x^{\+}_{31} x^{\+}_{32})\,.
\end{align}
Alternatively, condition \p{d1} can be reformulated as
\begin{align}\label{d2}
x^{\+}_{13} = \rho_1 x^{\+}_{12}
\end{align}
with an arbitrary complex factor $\rho_1$. Due to \p{e1'} 
this condition is equivalent to two others obtained by cyclic permutations of the indices,
\begin{align}\label{d2'}
&(x^{\+}_{23} x^{\+}_{21}) = 0 \quad \rightarrow \quad x^{\+}_{21} = \rho_2 x^{\+}_{23}\,, \quad \rho_2 = (1-\rho_1)^{-1}\,, \nt
&(x^{\+}_{31} x^{\+}_{32}) = 0 \quad \rightarrow \quad x^{\+}_{32} = \rho_3 x^{\+}_{31}\,, \quad \rho_3 = 1-\rho_1^{-1}\,.
\end{align}

The pole at $(123)=0$ appears in the diagrams containing $R(1;23i)$, $R(2;31j)$, $R(3;12k)$  with $i,j,k\neq 1,2,3$. Let us first evaluate the residue of the R-vertex $R(1;23i)$ in the singular regime \p{e1'}.  The calculation is greatly simplified by fixing a $Q-$supersymmetry gauge. Using the 8 parameters of the $Q-$shift $\delta \q^A_\a = \ep^A_\a$ we can choose the 
\begin{align}\label{e6}
\mbox{$Q$ SUSY gauge:} \qquad \q_{1+}=\q_{2+}=0\,.
\end{align}
Attaching the 
propagator factor $y_{12}^2y^2_{13}/(x^2_{12}x^2_{13})$ and using \p{Rinv}, \p{B}, \p{d2}, \p{d2'} and the gauge \p{e6}, we find
\begin{align}\label{d5}
{\rm Res}_{(123)=0}\, \frac{y^2_{12} y^2_{13}}{x^2_{12} x^2_{13}} R(1;23i) = 
  \frac{y^2_{12}\rho_1}{x_{12}^2 x_{13}^2}   \left( x^{\+}_{12} \q_{3+} \right)^2 \,.
\end{align}
Similarly,  for the two other residues we obtain
\begin{align}
&{\rm Res}_{(123)=0}\, \frac{y^2_{12} y^2_{23}}{x^2_{12} x^2_{23}} R(2;31j) 
= \frac{y^2_{12}\rho_2}{x^2_{12} x^2_{23}}   \left( x^{\+}_{23} \q_{3+} \right)^2  \nt
&{\rm Res}_{(123)=0}\, \frac{y^2_{13} y^2_{23}}{x^2_{13} x^2_{23}} R(3;12k) 
= \frac{y^2_{13} y^2_{23}}{x^2_{13} x^2_{23}} \rho_3 \left( x^{\+}_{31} \q_{3+} y_{31}^{-1} - x^{\+}_{31} \q_{3+} y_{32}^{-1}\right)^2
= \frac{ y^2_{12}\rho_3}{x^2_{13} x^2_{23}} \left( x^{\+}_{31} \q_{3+} \right)^2 . \notag
\end{align} 
Notice the independence of the residues from the fourth points $i,j,k$. 
Next, we remark that 
for the singular configuration specified in  \p{d2}, \p{d2'} the variables  $x_{12}^2$, $x_{13}^2$, $x_{23}^2$ 
become linearly dependent. This can be seen by writing $x^2 = x^{\+}\cdot x^{\-}$ and using the relations \p{d2} and \p{d2'}:
\begin{align} \label{xsqld}
\rho_1 x_{23}^2 + \rho_2^{-1} x_{13}^2 + \rho_2^{-2} \rho_3^{-1} x_{12}^2 = 0\,.
\end{align}
Summing up the residues and using \p{d2'}, \p{xsqld} we finally obtain
\begin{align}\label{e9}
{\rm Res}_{(123)=0} \Big[ \frac{y^2_{12} y^2_{13}}{x^2_{12} x^2_{13}} R(1;23i) 
+ \frac{y^2_{12} y^2_{23}}{x^2_{12} x^2_{23}} R(2;31j) + \frac{y^2_{13} y^2_{23}}{x^2_{13} x^2_{23}} R(3;12k)\Big]=0\,.
\end{align}
This result is equivalent to eq.~(4.35)  in \cite{Chicherin:2014uca}. The rest of the argument follows \cite{Chicherin:2014uca}.

Now, what happens in the non-chiral case? The effective Feynman rules amount to putting hats on the variables $\hat x^{\+}_{ij} = x^{\+}_{ij} + \q_{ij+} y^{-1}_{ij} \bq^{\+ }_{ij+}$. At first sight, this may invalidate the argument above, in particular the  identity $x_{12}+x_{23}=x_{13}$. However, this does not happen. Due to the full $Q$ and $\bQ$ supersymmetry,  in addition to the gauge \p{e6} we can fix the
\begin{align}\label{e6'}
\mbox{$\bQ$ SUSY gauge:} \qquad \bq_{1+}=\bq_{2+}=0\,.
\end{align}
The gauges \p{e6} and \p{e6'}  remove the hat from $\hat x^{\+}_{12}$ but not from, e.g., $\hat x^{\+}_{13} = x^{\+}_{13} + \q_{3+} y^{-1}_{13} \bq^{\+ }_{3+}$. Next, using the invariance of $\hat x_{ij}$ under R-symmetry  we can fix the
\begin{align}\label{e11}
\mbox{$SU(4)$ gauge:} \qquad y_3 = \infty \times \mathbb{I}\,,
\end{align}
so that $\hat x^{\+}_{13}$ and $\hat x^{\+}_{23}$ also loose their hats.\footnote{The R-symmetry  $SU(4)$ acts on the coordinates of the R-analytic basis of superspace as follows. $SU(4)$ inversion acts upon $\q_+,\bq_+,y$ in the familiar way, $I[\q_+] = \q_+ y^{-1}$, $I[\bq_+] = y^{-1} \bq_+$, $I[y]=y^{-1}$.  What is unusual is the transformation of $x_{\rm an}$ defined in \p{AnB}. Due to the presence of RHs in this definition, $x_{\rm an}$ is not inert under $SU(4)$ inversion, $I[x_{\rm an}] = x_{\rm an} + \q_+ y^{-1} \bq_+$. With this property one can show  that the hatted variable $\hat x_{ij} = x_{ij} + \q_{ij+} y^{-1}_{ij} \bq_{ij+}$ is invariant under $SU(4)$. } The hats survive in $\hat x_{1i}$ and $\hat x_{2i}$ (with $i\neq 1,2,3$) but this does not affect the argument. Then we can repeat all the steps form the chiral case. We have to be careful when taking the limit $y_3 \to \infty$ since $y^{-1}_3$ appears in the R-vertices  and $y_3$ in the propagator factors, giving non-vanishing limits. We obtain the same residues as above and the same final result \p{e9}.\footnote{We have checked with \emph{ Mathematica} that the residues cancel  in the sum without the  gauge \p{e11}. Instead, we chose  $x_1=y_1=0$, $x_2=y_2=\mathbb{I}$, $x_3,y_3$ diagonal and $\xi^{\+} = (\la, 1)$ and looked for the poles in $\la$. }

\section{Component calculation}\label{apE}

In order to support our general argument about the form of the non-chiral correlator, we carried out a 
calculation in the conventional Feynman graph approach and compared it with the LHC superspace results.
The full non-chiral correlator contains plenty of components.
We will stay at the level of the first $\bar{Q}$-variation of the chiral NMHV-like correlator
and we will consider only one of its components. 
Specifically we deal with the component $\theta^{\alpha a}_{1+}\bar{\theta}^{\dot\alpha a'}_{2+} (\theta_{3+})^4$
of the five- and six-point supercorrelators $\langle \cT(1)\cM(2)\cT(3) \ldots \cT(n) \rangle$  
(denoted $G^{\rm nMHV}_{1,1}$ in \p{7.1}), i.e.
we calculate the correlators ($n = 5, 6$)
\begin{align} \label{psibarpsiLOOO}
G_n = \left\langle \,\mathrm{tr} \bigl( \psi^{\alpha}_{+a} \phi_{++}\bigr)(1) \,
\mathrm{tr}\bigl( \bar{\psi}^{\dot\alpha}_{-a'} \phi_{++}\bigr)(2) \, L(3) \,
\cO(4) \cdots \cO(n) \,\right\rangle_{g^2}\,
\end{align}
 in the Born approximation (at order $g^2$). Here (recall~\p{5.2}, \p{4.48})
\begin{align}
&\cO = \mathrm{tr} \bigl( \phi_{++} \phi_{++}\bigr) \notag \\
\label{LSD}
&L = \mathrm{tr}  \left\{ -\frac12  F_{\alpha\beta}F^{\alpha\beta} 
+ {\sqrt{2}}  g \psi^{\alpha A} [\phi_{AB},\psi_\alpha^B] - \frac18 g^2 [\phi^{AB},\phi^{CD}][\phi_{AB},\phi_{CD}] \right\}\,.
\end{align}
$L$ is the on-shell chiral Lagrangian of $\mathcal{N} = 4$ SYM.
Let us emphasize that the interaction vertices are induced by the full Lagrangian of $\mathcal{N} = 4$ SYM, not by $L$. In \cite{Chicherin:2014uca} we performed a similar calculation for all the components of the chiral NMHV correlator, so here we will skip most of the details.

In principle, the relevant Feynman graphs produce a number of complicated Feynman integrals 
resulting in non-rational contributions (they are logarithms and dilogarithms since we are effectively 
at one-loop level). The sum of all graphs is supposed to be a rational function, so the non-rational pieces should sum up to zero.
Fortunately, we found the correlators $G_n$ without such laborious integrations.
The only integrals needed are the so-called $T$-blocks which are rational. 
They are the building blocks of some of the diagrams. In the chiral calculation \cite{Chicherin:2014uca} we
used $T$-blocks representing the interaction of a gluon with a pair of scalars and the Yukawa interaction of 
a scalar with a pair chiral fermions:
\begin{align} \label{T1}
\!\!\!\begin{array}{c}\includegraphics[width = 3 cm]{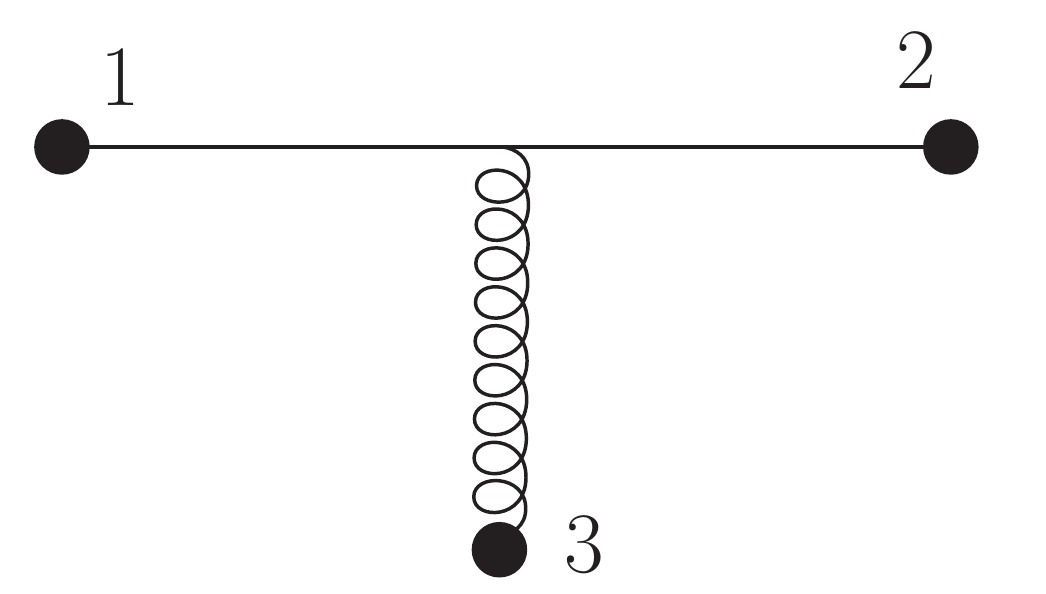}\end{array} = 
\langle \phi^{a}(1) \, F_{(\alpha\beta)}^{b}(3) \,\phi^{c}(2) \rangle , \!
\begin{array}{c}\includegraphics[width = 3 cm]{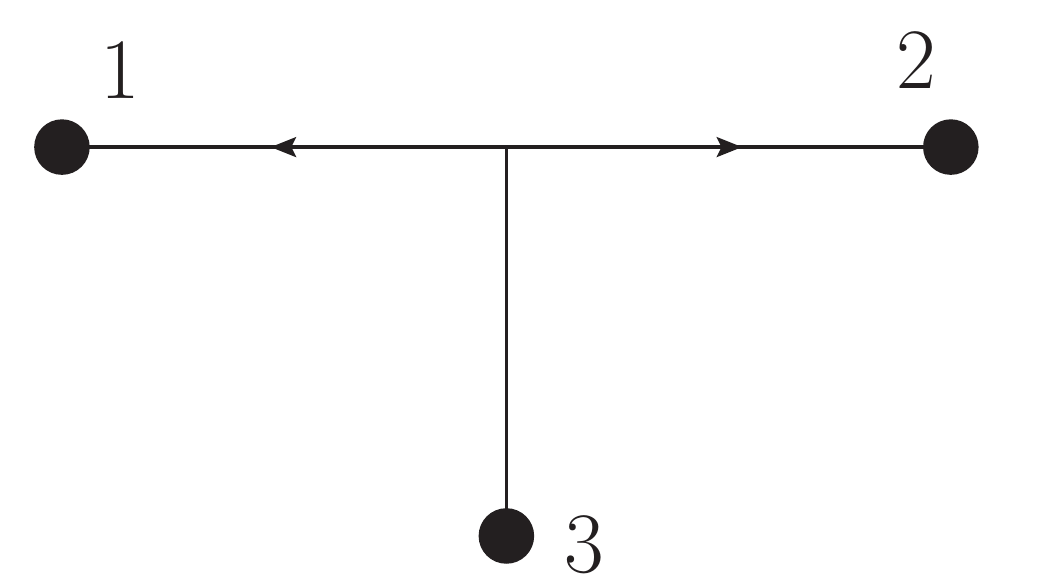}\end{array} = 
\langle \psi^{a,A}_{\alpha}(1) \, \phi_{++}^{b}(3) \,\psi^{c,B}_{\beta}(2) \rangle
\end{align}
Now we need the interaction of a gluon with a chiral and an anti-chiral fermions as well. This $T$-block
naturally decomposes as follows ($a,b,c$ are color indices)
\begin{align} \label{decomp}
\begin{array}{c}\includegraphics[width = 3 cm]{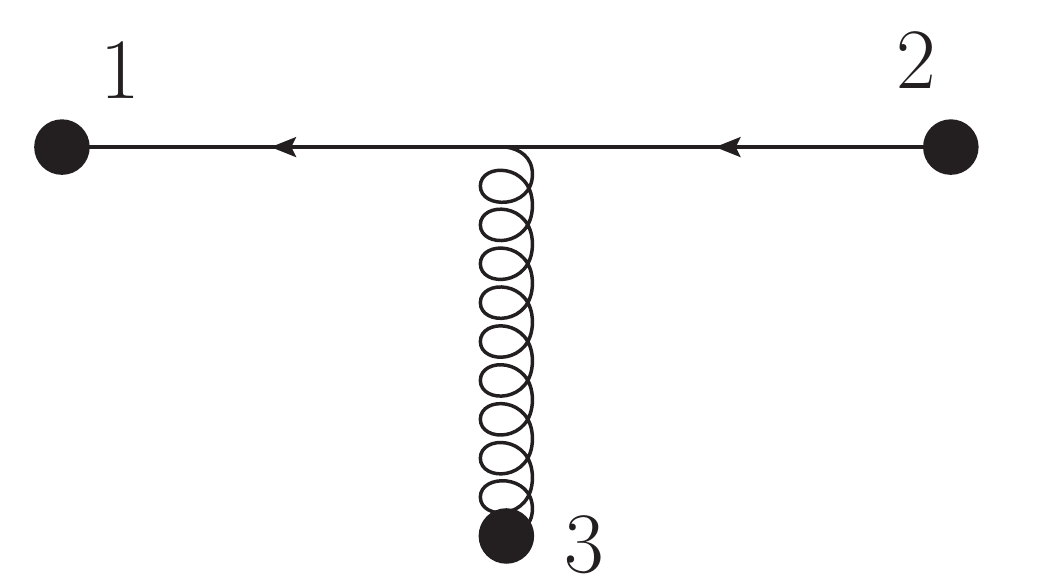}\end{array} = &
\langle \psi^{a,A}_{\alpha}(1) \, F_{(\beta\gamma)}^{b}(3) \,\bar\psi^{c}_{\dot\alpha,B}(2) \rangle =
\delta^{A}_{B} \frac{(x_{12})_{\a\da}}{x_{12}^2} \times \langle \phi^{a}(1) \, F_{(\beta\gamma)}^{b}(3) \,\phi^{c}(2) \rangle \notag\\
&- \frac{2g}{(2\pi)^4} f^{abc} \delta^{A}_{B} \frac{(x_{12} \tilde x_{23})_{\a(\b} (x_{23})_{\gamma)\da}}{x_{12}^2 x_{13}^2 x_{23}^4}\,.
\end{align}
Here we singled out the gluon $T$-block from \p{T1} multiplied by the propagator of fermions. 
This observation yields a substantial simplification of the calculation ($n = 4,5$), 
using the known chiral correlators
\begin{align} \label{chiralLOOO}
\left\langle \,\cO(1) \, \cO(2) \, L(3) \,
\cO(4) \cdots \cO(n) \,\right\rangle_{g^2}\,.
\end{align}
There are many gluon diagrams contributing to $G_n$ \p{psibarpsiLOOO}.
Each of them separately is not conformal, so we need to organize them in conformal  combinations.
The decomposition \p{decomp} allows us to do it.
Indeed, a diagram containing the $T$-block \p{decomp} can be split 
in a sum of two terms, one of which is a fermionic propagator times the corresponding diagram from the chiral correlator 
\p{chiralLOOO}. Thus the chiral correlator \p{chiralLOOO} captures the essential part of almost all gluon diagrams 
and provides a conformal expression for them. 

The rest of the gluon contributions independently sums up to a conformal combination.
Among such diagrams we encounter the one-loop correction of the fermionic propagator, i.e. 
$\left\langle \, \psi^{\alpha}_{+a}(1) \, \bar{\psi}^{\dot\alpha}_{-a'}(2) \, L(3) \,\right\rangle$.
It is a sum of several complicated Feynman integrals. 
In order not to perform these integration explicitly, we use the second trick that exploits the protectedness 
of the correlator of two half-BPS operators. It implies that the following three-point 
correlator calculated in the Born approximation vanishes,
\begin{align} \label{psibarpsiL}
\left\langle \,\mathrm{tr} \bigl( \psi^{\alpha}_{+a} \phi_{++}\bigr)(1) \,
\mathrm{tr}\bigl( \bar{\psi}^{\dot\alpha}_{-a'} \phi_{++}\bigr)(2) \, L(3) \,\right\rangle_{g^2} = 0\,.
\end{align}
Considering the Feynman graphs that contribute to \p{psibarpsiL},
we can express the one-loop correction of the fermionic propagator
in terms of simple Feynman graphs which are rational and are given by products of $T$-blocks.\footnote{ 
We need the one-loop correction to the scalar propagator as well, i.e. 
$\left\langle \, \phi_{++}(1) \, \phi_{++}(2) \, L(3) \,\right\rangle$. 
Its calculation follows the same strategy and is based on the protectedness of the two-point correlator of $\cO$.}

The results are as follows:
\begin{align}
G_5 = & \frac{c_5}{4} \frac{1}{x_{12}^2 x_{23}^2} \frac{x_{25}^2}{\prod_{i<j} x^2_{ij}}
\biggl[ - 2 y_{12}^2 y_{45}^2 (y_{15}\tilde{y}_{54}y_{42})_{aa'}
(x_{15}\tilde{x}_{53}x_{32})_{\alpha\dot\alpha} x_{12}^2 x_{14}^2 x_{24}^2 \nt
& + 2 y_{15}^2 y_{24}^2 (y_{14}\tilde{y}_{45}y_{52})_{aa'}
(x_{14}\tilde{x}_{43}x_{32})_{\alpha\dot\alpha} x_{12}^4 x_{45}^2
- y_{12}^2 y_{45}^4 (y_{12})_{aa'}(x_{12})_{\alpha\dot\alpha} x_{14}^2 x_{15}^2 x_{23}^2 x_{24}^2 \nt
& + 2 y_{15}^2 y_{24}^2 y_{45}^2 (y_{12})_{aa'} 
\Bigl( x_{14}^2 x_{23}^2 (x_{15}^2 x_{24}^2 - x_{14}^2 x_{25}^2) (x_{12})_{\alpha\dot\alpha}
- (x_{14}\tilde{x}_{45}x_{52})_{\alpha\dot\alpha} x_{12}^2 x_{14}^2 x_{23}^2 \nt 
& - (x_{15}\tilde{x}_{53}x_{32})_{\alpha\dot\alpha} x_{12}^2 x_{14}^2 x_{24}^2 
+ (x_{14}\tilde{x}_{43}x_{32})_{\alpha\dot\alpha} x_{12}^2 x_{14}^2 x_{25}^2 \Bigr) 
\biggr] + (4\rightleftarrows 5) \label{G5}
\end{align}
\begin{align}
G_6 = & \frac{c_6}{4} \frac{x_{25}^2 x_{26}^2 x_{46}^2}{x_{12}^2 x_{23}^2\prod_{i<j} x^2_{ij}}
\Bigl[ - 2 y_{12}^2 y_{45}^2 y_{56}^2 (y_{16}\tilde{y}_{64}y_{42})_{aa'}
(x_{16}\tilde{x}_{63}x_{32})_{\alpha\dot\alpha} x_{12}^2 x_{14}^2 x_{15}^2 x_{24}^2 x_{35}^2 \nt
& - 2 y_{14}^2 y_{24}^2 y_{56}^2 (y_{15}\tilde{y}_{56}y_{62})_{aa'}
(x_{15}\tilde{x}_{53}x_{32})_{\alpha\dot\alpha} x_{12}^4 x_{16}^2 x_{34}^2 x_{45}^2 \nt
& + 2 y_{16}^2 y_{24}^2 y_{56}^2 (y_{14}\tilde{y}_{45}y_{52})_{aa'}
(x_{14}\tilde{x}_{43}x_{32})_{\alpha\dot\alpha} x_{12}^4 x_{15}^2 x_{36}^2 x_{45}^2 \nt
& + 2 y_{16}^2 y_{24}^2 y_{45}^2 (y_{15}\tilde{y}_{56}y_{62})_{aa'}
(x_{15}\tilde{x}_{53}x_{32})_{\alpha\dot\alpha} x_{12}^4 x_{14}^2 x_{34}^2 x_{56}^2 \nt
& - y_{12}^2 y_{45}^2 y_{46}^2 y_{56}^2 (y_{12})_{aa'}
(x_{12})_{\alpha\dot\alpha} x_{14}^2 x_{15}^2 x_{16}^2 x_{23}^2 x_{24}^2 x_{35}^2 \nt
& - y_{14}^2 y_{24}^2 y_{56}^4 (y_{12})_{aa'}
(x_{12})_{\alpha\dot\alpha} x_{15}^2 x_{16}^2 x_{23}^2 x_{45}^2 \left( x_{14}^2 x_{23}^2 + x_{13}^2 x_{24}^2 \right) \nt
& + y_{16}^2 y_{24}^2 y_{45}^2 y_{56}^2 (y_{12})_{aa'} x_{14}^2 x_{15}^2 \Bigl( 2 (x_{14}\tilde{x}_{43}x_{32})_{\alpha\dot\alpha}
x_{12}^2 x_{25}^2 x_{36}^2 - 2 (x_{14}\tilde{x}_{45}x_{52})_{\alpha\dot\alpha}  x_{12}^2 x_{23}^2 x_{36}^2 \nt
&+ 2 (x_{15}\tilde{x}_{53}x_{32})_{\alpha\dot\alpha}  x_{12}^2 
( x_{26}^2 x_{34}^2 - x_{24}^2 x_{36}^2 ) 
- 2 (x_{16}\tilde{x}_{63}x_{32})_{\alpha\dot\alpha}  x_{12}^2 x_{25}^2 x_{34}^2 \nt
&- 2 (x_{15}\tilde{x}_{56}x_{62})_{\alpha\dot\alpha}  x_{12}^2 x_{23}^2 x_{34}^2 \nt
& - (x_{12})_{\alpha\dot\alpha} x_{23}^2 \bigl[ 
x_{13}^2\left( x_{25}^2 x_{46}^2 - x_{24}^2 x_{56}^2 - x_{26}^2 x_{45}^2 - 4 i \epsilon(x_{43},x_{53},x_{63},x_{23}) \right) \nt
& + x_{23}^2\left( x_{15}^2 x_{46}^2 - x_{16}^2 x_{45}^2 - x_{14}^2 x_{56}^2 - 4 i \epsilon(x_{53},x_{63},x_{13},x_{43}) \right)  \nt
& + x_{34}^2\left( x_{15}^2 x_{26}^2 + x_{12}^2 x_{56}^2 - x_{16}^2 x_{25}^2 - 4 i \epsilon(x_{63},x_{13},x_{23},x_{53}) \right) \nt
& + x_{35}^2\left( x_{14}^2 x_{26}^2 - x_{16}^2 x_{24}^2 - x_{12}^2 x_{46}^2 - 4 i \epsilon(x_{13},x_{23},x_{43},x_{63}) \right) \nt
& + x_{36}^2\left( x_{14}^2 x_{25}^2 + x_{12}^2 x_{45}^2 - x_{24}^2 x_{15}^2 - 4 i \epsilon(x_{23},x_{43},x_{53},x_{13}) \right) 
\bigr] \Bigl) \Bigr] +\mathrm{sym(4,5,6)} \label{G6}
\end{align}
with the normalization factor $c_n = g^2 N_c(N_c^2-1)/(2\pi)^{2n+2}$ and the parity-odd contribution  $\ep(x_1,x_2,x_3,x_4) \equiv \ep_{\mu \nu \rho \lambda} x^{\mu}_1 x^{\nu}_2 x^{\rho}_3 x^{\lambda}_4$ appearing in the case of six points.

Contrary to the harmonic superspace Feynman graph calculation now we work in the Feynman gauge instead of the light-cone gauge \p{417}.
Thus, the correlators $G_5$ and $G_6$ are manifestly conformal.
We would like to emphasize that the chiral correlator \p{chiralLOOO} captures a substantial part of $G_n$,
as it should be since the latter is the $\bar{Q}$-variation of the former.

We compared the correlators $G_5$, $G_6$ with the relevant component of the non-chiral supercorrelator 
$\langle \cT(1)\cM(2)\cT(3) \ldots \cT(n) \rangle$,
calculated by means of effective supergraphs according to Section \ref{snMHV}, and found perfect agreement.
The complexity of the expression produced by the LHC superspace Feynman graphs
and its artificial dependence on the gauge-fixing spinor $\xi^{\+}$ makes the comparison possible only by means of \texttt{Mathematica}.
We point out that only the agreement between the two expression for $G_6$ provides independent
support for the LHC superspace results since the five-point correlator (and correspondingly $G_5$) 
is fixed uniquely by superconformal symmetry.

\section{Quantization in a Lorentz-covariant gauge} \label{quantCG}

The light-cone gauge, eq.~\p{417}, is not the only possible gauge that yields simple  
 propagators in the LHC formulation of $\cN = 4$ SYM. 
An essential drawback of the light-cone gauge is that it breaks the Lorentz and conformal symmetries of the theory, still preserving the $SU(4)$ R-symmetry and $Q$-supersymmetry. Now we are going to impose 
a gauge which also preserves Lorentz invariance,
\begin{align}\label{landaug}
\pa^{-\da} A^+_{\da} = 0\,.
\end{align}
This Lorentz-covariant gauge is reminiscent of the usual Lorentz gauge in pure YM.
Roughly speaking, the gauge field $\cA_{\a\da}$ lives in the LH superfield $A^+_{\da}$, eq.~\p{25}, 
(in fact this statement is meaningful only in the Wess-Zumino gauge \p{23}). 
Then the constraint \p{landaug} implies $u^{-}_{\a} u^{+ \b} \pa^{\da \a}\cA_{\b\da} = 0$ where 
$\cA_{\a\da}(x,u)$ has an infinite harmonic expansion. 
Extracting the singlet part of the product of LHs, we 
obtain the  Lorentz gauge $\pa^{\da\a} \cA_{\a\da} = 0$.

The Lorentz-covariant gauge \p{landaug} is possible at least locally. 
Indeed, we can find an Abelian gauge transformation 
with an L-analytic parameter $\Lambda(x,\q^+,u)$ transforming to this gauge: $\pa^{-\da} \pa^+_{\da} \Lambda = \Box \Lambda = \pa^{-\da} A^+_{\da}$ since the d'Alembertian $\Box$ is invertible. We neglect the zero modes of $\Box$, so 
the gauge freedom is fixed completely by the condition \p{landaug}.

As in the light-cone gauge, we define the propagators using only the self-dual sector of the theory, 
which is described by the Chern-Simons Lagrangian $L_{\rm CS}$, eq.~\p{CS}. The bivalent vertex from the completion to the 
full theory $L_{\rm Z}$, eq.~\p{lint},
is treated as an interaction. 
We impose the gauge \p{landaug} 
by means of a Lagrangian multiplier $\rho^{+4}(x,\q^+,u)$, which is an L-analytic superfield carrying $U(1)$ charge $(+4)$.
Then the kinetic  Lagrangian  
\begin{align}
&L_{\rm kin} = \tr\Bigl( 
A^{++} \pa^{+ \da} A^+_{\da} - \frac{1}{2} A^{+ \da} \pa^{++} A^+_{\da} + \rho^{+4} \pa^{-\da} A^+_{\da} 
-\frac14 \tilde c^{++} \Box c^{++}
\Bigr)  \label{kinLandau}
\end{align}
becomes invertible,  allowing us to find the propagators.
Unlike the light-cone gauge, we have to introduce ghosts $\tilde c^{++}(x,\q^+,u)$ and $c^{++}(x,\q^+,u)$ as L-analytic superfields of $U(1)$ charge $(+2)$ with Fermi statistics. 
Another important difference from  the light-cone gauge is that the cubic interaction vertex in the Chern-Simons 
Lagrangian $L_{CS}$ does not vanish. Moreover, it is supplemented by a cubic interaction vertex for the ghost fields and $A^{+}_{\da}$,   
\begin{align}\label{intLandau}
&L_{\text{CS,int}} = \tr \Bigl(A^{++} A^{+ \da} A^+_{\da}  + [A^{+}_{\da},c^{++}] \pa^{-\da} \tilde c^{++}  \Bigr). 
\end{align}
So, the Chern-Simons Lagrangian $L_{CS}$  \p{CS} is replaced by the sum $L_{\rm kin} + L_{\text{CS,int}}$.
Inverting the bilinear form in eq.~\p{kinLandau} we find the set of propagators
\begin{align}
&\langle A^{+}_{\da}(x,\q^+,u_1) A^{+}_{\db}(0,0,u_2) \rangle = 0 \,, \label{p3}\\
&\langle A^{+}_{\da}(x,\q^+,u_1) A^{++}(0,0,u_2) \rangle = \pa^-_{\da} \,\langle A^{++}(x,\q^+,u_1) A^{++}(0,0,u_2)\rangle \,,\label{p2}\\
&\langle \tilde c^{++}(x,\q^+,u_1)\, c^{++}(0,0,u_2) \rangle = \langle A^{++}(x,\q^+,u_1) A^{++}(0,0,u_2)\rangle \,,\label{p4}\\
&\langle A^{++}(x,\q^+,u_1) A^{++}(0,0,u_2) \rangle = \frac{1}{\pi^2} \frac{1}{x^2} \,\delta(u_1,u_2)\, \delta^{(4)}(\q^+)\,. \label{p1}
\end{align}
The propagators \p{p4} and \p{p1} are even and \p{p2} is odd with respect to swapping the points. 
In contrast with the propagators in the light-cone gauge, eqs. \p{prop1}--\p{prop3}, the new ones have a more 
familiar form. Recalling that the LH analog $\phi_{AB}(x,u)$ of the scalar field lives in the superfield $A^{++}$
in front of $(\q^+)^2$, eq.~\p{23}, we extract from eq.~\p{p1} the familiar scalar propagator 
$\vev{\phi_{AB}(x,u_1) \phi_{CD}(0,u_2)} = \ep_{ABCD} x^{-2} \delta(u_1,u_2)$.
The space-time part of $\vev{A^{+}_{\da} A^{++}}$, eq.~\p{p2}, is the usual fermion propagator $\pa_{\a\da} x^{-2}$.
This is not a surprise since we can extract from there $\vev{\psi^{-A}(x,u_1) \bar\psi_{\da B}(0,u_2)}$, eqs. \p{23}, \p{25}.
  The propagators satisfy the gauge-fixing condition  \p{landaug} as a corollary of the identity $\pa^{-\da}\pa^{-}_{\da} = 0$.
In particular, $\pa^{-\da} \vev{A^+_{\da} A^{++}} = 0$, suggesting to call this gauge a {\it Landau gauge}. 
The propagators involving the Lagrangian multiplier, 
\begin{align}
& \langle \rho^{+4}(x,\q^+,u_1) A^{+}_{\da}(0,0,u_2) \rangle = \pa^+_{\da} \langle A^{++}(x,\q^+,u_1) A^{++}(0,0,u_2) \rangle \,, \nt
& \langle \rho^{+4}(x,\q^+,u_1) A^{++}(0,0,u_2) \rangle = \pa^{++} \langle A^{++}(x,\q^+,u_1) A^{++}(0,0,u_2) \rangle\,,\notag
\end{align}
do not participate in the calculations of gauge-invariant quantities.

The Feynman rules for this new  Landau gauge are similar to those in the light-cone gauge (see  \cite{PartI}), with the addition of ghosts appearing only in loops. The main difference is that the propagators are not contact terms in space-time.

As a simple application of these Feynman rules we show that there are no one-loop corrections to the bare propagators
induced by $L_{\rm CS,int}$, eq.~\p{kinLandau}. We do not consider quantum corrections coming from $L_{\rm Z}$ 
since they have a lower Grassmann degree.
There are two diagrams at order $O(g^2)$ giving corrections to the propagator $\langle A^{++} A^{++}\rangle$.
The first diagram contains two cubic vertices from $L_{\rm CS}$, and the second diagram
contains a ghost loop. They sum up to zero 
as a consequence of several facts: (i) the form of the propagators, i.e. 
$\langle A^+_{\da} A^{++}\rangle = \pa^-_{\da} \langle \tilde c^{++} c^{++}\rangle$, 
(ii) the cubic vertices from $L_{\rm CS,int}$ have a similar form, (iii) the ghost fields satisfy the Fermi statistics.
\begin{align}
\begin{array}{c}
\includegraphics[width = 4 cm]{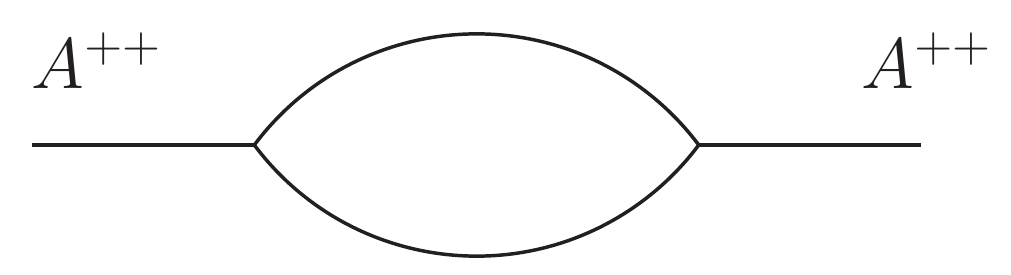}
\end{array} + \;
\begin{array}{c}  
\includegraphics[width = 4 cm]{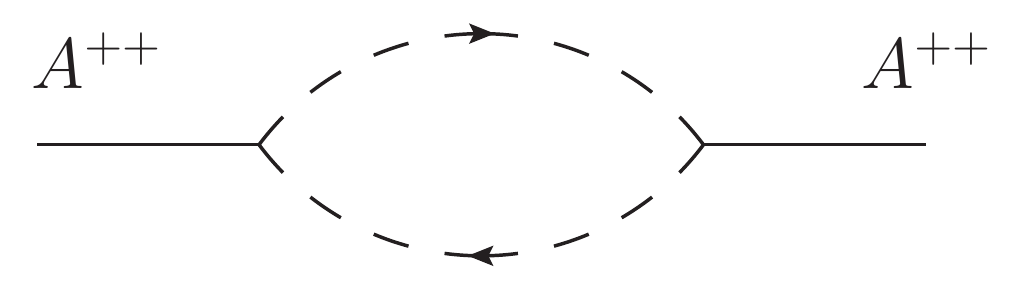} 
\end{array}\; = 0 \,. \label{zeroqc}
\end{align}
Then, there are no loop diagrams at order $O(g^2)$ made of the $L_{\rm CS,int}$ vertices 
and corresponding to quantum corrections of 
$\langle A^{+}_{\da} A^{++} \rangle$ and $\langle A^{+}_{\da} A^{+}_{\db} \rangle$.
The reason is the absence of the bare propagator $\langle A^{+}_{\da} A^{+}_{\db} \rangle$, eq.~\p{p3}.

\subsection{Calculations in the Landau gauge} \label{calCG}

Here we show a nontrivial application of the Feynman rules in the Landau gauge \p{landaug}. 
The MHV-like correlator $G_{n;0}$ \p{5.41}, which is given by graphs containing only bivalent vertices from $S_{\rm Z}$, 
is immediately reproduced in this formalism.
The NMHV-like $n$-point Born-level correlator $G_{n;1}$ is much more involved. 
Compared to the light-cone gauge Feynman rules in Section \ref{s7.2}, now there are more Feynman graphs 
and they have a more complicated form. This is the price of the explicit Lorentz invariance of the result.

Like before, we add to the Feynman rules the external vertex 
$\cT(x,\q_+,w)$, eq.~\p{439}. It involves the R-analyticity projector $\int d^4 \q_{-}$ and an 
LH factor without $\omega$. 
In the Born approximation we do not consider graphs with internal (interaction) vertices from $S_{\rm Z}$.
They involve eight Grassmann integrations, which shifts the balance between the order $O(g^{2p})$ in the coupling 
and the Grassmann degree $4p$ of the Born-level graphs. The interaction vertices from $S_{\rm Z}$ correspond 
to loop integrals.

The Born-level $n$-point NMHV-like supercorrelator $G_{n;1}$ is of order $O(g^2)$. 
For Born-level graphs we use the cubic internal vertices from $L_{\rm CS,int}$
and the $k$-valent external vertices from $\cT$. They can be chosen in four ways:
\begin{itemize} 
\item 
two interaction vertices from $S_{\rm CS}$ and $n$ bivalent vertices from $\cT$;
\item 
one internal vertex from $S_{\rm CS}$, one trivalent vertex and $n-1$ bivalent vertices from $\cT$;
\item 
no internal vertices, two trivalent vertices and $n-2$ bivalent vertices from $\cT$; 
\item
no internal vertices, one four-valent vertex and $n-1$ bivalent vertices from $\cT$.
\end{itemize}

In the following we omit the quantum corrections to the propagator $\langle A^{++} A^{++} \rangle$ , shown to vanish in   eq.~\p{zeroqc}. At order $O(g^2)$ the ghost loop can appear only in diagrams of this type. So, ghosts will not appear in our calculation at all.

In drastic contrast with the light-cone gauge, 
now the Born-level calculation is not a free field theory calculation.
We have to consider diagrams with internal (interaction) vertices from $S_{CS}$ 
which involve genuine Feynman integrals $\int d^4 x$. At least for NMHV-like correlators 
these integrals can be evaluated explicitly and the result is rational.

\begin{figure}
\begin{center}
\begin{tabular}{c}
\includegraphics[width = 3.5 cm]{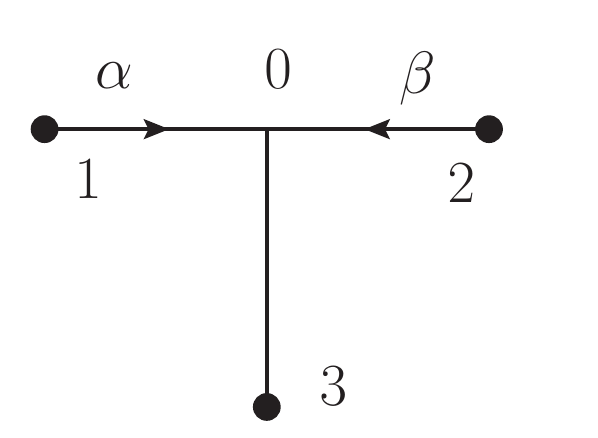}
\end{tabular} \qquad\qquad
\begin{tabular}{c}
\includegraphics[width = 3.5 cm]{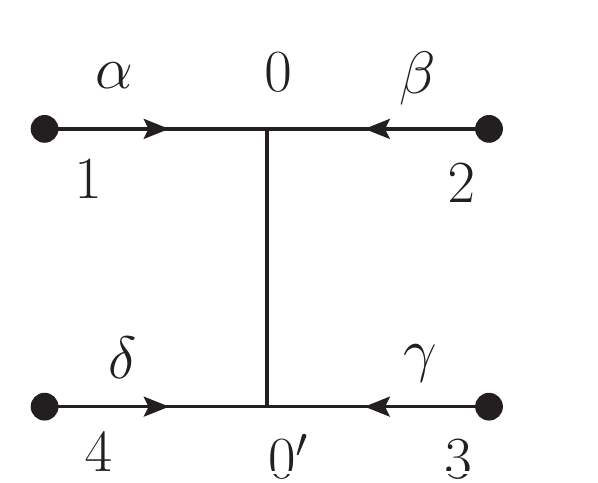}
\end{tabular} +
\begin{tabular}{c}
\includegraphics[width = 3.5 cm]{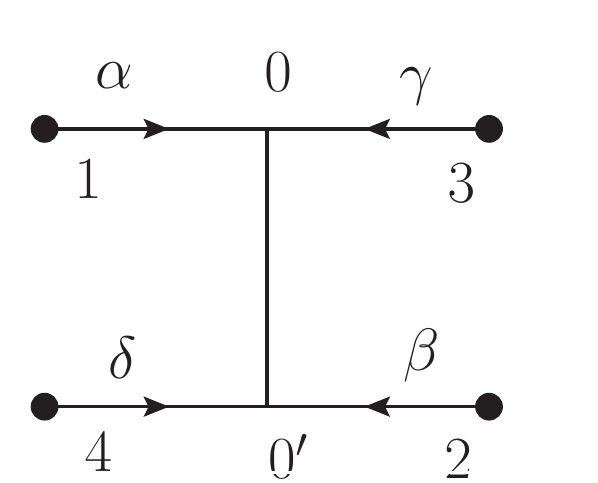}
\end{tabular}
\end{center}
\caption{The one-loop and two-loop integrals appearing in the Landau gauge calculation 
of the Born-level NMHV-like correlators.}\label{Loopint}
\end{figure}

We will need two Feynman integrals of this type. The one-loop integral (known as the star-triangle relation)
\begin{align}
- \frac{1}{\pi^2} \int d^4 x_0\; x_{30}^{-2}\, \pa_{\a\da} x_{10}^{-2}\, \pa^{\da}_{\b} x_{20}^{-2} 
= \frac{(x_{132})_{\a\b}}{x_{12}^2 x_{13}^2 x_{23}^2}\,, \label{strtriang}
\end{align}
where we use the shorthand notation $(x_{132})_{\a\b} = (x_{13})_{\a\da} (x_{32})^{\da}_{\b}$ ,
describes the Yukawa interaction of two fermions of the same chirality and a scalar, Fig.~\ref{Loopint}, eq. \p{T1}. 
The sum of two two-loop H-shape integrals also has a simple rational expression 
\begin{align}
&\frac{1}{\pi^4} \int d^4 x_0 d^4 x_{0'}\; x_{00'}^{-2} \,
\pa_{\da \a} x_{10}^{-2}\, \pa^{\da}_{\b} x_{20}^{-2}\,
\pa_{\db \gamma} x_{30'}^{-2}\, \pa^{\db}_{\delta} x_{40'}^{-2} + (2\; \beta \leftrightarrows 3 \; \gamma) \nt
& = x_{12}^{-2} x_{13}^{-2} x_{14}^{-2} x_{23}^{-2} x_{24}^{-2} x_{34}^{-2} 
\Bigl[ x_{23}^2 (x_{143})_{\a\gamma}(x_{214})_{\beta\delta} 
+ x_{14}^2 (x_{132})_{\a\b}(x_{324})_{\gamma\delta} \Bigr]. \label{Hint}
\end{align}
The right hand side of eq.~\p{Hint} is rational due to the symmetrization. 
One of the loop integrals in eq.~\p{Hint} can be evaluated with the help of eq.~\p{strtriang}.
The remaining one-loop integral can be reduced to scalar integrals 
(the four-point cross-integral and its three-point truncation, which are known in terms of logs and
dilogs) acted upon by two derivatives. After simplification all transcendental pieces cancel out.

At higher non-MHV levels the situation is much more involved. 
The Born-level correlators are also rational, but certain individual Feynman graphs 
can produce highly transcendental expressions. 
The sum of all these intricate Feynman integrals has to simplify to a rational expression.
The coupling dependence of the correlator is due to the nonpolynomial form of the operator $\cT$ \p{439}
and to the cubic interaction vertex from $S_{\rm CS}$.
Thus we see that the calculation in the Landau gauge is more in the spirit of the 
usual perturbative calculations in terms of component fields of $\cN = 4$ SYM in Appendix~\ref{apE}.
The advantage of the present approach  is the manifest $Q$-supersymmetry.

\begin{figure}
\begin{center}
\begin{tabular}{c}
\includegraphics[width = 2.5 cm]{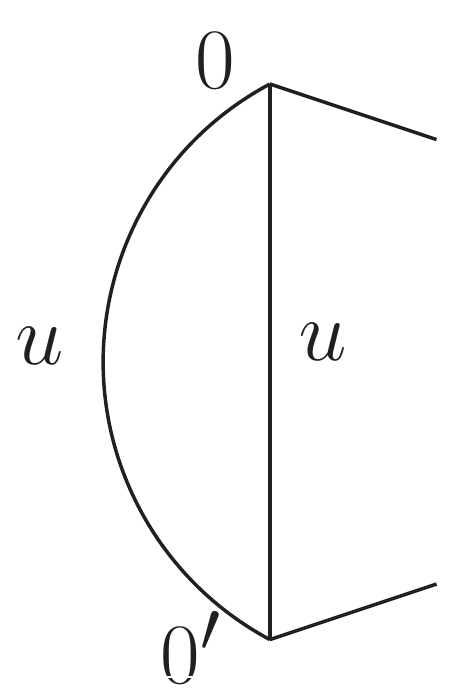}
\end{tabular} \qquad\qquad
\begin{tabular}{c}
\includegraphics[width = 2.5 cm]{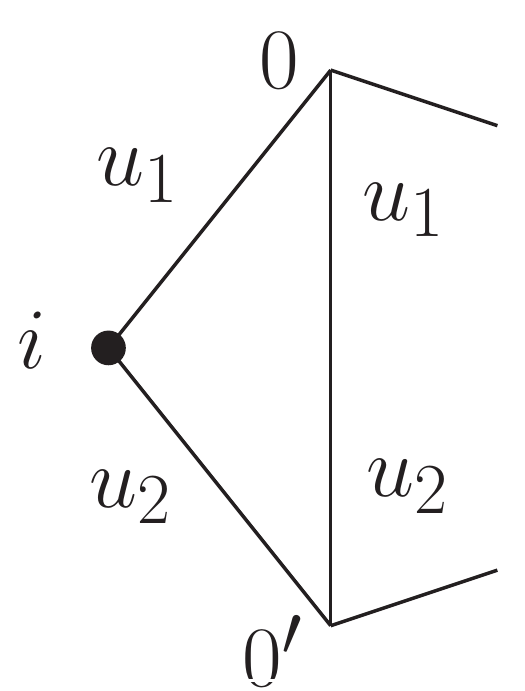}
\end{tabular} \qquad\qquad
\begin{tabular}{c}
\includegraphics[width = 2.5 cm]{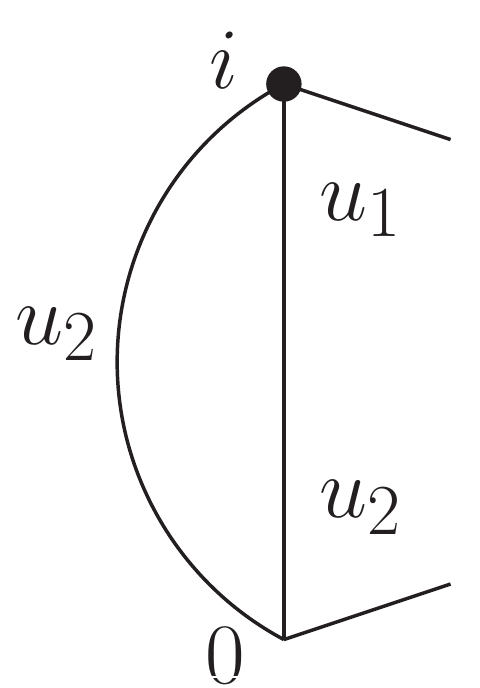}
\end{tabular}
\end{center}
\caption{Vanishing super subgraphs in the Landau gauge, eq.~\p{landaug}.}\label{vanishsg}
\end{figure}

There are several subgraphs  in an NMHV graph  that can make the whole graph vanish, Fig.~\ref{vanishsg}.
The first diagram in Fig.~\ref{vanishsg} contains a pair of interaction cubic vertices from $S_{\rm CS}$ linked by two propagators.
The vertices are local in the LH space. The diagram vanishes due to the identity
$\delta^4(\q_{00'} \cdot u^+) \delta^4(\q_{00'} \cdot u^+) = 0$, eq.~\p{p1}.
The second diagram in Fig.~\ref{vanishsg} contains two cubic vertices from $S_{\rm CS}$ and one external bivalent vertex from $\cT(i)$
linked to form a triangle. It is proportional to 
\begin{align} 
\int d u_1 \frac{1}{(u_1^+ u_2^+)^2}
\delta^4(\q_{10} u_1^+ ) \delta^4(\q_{10'} u_2^+ ) \delta^4(\q_{00'} u_1^+) \delta(u_1,u_2) = 0\,.
\end{align}
In the third diagram in Fig.~\ref{vanishsg} the interaction vertex from $S_{\rm CS}$ and 
the external vertex $\cT(i)$ are linked by two propagators, so again the vanishing is due to the Grassmann delta functions,
\begin{align} 
\int d u_1 \frac{1}{(u_1^+ u_2^+)}
\delta^4(\q_{10} u_1^+ ) \delta^4(\q_{10} u_2^+ ) \delta(u_1,u_2) = 0\,.
\end{align}
In the following we  omit the topologies containing the subgraphs from Fig.~\ref{vanishsg}.
In particular one can easily see that all the supergraphs representing the two-point NMHV-like Born-level correlator 
$G_{2;1}$ contain these subgraphs.

\subsubsection{Three-point NMHV-like correlator}

\begin{figure}
\begin{tabular}{cccc}
\begin{tabular}{c}
\includegraphics[width = 3 cm]{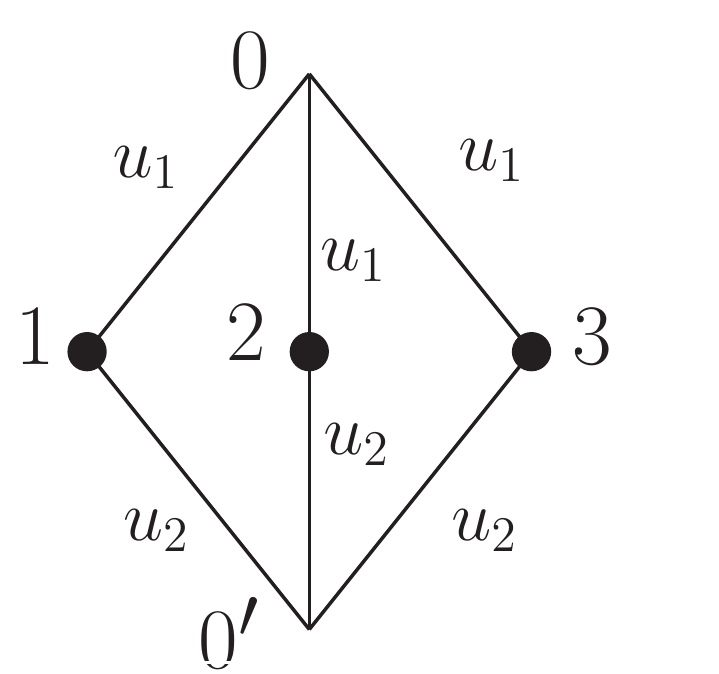}
\end{tabular} & \!\!\!\!\!
\begin{tabular}{c}
\includegraphics[width = 3.5 cm]{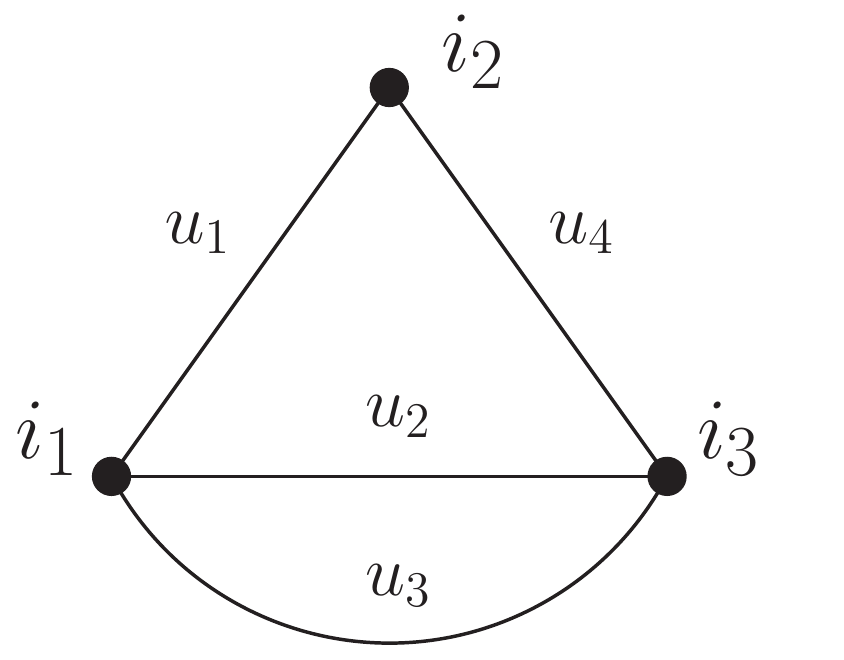}
\end{tabular} & \!\!\!\!\!
\begin{tabular}{c}
\includegraphics[width = 2.2 cm]{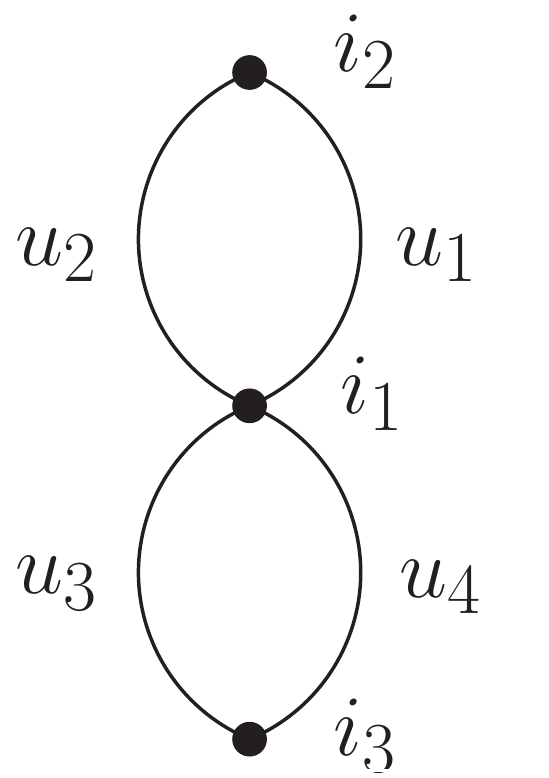}
\end{tabular} & \!\!\!\!\!
\begin{tabular}{c}
\includegraphics[width = 4.0 cm]{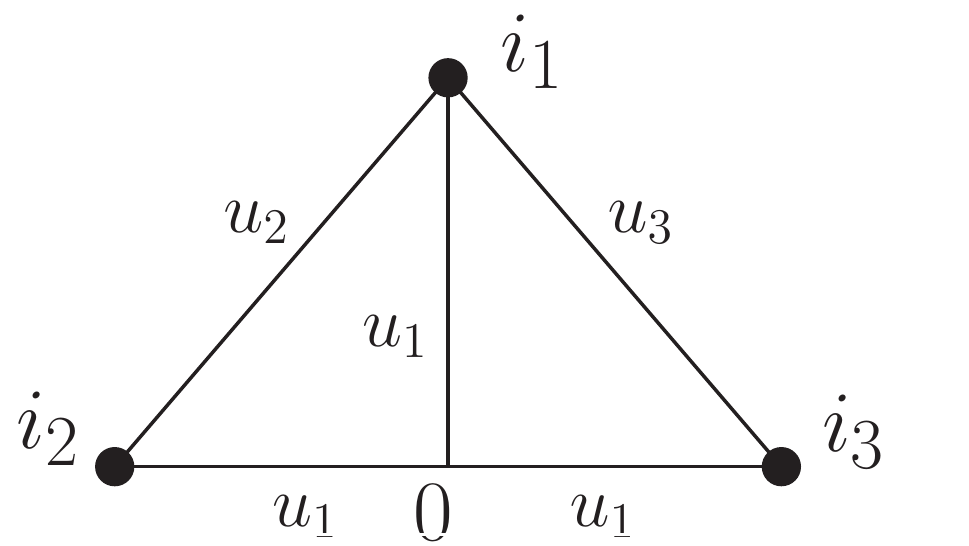}
\end{tabular} 
\\ $(a)$ & $(b)$ & $(c)$ & $(d)$ 
\end{tabular}
\caption{Four topologies contributing to the Born-level three-point NMHV-like correlator $G_{3;1}$ in the Lorenz-covariant gauge.}
\label{3NMHVLCG}
\end{figure}

The three-point NMHV-like Born-level correlator $G_{3;1}$ has to vanish. This is obvious in the light-cone gauge \p{417},
however in the Landau gauge there are four topologies giving nontrivial contributions, Fig.~\ref{3NMHVLCG}. 
As a simple illustration of the new rules in this gauge we show that they sum up to zero.
In the light-cone gauge  we are not allowed to draw more that one propagator between 
a pair of external vertices because of the mixed delta functions in the propagator $\vev{A^{++}A^{++}}$, eq.~\p{prop1}.
In the Landau gauge we can link a pair of external vertices by two propagators $\vev{A^{++}A^{++}}$, eq.~\p{p1}.  

Firstly, let us show that topology $(d)$ in Fig.~\ref{3NMHVLCG} vanishes because of the harmonic integrations.
The Grassmann integration $\int d^4 \q^+_0$ produces the harmonic factor $(u_1^+ u_2^+)^4 (u_1^+ u_3^+)^4$. 
Then we collect the harmonic factors from the trivalent vertex $\cT(i_1)$, 
from the bivalent vertices at points $i_2,i_3$, and the harmonics $u^-_{1\a} u^-_{1\b}$ 
coming from the two propagators $\langle A^+_{\da} A^{++}\rangle$, and we perform the harmonic integrations,
\begin{align}
&\int d u \,  u^-_{1\a} u^-_{1\b}(u_1^+ u_2^+) (u_1^+ u_3^+)(u_2^+ u_3^+)^{-1} = 0\,.
\end{align}

The remaining three topologies $(a)$, $(b)$, $(c)$ are nontrivial. 
The integral $\int d^4 \q_0^+ d^4 \q_{0'}^+ $ of the product of six Grassmann delta functions 
in the diagram $T_{(a)}$  results in \\ $ (u_1^+ u_2^+)^8 \delta^8(\q_{12}) \delta^8(\q_{13})$. 
To implement the R-analytic projections at points $1,2,3$ we decompose the Grassmann delta function by means 
of RHs according to eq.~\p{5.36}, 
\begin{align}
\int d^4 \q_{1-} d^4 \q_{2-} d^4 \q_{3-} \; \delta^8(\q_{12}) \delta^8(\q_{13}) = y_{12}^4 y_{13}^4\, 
\delta^4(A_{1;2 3})\,,
\end{align}
where we introduced the three-point $Q$-supersymmetry invariant $A_{i;j,k}$. 
It is made of two two-point $Q$-supersymmetry covariants $B_{ij}$ and $B_{ik}$, eq.~\p{B},
\begin{align} \label{Ainv}
A^{a'\a}_{i;j k} \equiv B^{a'\a}_{ij} - B^{a'\a}_{ik}\,.
\end{align}
The Feynman integral in $T_{(a)}$ factorizes into two one-loop integrals of the star-triangle type \p{strtriang}. 
The LH integrals are done with the help of \p{6}. 
So, topology $(a)$ contributes
\begin{align} \label{Ta}
T_{(a)}(1,2,3) = \frac{y_{12}^4 y_{13}^4\delta^4(A_{1;23})}{x_{12}^4 x_{13}^4 x_{23}^4}
(- x_{12}^4 - x_{13}^4 - x_{23}^4 + 2 x_{12}^2 x_{13}^2 + 2 x_{12}^2 x_{23}^2 + 2 x_{13}^2 x_{23}^2)\,.
\end{align}
Topologies $(b)$ and $(c)$ in Fig.~\ref{3NMHVLCG} do not involve Feynman integrals, however they require
nontrivial harmonic integrations, for example \cite{Galperin:2001uw}
\begin{align}
\int d u \frac{(u_1^+ u_2^+) (u_3^+ u_4^+)}{(u_1^+ u_4^+)(u_2^+ u_3^+)} = -\frac{1}{2}\,.
\end{align}
So, the contributions of the two topologies are  
\begin{align} \label{Tbc}
T_{(b)}(i_1,i_2,i_3) = -2\frac{y_{i_1 i_2}^4 y_{i_1 i_3}^4}{x_{i_1 i_2}^2 x_{i_1 i_3}^4 x_{i_2 i_3}^2}\delta^4(A_{i_1;i_2 i_3}) 
\;, \quad
T_{(c)}(i_1,i_2,i_3) = \frac{y_{i_1 i_2}^4 y_{i_1 i_3}^4}{x_{i_1 i_2}^4 x_{i_1 i_3}^4}\delta^4(A_{i_1;i_2 i_3})\,.
\end{align}
Summing $T_{(a)}$, $T_{(b)}$, $T_{(c)}$ over all nonequivalent permutations of points $1,2,3$ we get zero, i.e. 
$G_{3;1} = 0$ at Born level, as expected. 

The contributions $T_{(a)}$, $T_{(b)}$, $T_{(c)}$, eqs. \p{Ta}, \p{Tbc}, to the three-point correlator give us an idea how 
multipoint diagrams look like. Their $SU(4)$ R-symmetry and 
$Q$-supersymmetry are manifest in the three-point supersymmetry invariant $A_{i;jk}$, eq.~\p{Ainv}.
Lorentz invariance is also respected, but each particular diagram is not conformally covariant. 
There are no spurious poles like those in the light-cone gauge produced by the gauge-fixing spinor $\xi^{\-}$. However, $T_{(a)}$, $T_{(b)}$, $T_{(c)}$ contain higher order poles $x_{ij}^{-4}$. 
The allowed singularities of the correlator are simple poles $x_{ij}^{-2}$, so the higher order
poles have to disappear in the sum of all diagrams.

\subsubsection{$n$-point NMHV-like correlator}

The four-point NMHV-like Born-level correlator $G_{4;1}$ has to vanish as a consequence of supersymmetry, eq.~\p{7.1}. 
In the light-cone gauge   we have six contributing diagrams summing up to zero due to complicated algebraic identities.
Working in the Landau gauge  also requires a nontrivial calculation to show that $G_{4;1} = 0$.
This calculation is very helpful, it allows us to get rid of the higher order poles in the $n$-point correlator. 

Along with the three-point $Q$-supersymmetry invariant $A_{i;jk}$, eq.~\p{Ainv}, we will need the 
four-point invariant
\begin{align} \label{F19}
C^{\a\b\gamma\, a'}_{i;jkl} \equiv (B_{ij})^{\a a'} \ep^{\b\gamma} + (B_{ik})^{\b a'} \ep^{\gamma\a} + (B_{il})^{\gamma a'} \ep^{\b\a} 
= (A_{i;jk})^{\a a'} \ep^{\b\gamma} + (A_{i;kl})^{\gamma a'} \ep^{\a\b} \,.
\end{align}
The Grassmann delta function $\delta^4$ contains the $\q_+$-dependence of the three-point diagrams, eqs. \p{Ta}, \p{Tbc}.
Multipoint diagrams involve squares of three- and four-point invariants,
\begin{align}
A_{i;jk}^{(\a\b)} \equiv \frac{1}{2} \ep_{a' b'} A_{i;jk}^{\a a'} A_{i;jk}^{\b b'} \quad,\quad 
C^{\left(\substack{\a_1\b_1\gamma_1 \\ \a_2\b_2\gamma_2}\right)}_{i;jkl} 
\equiv \frac{1}{2} \ep_{a' b'} C^{\a_1\b_1\gamma_1\, a'}_{i;jkl} C^{\a_2\b_2\gamma_2\, b'}_{i;jkl} \,.
\end{align}
We will profit from the following properties of the three-point invariants:
\begin{align} \label{ACsq}
A^{a'\a}_{i;jk} (y_{ik})_{a'a} = A^{a'\a}_{j;ik} (y_{jk})_{a'a} \quad, \quad
y_{ik}^2 A_{i;jk}^{(\a\b)} = y_{jk}^2 A_{j;ik}^{(\a\b)}\,.
\end{align}
Further, we deal with products of three-  and four-point invariants with symmetrized Lorentz indices.
Such combinations can be rewritten in many different ways. Indeed, the identity
 $\delta^2(A_{i;jk} \cdot u^{+})\delta^2(A_{i;jl} \cdot u^{+}) = \delta^2(A_{i;jk} \cdot u^{+})\delta^2(A_{i;kl} \cdot u^{+})$
is equivalent to
\begin{align}
\underset{(\a_1 \ldots \a_4)}{\text{Sym}} A_{i;jk}^{(\a_1\a_2)} A_{i;jl}^{(\a_3\a_4)} 
= \underset{(\a_1 \ldots \a_4)}{\text{Sym}} A_{i;jk}^{(\a_1\a_2)} A_{i;kl}^{(\a_3\a_4)}  
\end{align}
due to the symmetrization by means of the LH $u^+$. Similarly, we have
\begin{align}
\underset{(\a_1 \ldots \a_4)}{\text{Sym}} A_{i;jk}^{(\a_1\a_2)} C_{i;jlm}^{\left(\substack{\a_3 \b_1 \gamma_1\\ \a_4 \b_2 \gamma_2}\right)} 
= \underset{(\a_1 \ldots \a_4)}{\text{Sym}} A_{i;jk}^{(\a_1\a_2)} C_{i;klm}^{\left(\substack{\a_3 \b_1 \gamma_1\\ \a_4 \b_2 \gamma_2}\right)} \,. 
\end{align}

\begin{figure}
\begin{center}
\begin{tabular}{cccc} \!\!\!\!\!\!
\begin{tabular}{c}
\includegraphics[width = 3.5 cm]{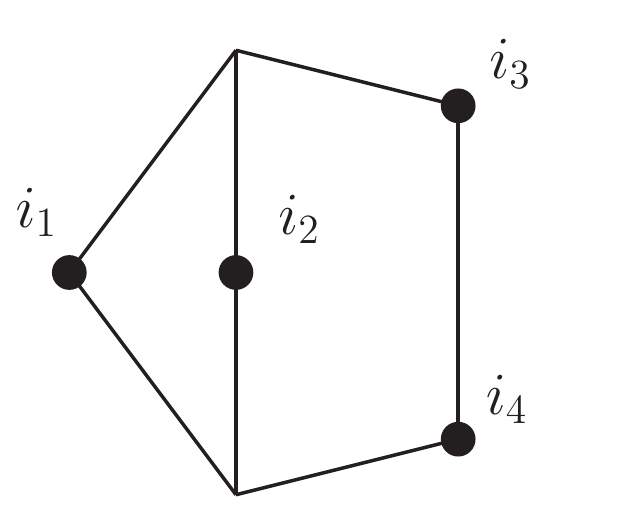}
\end{tabular} & \!\!\!\!\!\!\!
\begin{tabular}{c}
\includegraphics[width = 3.5 cm]{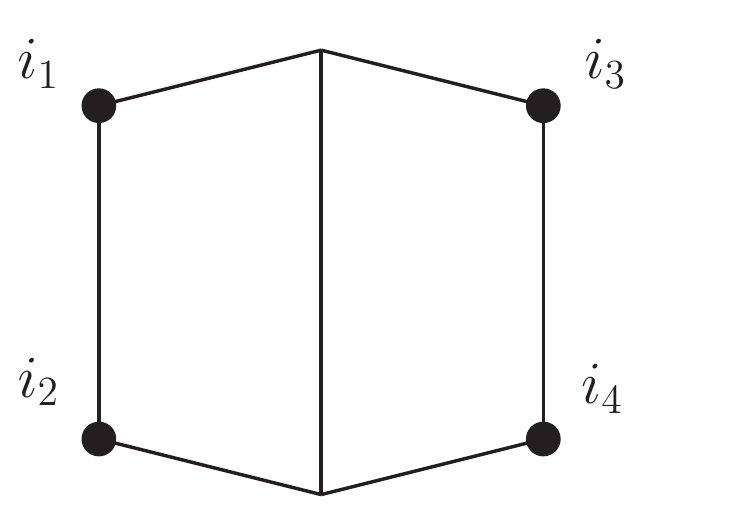}
\end{tabular} & \!\!\!\!\!\!\!
\begin{tabular}{c}
\includegraphics[width = 3.5 cm]{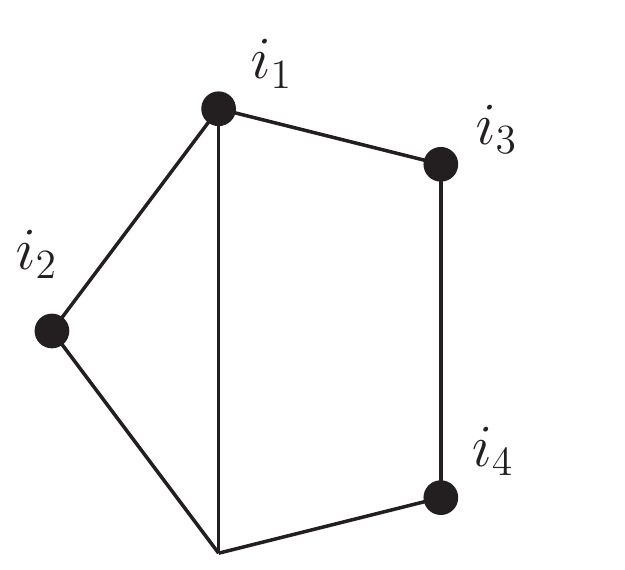}
\end{tabular} & \!\!\!\!\!\!\!
\begin{tabular}{c}
\includegraphics[width = 3.0 cm]{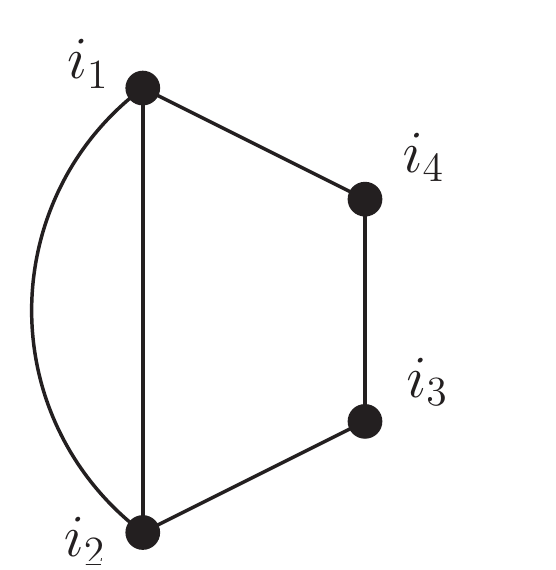}
\end{tabular} \\
$(a)$ & $(b)$ & $(c)$ & $(d)$
\end{tabular}
\end{center}
\begin{center}
\begin{tabular}{ccc}
\begin{tabular}{c}
\includegraphics[width = 3.5 cm]{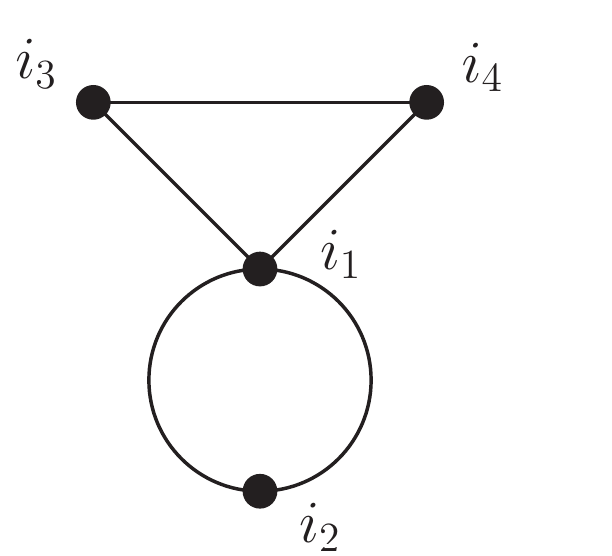}
\end{tabular} &
\begin{tabular}{c}
\includegraphics[width = 3.9 cm]{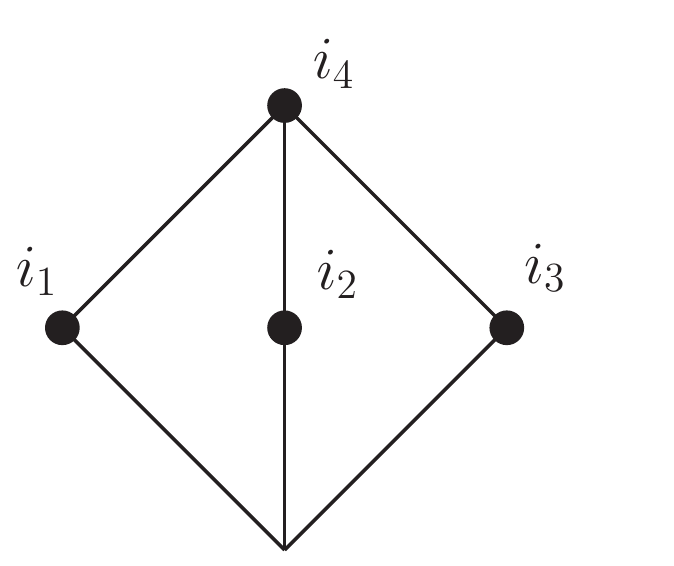}
\end{tabular} &
\begin{tabular}{c}
\includegraphics[width = 3.7 cm]{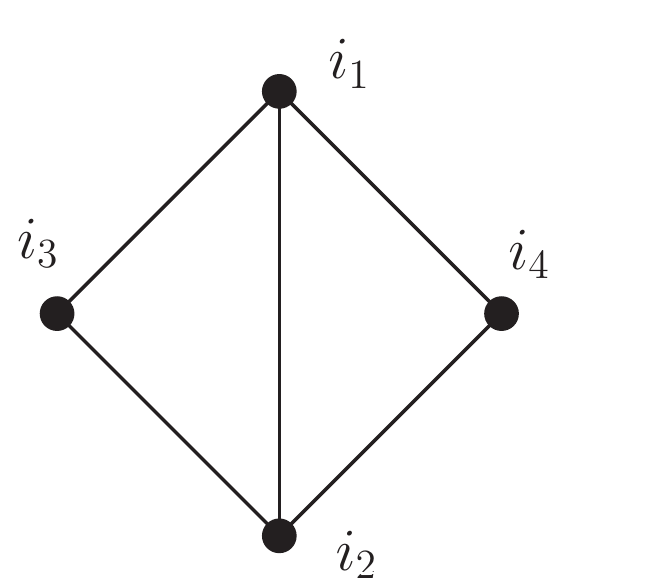}
\end{tabular}  \\
$(e)$ & $(f)$ & $(g)$
\end{tabular}
\end{center}
\caption{LHC supergraphs contributing to the four-point NMHV-like Born-level correlator 
in the Lorenz-covariant gauge.}\label{4NMHVLCG}
\end{figure}

There are seven topologies $(a)-(g)$ contributing to $G_{4;1}$, Fig.~\ref{4NMHVLCG}.
The result of the diagram calculation is as follows:
\begin{align}
&T_{(a)}= 6 \times \frac{1}{4} \times d_{i_1 i_2}^2 d_{i_1 i_3} d_{i_1 i_4} d_{i_3 i_4} x_{i_2 i_3}^{-2} x_{i_2 i_4}^{-2} 
A^{(\a_1\a_2)}_{i_1;i_2 i_3} (x_{i_1 i_2 i_3})_{\a_1\a_2} 
A^{(\b_1\b_2)}_{i_1;i_2 i_4} (x_{i_1 i_2 i_4})_{\b_1\b_2}  \nt
&T_{(b)} = -6 \times \frac{1}{4} \times d_{i_1 i_2}^2 d_{i_1 i_3} d_{i_1 i_4} d_{i_3 i_4} x_{i_2 i_3}^{-2} x_{i_2 i_4}^{-2} 
A^{(\a_1\a_2)}_{i_1;i_2 i_3} A^{(\a_3\a_4)}_{i_1;i_3 i_4}
(x_{i_1 i_2 i_4})^{\b_1\b_3} (x_{i_3 i_1 i_2})^{\b_2\b_4} \nt 
& \times \text{Sym} \, \ep_{\a_1\b_1}\ep_{\a_2\b_2}\ep_{\a_3\b_3}\ep_{\a_4\b_4} \nt
&T_{(c)} = - 3 \times 1 \times d_{i_1 i_2}^2 d_{i_1 i_3} d_{i_1 i_4} d_{i_3 i_4} x_{i_2 i_4}^{-2} 
A^{(\a_1\a_2)}_{i_1;i_2 i_3} A^{(\b_1\b_2)}_{i_1;i_2 i_4} (x_{i_2 i_4 i_1})^{\gamma_1\gamma_2} 
\text{Sym} \, \ep_{\a_1\gamma_2}\ep_{\a_2\b_1}\ep_{\b_2\gamma_1}  \nt
&T_{(d)} = -1 \times \frac{1}{2} \times d_{i_1 i_2}^2 d_{i_1 i_4} d_{i_2 i_3} d_{i_3 i_4} 
A^{(\a_1\a_2)}_{i_1;i_2 i_4} A_{i_2;i_1 i_3\, (\a_1\a_2)}  \nt
&T_{(e)} = 1 \times \frac{1}{2} \times d_{i_1 i_2}^2 d_{i_1 i_3} d_{i_1 i_4} d_{i_3 i_4} A^{(\a_1\a_2)}_{i_1;i_2 i_3} A_{i_1;i_2 i_4\, (\a_1\a_2)} \nt
&T_{(f)} = - 2 \times \frac{1}{6} \times d_{i_1 i_2} d_{i_1 i_3} d_{i_1 i_4} d_{i_2 i_4} d_{i_3 i_4} x_{i_2 i_3}^{-2} 
A^{(\delta_1\delta_2)}_{i_1;i_2 i_3} (x_{i_1 i_2 i_3})_{\delta_1\delta_2} 
C_{i_4;i_1 i_2 i_3}^{\left(\substack{\a_1\b_1\gamma_1\\ \a_2\b_2\gamma_2}\right)}
\text{Sym} \, \ep_{\a_1\gamma_2}\ep_{\a_2\b_1}\ep_{\b_2\gamma_1}  \nt
&T_{(g)} = - \frac{6}{5} \times \frac{1}{4} \times d_{i_1 i_2} d_{i_1 i_3} d_{i_1 i_4} d_{i_2 i_3} d_{i_2 i_4} 
C_{i_1;i_2 i_3 i_4}^{\left(\substack{\a_1\b_1\gamma_1\\ \a_2\b_2\gamma_2}\right)} 
C_{i_2;i_1 i_3 i_4}^{\left(\substack{\a_3\delta_1\varepsilon_1\\ \a_4\delta_2\varepsilon_2}\right)} \nt
&\times  \text{Sym} 
\Bigl[ \ep_{\a_1\varepsilon_1}\ep_{\a_2\varepsilon_2}\ep_{\a_3\b_2}\ep_{\a_4\delta_2}\ep_{\b_1\gamma_1}\ep_{\gamma_2\delta_1} 
-  \ep_{\a_1\delta_1}\ep_{\a_2\delta_2}\ep_{\a_3\b_2}\ep_{\a_4\varepsilon_2}\ep_{\b_1\gamma_1}\ep_{\gamma_2\varepsilon_1} \Bigr] \label{T4NMHV}
\end{align}
Here we omit the color factor $N_c(N_c^2-1)$. The second numerical factors in eq.~\p{T4NMHV} 
count the dimensions of the graph automorphism groups.
We imply that the symmetrization $\rm Sym$ acts only on the numerated indices carried by the same Greek letter, i.e.
in the expression for $T_{(c)}$ and $T_{(f)}$ we symmetrize $(\a_1,\a_2)$, $(\b_1,\b_2)$, and $(\gamma_1,\gamma_2)$;
in the expression for $T_{(g)}$ we symmetrize $(\a_1,\a_2)$, $(\b_1,\b_2)$, $(\gamma_1,\gamma_2)$, $(\delta_1,\delta_2)$, 
and $(\varepsilon_1,\varepsilon_2)$. The one-loop integrals in graphs $(c)$ and $(f)$ are of the star-triangle type, eq.~\p{strtriang}.
The two-loop integral in graph $(a)$ is a product of one-loop integrals of the same type, and the 
two-loop integral in graph $(b)$ is given in eq.~\p{Hint}. The expression for $T_{(b)}$ 
contains only the first term from eq.~\p{Hint}, since the second one is restored by   permutations.

The nontrivial Lorentz-tensor structures in eq.~\p{T4NMHV} constructed out of the skew-symmetric $\ep_{\a\b}$ arise after 
multiple harmonic integrations. The four-point $Q$-supersymmetry invariant $C$, eq. \p{F19}, comes from the Grassmann $\delta^2$ 
constituting the trivalent $R$-vertex, eq.~\p{Rinv}, integrated over the LHs. 
In the light-cone gauge all harmonic integrals are trivial 
since the harmonics are localized on the support of the mixed delta functions coming from the propagators.
The nontrivial harmonic integrations in the Landau gauge is a kind of averaging over 
the gauge-fixing spinor $\xi^{\+}$. 
 
Only graphs $(a)-(e)$ contain a pole $1/x_{i_1 i_2}^4$. In the sum of all the permutations $\sigma$  the pole $1/x_{i_1 i_2}^4$ drops out due to the identity
\begin{align}
&\sum_{l=a,b,c,d,e} \sum_{\sigma^{(l)}} T_{(l)}(\sigma_{i_1},\sigma_{i_2},\sigma_{i_3},\sigma_{i_4}) 
= x_{i_1 i_2}^2 d^2_{i_1 i_2} d_{i_1 i_3} d_{i_1 i_4} d_{i_3 i_4} x_{i_2 i_3}^{-2} x_{i_2 i_4}^{-2}\;
 p(i_1,i_2,i_3,i_4) \label{sum4pt}
\end{align}
with the polynomial 
\begin{align}
&p(i_1,i_2,i_3,i_4) = \frac{1}{2}
\bigl( 2x_{i_3 i_4}^2 - 2x_{i_1 i_2}^2 + 5 x_{i_1 i_3}^2 + 5 x_{i_2 i_3}^2 - x_{i_1 i_4}^2 - x_{i_2 i_4}^2 \bigr) 
A^{(\a_1\a_2)}_{i_1;i_2 i_3} A_{i_1;i_2 i_4\, (\a_1\a_2)} \nt
&+ 3 \ep_{\a_2 \a_4} A^{(\a_1\a_2)}_{i_1;i_2 i_3} A^{(\a_3\a_4)}_{i_1;i_2 i_4} 
\bigl((x_{i_1 i_2 i_4})_{\a_1 \a_3} - (x_{i_1 i_2 i_3})_{\a_1 \a_3} - 2(x_{i_1 i_3 i_4})_{\a_1 \a_3} \bigr)\ .  \label{ppol}
\end{align}
To prove this formula we use $Q$-supersymmetry to gauge away 
$\q_{i_1+} = \q_{i_2+}=0$, so the three-point invariants  simplify to 
$A_{i_1;i_2i_3}^{(\a_1\a_2)} \to y_{i_1i_3}^{-2} (\q_{i_3+})^{(\a_1\a_2)}$ 
and $A_{i_1;i_2i_4}^{(\a_1\a_2)} \to y_{i_1i_4}^{-2} (\q_{i_4+})^{(\a_1\a_2)}$,
and the Grassmann part of the sum in eq.~\p{sum4pt} takes the form $(\q_{i_3+})^{(\a_1\a_2)}(\q_{i_4+})^{(\a_3\a_4)}$.
Consequently, eq.~\p{sum4pt} turns into an identity involving only $x$-structures.  

After the elimination of the higher-order poles one can show that the sum over all topologies $(a)-(g)$ 
along with their $S_4$ permutations equals zero.

\begin{figure}
\begin{center}
\begin{tabular}{ccccc}
\includegraphics[width = 2.5 cm]{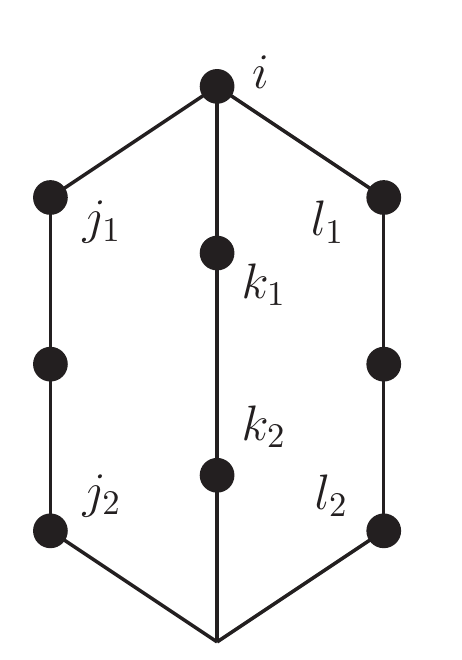} &
\includegraphics[width = 2.5 cm]{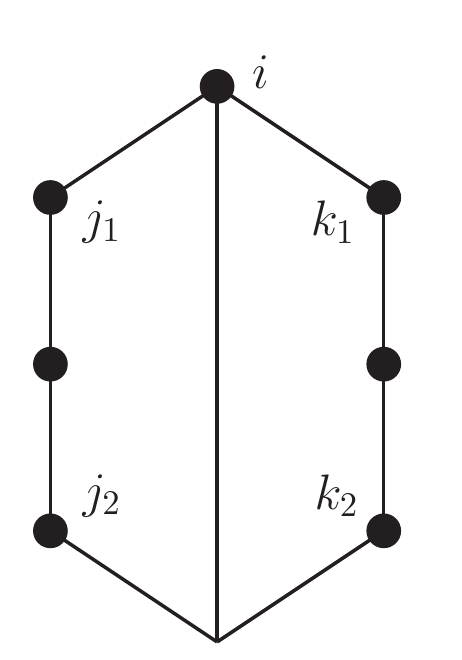} &
\includegraphics[width = 2.5 cm]{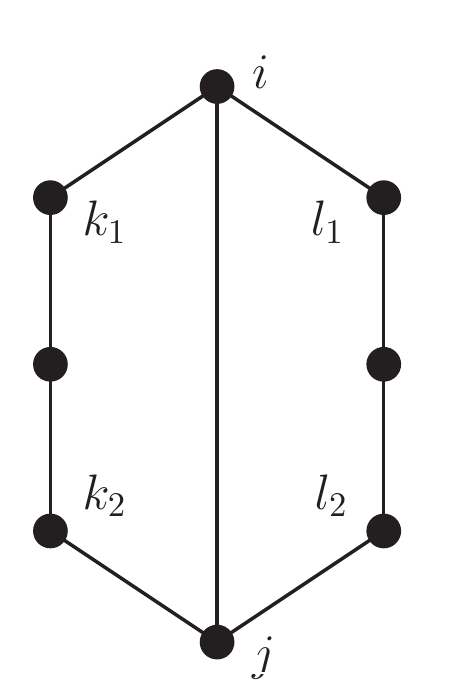} &
\includegraphics[width = 2.5 cm]{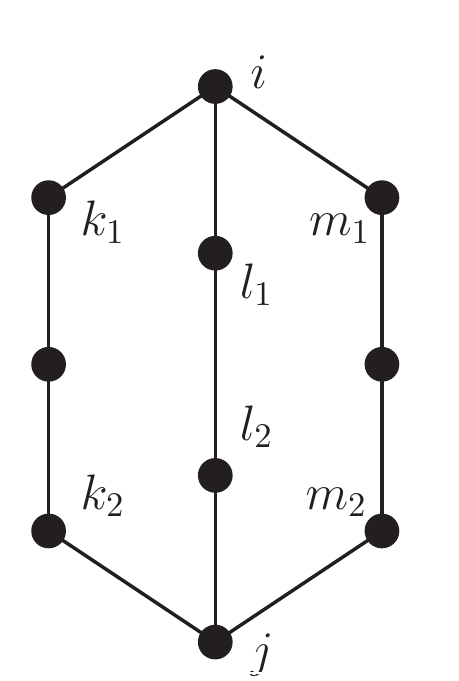} &
\includegraphics[width = 2.5 cm]{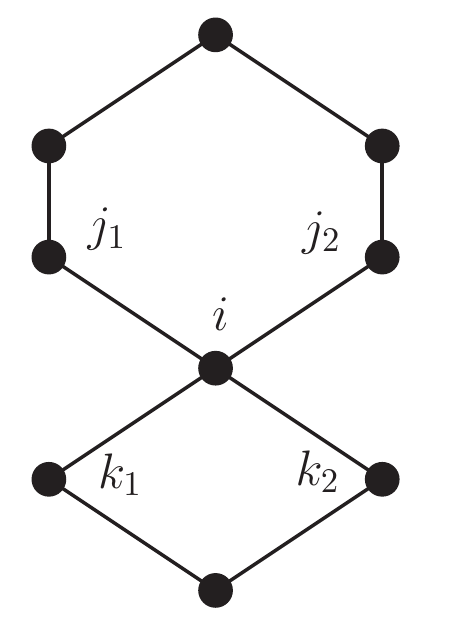}
\\
$(a)$ & $(b)$ & $(c)$ & $(d)$ & $(e)$
\\
\includegraphics[width = 2.5 cm]{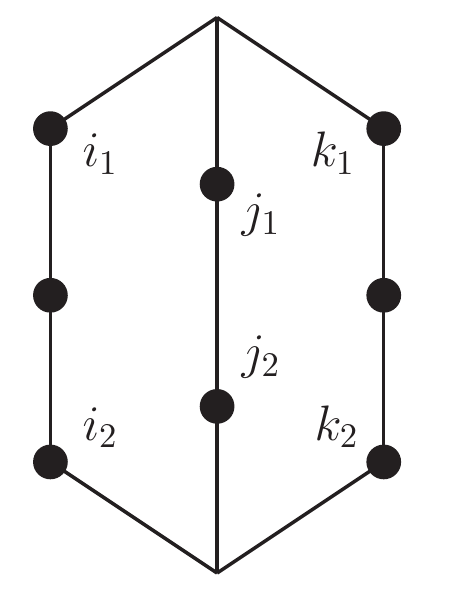} &
\includegraphics[width = 2.5 cm]{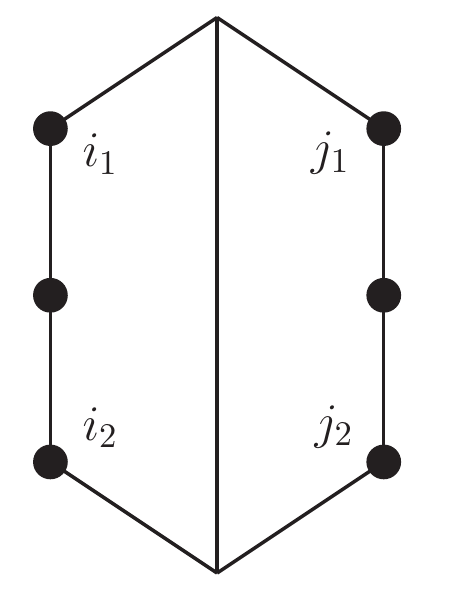} &
\includegraphics[width = 2.5 cm]{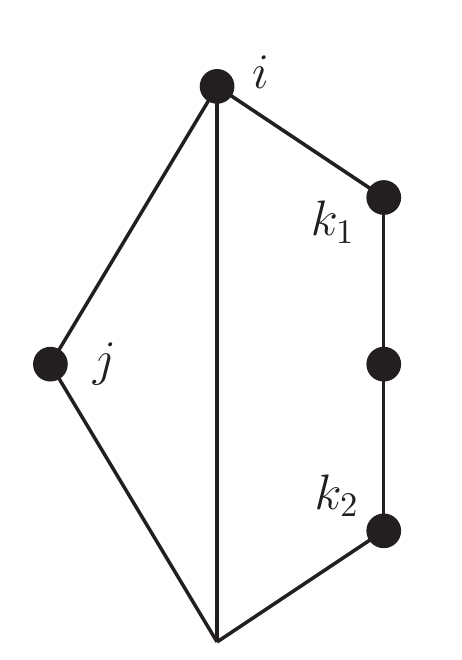} &
\includegraphics[width = 2.5 cm]{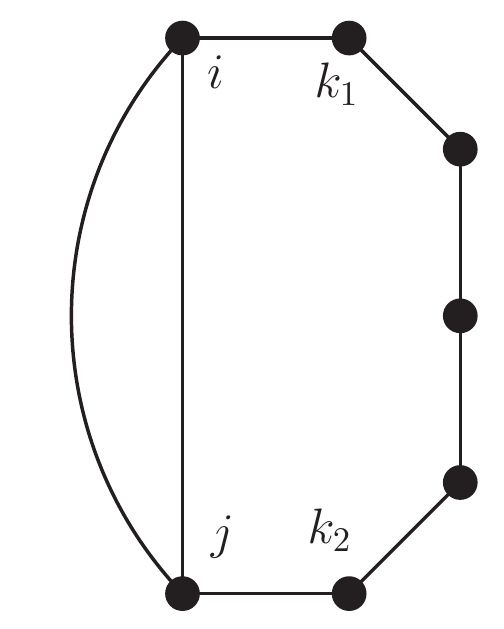} &
\includegraphics[width = 2.5 cm]{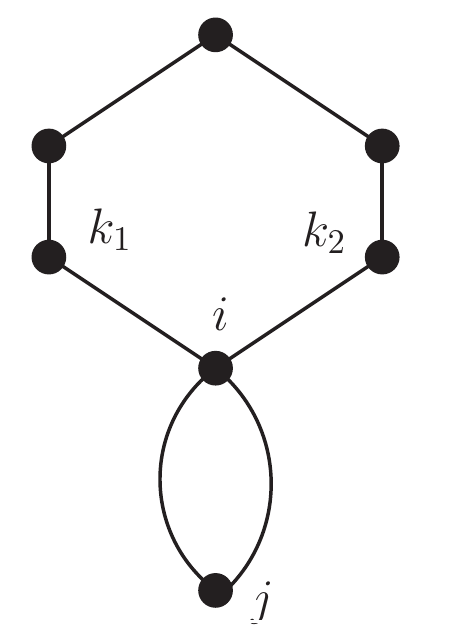}
\\
$(f)$ & $(g)$ & $(h)$ & $(p)$ & $(q)$
\end{tabular}
\end{center}
\caption{Harmonic supergraphs contributing to the $n$-point NMHV-like Born-level correlator 
in the Lorenz-covariant gauge.}\label{nNMHVLCG}
\end{figure}

The NMHV-like Born-level correlators are nontrivial quantities starting with $n \geq 5$ points.
In the harmonic supergraph approach the case of the six-point correlator can be viewed as generic.
Adding new external points to the six-point correlator 
but staying at NMHV level corresponds to the insertion of spectator scalar propagators $d_{ij}$ in the harmonic supergraphs,  
which modify the result rather trivially. We have already observed this phenomenon in the light-cone gauge 
calculation. The topologies contributing to the $n$-point correlator are depicted in Fig.~\ref{nNMHVLCG}. 
The corresponding expressions are rather similar to the four-point case. 
They are constructed out of the three- and four-point $Q$-supersymmetry invariants, eq.~\p{ACsq}.

Let us first consider the topologies that produce higher order poles $x_{i j}^{-4}$.
The  cancellation mechanism is the same as in the four-point correlator  example. The relevant diagrams are obtained from the four-point ones by inserting spectator propagators. 
The relevant topologies are $(h)$, $(p)$, $(q)$; $(f)$ at $i_1=i_2,\,j_1=j_2$;
and $(g)$ with $i_1$, $i_2$ being adjacent, Fig.~\ref{nNMHVLCG}. 
Partially summing the diagrams we obtain an analog of eq.~\p{sum4pt}, 
\begin{align}
&(T_{f})_{i_1=i_2,j_1=j_2,k_1 \neq k_2} + (T_{g})_{i_1 , i_2 - \text{adjacent}} + T_{(h)} + T_{(p)} + T_{(q)} + \text{perm} \nt 
& = x_{ij}^2 d_{ij}^2\, d_{i k_1} d_{i k_2} d_{k_1 \ldots k_2} x_{j k_1}^{-2} x_{j k_2}^{-2}\, p(i,j,k_1,k_2) \label{nosing}
\end{align} 
with the polynomial $p$ from eq.~\p{ppol}. Here we sum over all permutations containing the pole $x_{ij}^{-4}$. 
We see that the singularity on the right-hand side of eq.~\p{nosing} is softened.

The remaining topologies, $(a)-(e)$, $(f)$ at $j_1 \neq j_2$, $k_1 \neq k_2$, and $(g)$
with neither $i_1, i_2$ nor $j_1, j_2$ adjacent, have the correct singularity structure. 
Although they contain only one-loop integrals of the star-triangle type, eq.~\p{strtriang}, 
they involve bulky harmonic integrations.
The result of the diagram calculation is as follows:
\begin{align}
&T_{(a)} = - 2 \times d_{i j_1} d_{i k_1} d_{i l_1} d_{j_2 k_2} d_{j_2 l_2} d_{j_1 \ldots j_2} d_{k_1 \ldots k_2} 
d_{l_1 \ldots l_2} x_{k_2 l_2}^{-2} \nt & \times
A^{(\delta_1\delta_2)}_{j_2;k_2 l_2} (x_{j_2 k_2 l_2})_{\delta_1\delta_2} 
C_{i;j_1 k_1 l_1}^{\left(\substack{\a_1\b_1\gamma_1\\ \a_2\b_2\gamma_2}\right)}
\text{Sym} \, \ep_{\a_1\b_1}\ep_{\a_1\gamma_1}\ep_{\b_2\gamma_2} \nt
&T_{(b)} = -\frac{18}{5} \times d_{i j_1} d_{i k_1} d_{i j_2} d_{i k_2} d_{j_1 \ldots j_2} d_{k_1 \ldots k_2} x_{j_2 k_2}^{-2}
A^{(\a_3\a_4)}_{i;j_2 k_2} C_{i;k_2 j_1 k_1}^{\left(\substack{\a_1\b_1\gamma_1 \\ \a_2\b_2\gamma_2 }\right)} 
(x_{j_2 k_2 i})^{\delta_1\delta_2} \nt
&\times \text{Sym} \, \ep_{\a_1\delta_1}\ep_{\a_2\delta_2}\ep_{\a_3\b_1}\ep_{\a_4\gamma_1}\ep_{\b_2\gamma_2} \nt
&T_{(c)} = - \frac{6}{5} \times d_{ij} d_{i k_1} d_{i l_1} d_{j k_2}  d_{j l_2} d_{k_1 \ldots k_2} d_{l_1 \ldots l_2} 
C_{i;j k_1 l_1}^{\left(\substack{\a_1\b_1\gamma_1\\ \a_2\b_2\gamma_2 }\right)} 
C_{j;i k_2 l_2}^{\left(\substack{\a_3\delta_1\varepsilon_1\\ \a_4\delta_2\varepsilon_2 }\right)} \nt
&\times  \text{Sym} \,\ep_{\a_3\b_2}\ep_{\b_1\gamma_1}
\Bigl[ \ep_{\a_1\varepsilon_1}\ep_{\a_2\varepsilon_2}\ep_{\a_4\delta_2}\ep_{\gamma_2\delta_1} 
-  \ep_{\a_1\delta_1}\ep_{\a_2\delta_2}\ep_{\a_4\varepsilon_2}\ep_{\gamma_2\varepsilon_1} \Bigr] \nt
&T_{(d)} = \frac{2}{3} \times d_{i k_1} d_{i l_1} d_{i m_1} d_{j k_2} d_{j l_2} d_{j m_2} 
d_{k_1 \ldots k_2} d_{l_1 \ldots l_2} d_{m_1 \ldots m_2}
C_{i;k_1 l_1 m_1}^{\left(\substack{\a_1\b_1\gamma_1\\ \a_2\b_2\gamma_2}\right)} 
C_{j;k_2 l_2 m_2}^{\left(\substack{\delta_1\varepsilon_1\zeta_1\\ \delta_2\varepsilon_2\zeta_2}\right)} \nt
&\times \text{Sym} \,\ep_{\a_1\b_1}\ep_{\a_2\gamma_1}\ep_{\b_2\gamma_2} 
\text{Sym} \,\ep_{\delta_1\varepsilon_1}\ep_{\delta_2\zeta_1}\ep_{\varepsilon_2\zeta_2} \nt
&T_{(e)} = -\frac{12}{7} \times d_{i j_1} d_{i j_2} d_{i k_1} d_{i k_2} d_{j_1 \ldots j_2} d_{k_1 \ldots k_2}
C_{i;k_1 k_2 j_2}^{\left(\substack{\a_1\b_1\gamma_1 \\ \a_2\b_2\gamma_2}\right)} 
C_{i;k_1 j_2 j_1}^{\left(\substack{\a_3\gamma_3\delta_1 \\ \a_4\gamma_4\delta_2}\right)} \nt
&\times \text{Sym}\, \ep_{\a_1\gamma_1}\ep_{\a_2\gamma_2} \Bigl[ 
\frac{1}{15} \ep_{\a_3\gamma_3}\ep_{\a_4\gamma_4}\ep_{\b_1\delta_1}\ep_{\b_2\delta_2} 
+ \ep_{\a_3\b_1}\ep_{\a_4\delta_1}\ep_{\b_2\gamma_3}\ep_{\gamma_4\delta_2} \Bigr] \nt
&T_{(f)}= 6 \times d_{i_1 j_1} d_{i_1 k_1} d_{i_2 j_2} d_{i_2 k_2} d_{i_1 \ldots i_2} 
d_{j_1 \ldots j_2} d_{k_1 \ldots k_2} 
x_{j_1 k_1}^{-2} x_{j_2 k_2}^{-2} A^{(\a_1\a_2)}_{i_1;j_1 k_1} (x_{i_1 j_1 k_1})_{\a_1\a_2}\times \nt
&\hspace{1.2cm} A^{(\b_1\b_2)}_{i_2;j_2 k_2} (x_{i_2 j_2 k_2})_{\b_1\b_2}\nt
&T_{(g)} = -6 \times d_{i_1 i_2} d_{j_1 j_2} d_{i_1 j_1} d_{i_1 \ldots i_2} d_{j_1 \ldots j_2}
x_{i_2 j_2}^{-2} x_{i_1 j_2}^{-2} x_{i_2 j_1}^{-2} A^{(\a_1\a_2)}_{i_1;j_1 i_2} A^{(\a_3\a_4)}_{j_1;i_1 j_2} 
\text{Sym} \, \ep_{\a_1\b_1}\ep_{\a_2\b_2}\ep_{\a_3\b_3}\ep_{\a_4\b_4} \nt
&\times \Bigl[ x_{i_1 i_2}^2 (x_{i_1 j_2 j_1})_{\b_1\b_2} (x_{j_2 j_1 i_2})_{\b_3\b_4} 
+ x_{j_1 j_2}^2 (x_{i_1 i_2 j_2})_{\b_1\b_3} (x_{j_1 i_1 i_2})_{\b_2\b_4} \Bigr]\,. \label{LCGres}
\end{align}
Summing these expressions over the nonequivalent permutations along with 
the simplified sum from eq.~\p{nosing} we obtain the $n$-point NMHV Born-level correlator.

This expression is to be compared with the result of the light-cone gauge calculation.
Both are $Q$- and R-symmetric. Eq. \p{LCGres} is more complicated than the 
$R$-vertex expression, eq.~\p{Rinv}. This is the price for explicit Lorentz invariance.
The simplicity of eq.~\p{Rinv} is due to the auxiliary harmonic $\xi^{\+}$ 
which breaks Lorentz invariance and creates spurious poles.
The new result \p{LCGres}   has the correct singularity structure.
The correlator as a whole is superconformal,
but the symmetry is restored only in the sum of all diagrams.

Although the results of the calculations in the two gauges look rather different, they do coincide. 
We checked it numerically for the five-point and six-point NMHV correlators.

In the light-cone gauge we can construct the correlator at any non-MHV level in the Born approximation 
as a sum of products of trivalent $R$-vertices, eq.~\p{Rinv}. The calculation is purely algebraic.
We cannot proceed so easily to higher non-MHV levels in the Landau gauge. 
There are several obstacles. Firstly, at the N$^{k}$MHV level we have to deal with $2k$-loop integrals.
Some of them are very complicated, but the Born-level correlator is always rational. 
Secondly, the harmonic integrations become very involved.
Although there is a systematic procedure to their evaluation, the result can be rather messy.



\end{document}